\RequirePackage{fix-cm}
\documentclass[natbib,smallextended]{svjour3}       % onecolumn (second format)
\smartqed  % flush right qed marks, e.g. at end of proof
\usepackage{graphicx}
\usepackage{amsmath}
\usepackage{amssymb}
\usepackage[colorlinks=true,citecolor=blue,urlcolor=blue,breaklinks]{hyperref}

\defcitealias{Chiappini1997}{CMG97} %FS: not sure we should use these
\defcitealias{Grisoni2017}{G17}
\defcitealias{Tinsley1980}{T80}
\journalname{The Astronomy and Astrophysics Review}
\begin{document}

\title{Modelling the chemical evolution of the Milky Way}

%\titlerunning{Short form of title}        % if too long for running head

\author{Francesca Matteucci}

%\authorrunning{F. Matteucci} % if too long for running head

\institute{Francesca Matteucci \at
              Department of Physics, Trieste University, Trieste, Italy\\
              \email{francesca.matteucci@inaf.it}
}

\date{Received: date / Accepted: date}

% The correct dates will be entered by the editor

\maketitle

\begin{abstract}
In this review I will discuss the comparison between model results and observational data for the Milky Way, the predictive power of such models as well as their limits. Such a comparison, known as Galactic archaeology, allows us to impose constraints on stellar nucleosynthesis and timescales of formation of the various Galactic components (halo, bulge, thick disk and thin disk).
\end{abstract}

\setcounter{tocdepth}{3} % TOC subsubsections
\tableofcontents

\section{Introduction}
\label{sec:intro}

Our Galaxy, the Milky Way, is by far the best studied stellar system. At the present time there is a large flood of data coming from large spectroscopic Galactic surveys, such as Gaia-ESO Public Spectroscopic Survey (GES, \citealt{Gilmore2012}), the Apache Point Observatory Galactic Evolution Experiment (APOGEE, \citealt{Majewski2017}) and the GALactic Archaeology with HERMES (GALAH, \citealt{desilva2015}), providing data for thousands of stars in all Milky Way stellar components: halo, thick disk, thin disk and bulge. At the same time, a great deal of chemical evolution models, trying to reproduce the observed abundance patterns,  have appeared. Galactic chemical evolution tries to explain how the chemical elements have formed and their abundances have evolved in time and distributed in space. It is well known that during the Big Bang only light elements (H, D, $^{3}$He, $^{4}$He, $^{7}$Li) were formed, whereas all the elements from $^{12}$C to uranium and beyond have been formed inside stars. In particular, chemical evolution models aim at following the evolution of the gas and its chemical composition inside galaxies. Such an evolution depends on the star formation history, stellar nucleosynthesis and possible gas flows. The very first seminal paper on chemical evolution is from \cite{Schmidt1963}, followed by \cite{LyndenBell1975}, \cite{Pagel1975}, and \cite{Tinsley1980}.
The simplest model for the chemical evolution is called the \emph{Simple Model} because of its basic (although unrealistic) assumptions, such as the fact that the studied system behaves as a closed-box. The solution of this model can be obtained analytically  if the stellar lifetimes are neglected, and provides a relation between the global gas metallicity $Z$ (all the elements heavier than He) and the fraction of gas in the studied system. The first pioneering chemical papers from \cite{LyndenBell1975} and \cite{Pagel1975} discussed how to solve the  \emph{G-dwarf problem}, namely the excess of low metallicity stars, as predicted by the Simple Model, when compared to observations. The G-dwarfs are those in the solar vicinity, a cylinder centered in the Sun with 1 kpc radius, and belong to the thin disk. 
The most accepted solution to the G-dwarf problem consists in the fact that the solar vicinity should not have evolved as a closed-box, but rather suffered gas flows, in particular gas infall.
\cite{Tinsley1980} produced a seminal review paper on galactic chemical evolution which represents the foundation of this field: in that paper, she summarizes the basic ingredients and the basic equations to build  chemical evolution models.
Since these fundamental papers, a great deal of work has appeared on this subject and in particular concerning the Milky Way. Many models solving numerically the basic equations of chemical evolution appeared, and in this way it was possible to relax the so-called instantaneous recycling approximation (I.R.A.), which neglects the stellar lifetimes by assuming that stars of masses $M \ge 1\,M_{\odot}$ die instantaneously, while those with $M< 1\,M_{\odot}$ live forever. In fact, by relaxing I.R.A.\ we can follow the evolution of the abundance of every single chemical species. This is very important, since different chemical elements are produced on different timescales by stars of different masses. Among the first numerical models relaxing I.R.A., we recall those of \cite{Chiosi1980} and \cite{Tosi1982}. Generally, in such models, it is possible to follow the chemical enrichment by different supernova (SN) types and by single stars ending their lives as white dwarfs (WDs). The first attempt to include the chemical enrichment by Type Ia SNe (assumed to be WDs in binary systems) was by \cite{Greggio1983, MatteucciT1985} and \cite{MatteucciGreggio1986}: in particular, in the latter two papers the SNIa rate was included in a detailed chemical evolution model relaxing I.R.A., and following the evolution of several chemical species, such as He, C, N, O, $\alpha$-elements and Fe. This allowed the authors to predict the evolution of several [$\alpha$/Fe]\footnote{(${\rm [X/Fe]} = \log(X/H)_* -\log (X/H)_{\odot}$ for any generic X element)} ratios ($\alpha$=O, Si, Mg) resulting positive and around $\sim + 0.4$\,dex in halo stars (with ${\rm [Fe/H]} < -1.0$\,dex), and decreasing in disk stars until reaching the solar value ([Fe/H]=0 dex). At that time,  there were data only for [O/Fe] in halo stars, and so the results for Si and Mg were just predictions. Those predictions were then confirmed by \cite{Francois1986}, who found indeed overabundances of Si and Mg relative to Fe  in halo stars. The interpretation of the behaviour of the [$\alpha$/Fe] ratios was that in the early phases of Galactic evolution only $\alpha$-elements are produced by short living massive stars (core-collapse SNe, hereafter CC-SNe), thus reflecting the production ratios of these elements by massive stars, then when supernovae Ia (SNeIa) start to die, with a time delay relative to CC-SNe, and produce the bulk of Fe, the [$\alpha$/Fe] ratios decrease, during the disk phase, until they reach the solar value. This interpretation is now known as \emph{time-delay model}.

In the late 1980s and 1990s other papers appeared and they were dealing with the chemical evolution of the Milky Way by means of numerical models (\citealt{Tosi1988a, Matteucci1989, Matteucci1992, Prantzos1993, Chiappini1997, Boissier1999}; among others). It is worth reminding also a review paper by \cite{Matteucci1996} and two monographs by \cite{Pagel1997} and \cite{Matteucci2001} on chemical evolution.
Most of the models in the 1980s and 1990s were assuming that the Galactic disk formed by infall of gas, and in order to reproduce the G-dwarf metallicity distribution they all suggested infall timescales larger than 3--4\,Gyr at the solar vicinity.
\cite{Matteucci1989} computed a model in which  the disk of the Milky Way formed \emph{inside out}, namely on much shorter timescales in the inner than the outer regions, in order to reproduce the negative abundance gradients found along the thin disk. Such a mechanism has been found later also in complex cosmological simulations of galaxy formation (\citealt{Samland2003, Kobayashi2011, Minchev2014, Grand2017, dominguez2017, Vincenzo2020}). The inside-out disk formation was originally suggested by \cite{Larson1976} by means of chemo-dynamical models: he suggested that the disk forms by gas accretion occurring faster in the inner than in outer regions.

\cite{Chiappini1997} proposed the so-called ``two-infall'' model, where the Galaxy formed by means of two major episodes of gas infall, the first giving rise to the stellar halo and thick disk, while the second forming the thin disk on a much longer timescale (7--8 Gyr in the solar neighbourhood); moreover, the model predicts a gap in the star formation due to an assumed gas threshold, as suggested by \cite{Kennicutt1989}.  The two-infall scenario was inspired by the [$\alpha$/Fe] vs.\ [Fe/H] relation observed in solar vicinity stars by \cite{Gratton1996, Gratton2000}, and subsequently confirmed by  \cite{Fuhrmann1998}. In such studies, a gap appeared in the star formation process between the formation of the halo and disk stars. The long infall timescale for the assembly of the thin disk at the solar ring, was later confirmed by subsequent papers such as \cite{Boissier1999, Chiappini2001} and \cite{Alibes2001}. This long timescale contrasted with the conclusions of the seminal paper of \cite{Eggen1962}, where the authors inferred, from the motions and chemical composition of stars, the mechanism of the formation of the Milky Way and suggested that it happened by gas collapse on a timescale no longer than 300 Myr. However, the more precise data and in particular the G-dwarf metallicity distribution appeared in the 1980s, 1990s and early 2000s did not allow us to suggest such a short collapse timescale. On the other hand, \cite{Searle1978} had already challenged the short timescale of \cite{Eggen1962} for the formation of the outer stellar halo,  by suggesting that it should have assembled on a timescale larger than the inner halo. Moreover, the timescale of 300 Myr is too short even for the formation of the inner halo-thick disk, which appears to be rather $\sim$1 Gyr as suggested by \cite{MatteucciGreggio1986} and \cite{Chiappini1997}.
In recent times, it has been possible to distinguish not only among halo and disk stars but also among thick and thin disk stars. The existence of a structure called ``thick disk'' was suggested by \cite{Yoshii1982} and \cite{Gilmore1983}, who found two distinct stellar density distributions both exponential and related to the thick and thin disk, respectively. The stars in the thick disk are characterized by high [$\alpha$/Fe] ratios, similar to halo stars, but kinematically separated either from the halo or the thin disk stars. Models including the thick disk explicitely (in \citealt{Chiappini1997} model the thick disk was considered together with the halo) have appeared such as that of \cite{Pardi1995} and then  \cite{Micali2013}; these latter suggested a ``three-infall model'' with the halo forming on a time scale of $\sim$0.2 Gyr, the thick disk on $\sim$ 1.25 Gyr and the thin disk at the solar vicinity on $\sim$6 Gyr. In this model and all of the previous ones,  the star formation rate (SFR) in halo and thick disk is assumed to be more intense than in thin disk (for an extensive discussion of the models of the 2000s, see \citealt{Prantzos2008a} and \citealt{Matteucci2012}).
Other approaches to the chemical evolution of the Milky Way halo, such as the stochastic inhomogeneous mixing and the accretion of extant stellar systems will be discussed in Sect.~\ref{sec:history}.

More recently, it has been evident that the [$\alpha$/Fe] vs.\ [Fe/H] diagram shows a clear bimodal distribution between stars with high and low [$\alpha$/Fe] ratios; the former are attributed to the thick disk while the latter to the thin disk. Such a bimodality is particularly evident in APOGEE data (\citealt{Anders2014, Nidever2014, Hayden2015, Queiroz2020}).
By analyzing the High Accuracy Radial velocity Planet Searcher (HARPS) spectra of local solar twin stars, \cite{Nissen2020} found that the age-metallicity distribution has two distinct populations with a clear age dissection. The authors suggested that these two sequences may be interpreted as evidence of two episodes of accretion of gas onto the Galactic disk with quenching of star formation in between them. 
In order to explain that, the two-infall model has been applied to the thick and thin disk. In this scenario, the thick disk forms first by accretion of primordial gas with a high star formation efficiency, whereas the thin disk forms with a time delay, due to the stop in the star formation induced by the threshold in the gas density, on a much longer timescale and from accretion of fresh primordial gas together with the gas leftover from the thick disk formation (\citealt{Grisoni2017}; \citealt{Spitoni2019, Spitoni2020}). In particular, in \cite{Grisoni2017}, aiming at reproducing AMBRE (Arch\'eologie avec Matisse Bos\'ee sur les aRchives de l'ESO) data (\citealt{delaverny2013}), the gap in the star formation between the assembly of the two disks occurs on a timescale no longer than 1 Gyr, whereas in \cite{Spitoni2019}, aimed at reproducing the chemical and age data of \cite{Silva2018}, a gap of $\sim$ 4.3 Gyr is suggested. On the other hand, \cite{Grisoni2017} presented also a parallel scenario, where the two Galactic components formed and evolved in parallel in a completely independent way. In both the approaches presented by \cite{Grisoni2017} (two-infall and parallel), the thick disk formed on a timescale of $\sim$0.1 Gyr and the thin disk at the solar ring of $\sim$7 Gyr.

Cosmological models  have also suggested infall of gas as a solution to explain the observed bimodality in thick and thin disk stars (e.g., \citealt{Calura2009, Buck2020}). However, it is worth noting that the existing differences in data of different surveys may indeed lead to different conclusions. What appears as a common feature to all chemical models is that the thick disk should have formed quickly and with intense star formation, in order to produce $\alpha$-enhanced stars.

Concerning the formation and evolution of the stellar halo, we remind the pioneering work of \cite{Hartwick1976}, who studied the metallicity distribution function (MDF) of globular clusters (GCs) and found that is different from the G-dwarf metallicity distribution, since it contains more metal poor objects. He devised a simple model for the Galactic halo to reproduce the MDF for GCs, where removal of gas from the halo and consequent formation of the disk was assumed. In the following years, it was suggested that the inner and outer halo might have formed on different timescales (\citealt{Searle1978}) and that part of the halo could have formed by accretion of stars from small satellites of the Galaxy such as dwarf sheroidals (dSphs)  or dwarf irregulars (DIs)  or Ultra faint Dwarfs (UfDs). In particular, the accretion of satellite galaxies is predicted by the $\Lambda$CDM paradigm, assuming that dark matter halos form hierarchically via a series of mergers with smaller halos. As a consequence, it is expected that the stellar Galactic halo might have formed from disrupted and accreted satellites (\citealt{Johnston1996, HelmiWhite1999, Bullock2001, Font2005, Robertson2005, Helmi2008}, amongst others). The comparison between abundances and abundance patterns between Galactic halo stars and stars in dwarf satellites can help in understanding the formation of the halo (see \citealt{Geisler2007}, \citealt{Matteucci2012, Spitoni2016}).  In particular, all halo stars show overabundances of $\alpha$-elements relative to Fe, while stars in dSphs and UfDs present a small fraction of $\alpha$-enhanced stars and many stars with low [$\alpha$/Fe] ratios at low metallicity (e.g., \cite{Tolstoy2009}). Therefore, the situation is still unclear. Very recently, \cite{Helmi2018} have suggested that the inner Galactic halo is dominated by debris from an object that infalled 10 Gyr ago and it was as large as the Small Magellanic Cloud (SMC) and they called this object Gaia-Enceladus. This conclusion was deduced from the chemical and dynamical analysis of two large stellar surveys (Gaia-ESO and APOGEE). These studies hinted that this accretion event might have been responsible also for the formation of the thick disk.

The Galactic bulge is made mainly by old stars with a large range of metallicities, but with lack of very metal poor objects. Its characteristics are intermediate between a ``classical bulge'' and a ``pseudo-bulge''. There is evidence that in the bulge there are at least two stellar populations (e.g.,  \citealt{Hill2011, Zoccali2017}), one showing the characteristics of a classical bulge and the other compatible with secular evolution of the inner disk through the formation of a bar and  boxy/peanut structure. Other studies (\citealt{Bensby2011, Bensby2013, Bensby2017, Ness2016}) suggested even multiple stellar populations in the bulge. 

The first detailed chemical evolution model of the Galactic bulge was by \cite{Matteucci1990}, who predicted that the bulk of bulge stars should have formed quickly during a  burst of star formation, and should exhibit [$\alpha$/Fe]$>0$\,dex for a large range of metallicities, as expected from the time-delay model, a result later confirmed by the first very detailed abundances derived by \cite{McWilliam1994}. In the following years, a great deal of chemical models appeared and they all confirmed, after comparison with the data, the fast formation of the majority of bulge stars (\citealt{Ballero2007, Cescutti2011, Grieco2012b, Matteucci2019}, among others). Chemo-dynamical models based on accretion of substructures in the framework of $\Lambda$CDM, have also concluded that a fast bulge formation should be preferred (\citealt{Samland2003, Immeli2004, Kobayashi2011}). In particular, \cite{Samland2003} predicted that the bulge should contain two stellar populations, an old one formed during the proto-Galactic collapse and a young bar population.
Other, dynamical simulations have suggested that the old bulge population originated from the thick disk stars and those belonging to the peanut structure (the bar population) from the thin disk (e.g.\, \citealt{dimatteo2015, Bekki2011}).
Abundance data for stars in the very inner bulge have also become available
(\citealt{Ryde2016}) and the derived abundance patterns seem to confirm a fast formation with intense SFR of this part of the bulge, when compared to theoretical models.
Finally, we recall a recent and exhaustive review on cosmic chemical evolution by \cite{Maiolino2019}.

In this review, in Sect.~\ref{sec:ingr} we will remind the basic assumptions and the main ingredients necessary to build chemical evolution models. In Sect.~\ref{sec:anal}, we will present analytical solutions for chemical evolution models.
In Sect.~\ref{sec:numer}, we will discuss the basic equations of numerical models for the chemical evolution of the Milky Way. Section~\ref{sec:history} will be dedicated to the description of historical model approaches. Section~\ref{sec:heavyhd} will present the results of numerical models for the evolution of heavy elements in halo and disk. Section~\ref{sec:abgrad} will contain results on abundance and abundance ratio gradients along the thin disk. In Sect.~\ref{sec:migr} we will discuss stellar migration and in Sect.~\ref{sec:light} the evolution of light elements. Section~\ref{sec:bulge} will present results for the Galactic bulge, while Sect.~\ref{sec:halo} will describe the results for the Galactic stellar halo. Section~\ref{sec:cosmo} will be dedicated to the chemo-dynamical models of the Milky Way in a cosmological context and comparison with pure chemical models. Finally, in Sect.~\ref{sec:concl} we will present a discussion and conclusions.

%\section{Observational constraints}
%Here we will summarize the most important observational constraints for the main Galactic components.%
%\subsection{Galactic halo}

%\subsection{Galactic thick disk}

%\subsection{Galactic thin disk}

%\subsection{Galactic bulge}

\section{Basic ingredients to build chemical models}
\label{sec:ingr}

The basic ingredients necessary to build a chemical evolution model are: 

\subsection{Initial conditions}

As initial conditions we can assume an open or closed model, namely if the total mass of the system is constant or variable in time. In other words, we can assume that all the gas, out of which stars will form, is present at the time $t=0$, or that it will be accreted in time.
Moreover, we should assume whether the initial gas has a primordial (only light elements from the Big Bang) or metal enriched chemical composition. This latter case is known as Prompt Initial Enrichment (PIE), and it can be created by an initial generation of massive zero metal stars (Population III stars).

\subsection{Stellar birthrate function (SFR\,x\,IMF)}

The stellar birthrate function is the history of star formation in a galaxy and it can be expressed as the product of the SFR times the initial mass function (IMF). In other words, the stellar birthrate function, namely the number of stars formed in the time interval, $(t, t+dt)$ and in the mass interval $(m, m+dm)$, can be written as:
\begin{equation}
B(m,t)dm\,dt= \psi(t) \cdot \phi(m)dm\,dt,
\label{eq:birthrate}
\end{equation}
where the function $\psi(t)$ represents the SFR and it is generally assumed to be only a function of time, whereas $\phi(m)$ is the IMF which is assumed to be only a function of mass. The SFR represents how many solar masses go into stars per unit time, while the IMF describes the distribution of stars at birth as a function of stellar mass. Clearly, these hypotheses are semplifications and we do not know whether the SFR is independent of mass and the IMF independent of time. Besides that, there is a sort of indetermination principle in the definition of $B(m,t)$, since in order to know the SFR we need to assume an IMF, and viceversa.

\subsubsection{Parametrization of the SFR}

The most common parametrization is the \cite{Schmidt1959} law, where the SFR is proportional to some power $k$ of the gas volume density. \cite{Kennicutt1998} suggested a star formation law depending on the surface gas density, as deduced by data relative to local star forming galaxies.
Other important parameters such as gas temperature, viscosity and magnetic field are usually ignored.

In general, the SFR can be written as:
\begin{equation}
\psi(t)= \nu \sigma_{\rm gas}(t)^{k},
\label{eq:SFR}
\end{equation}
where $\sigma_{\rm gas}^{k}$ is the gas surface mass density and $\nu$ is the efficiency of star formation, namely the SFR per unit mass of gas, and is expressed in units of $t^{-1}$.

In particular, the formula suggested by Kennicutt is:
\begin{equation}
  \psi(t)=(2.5 \pm 0.7) \cdot 10^{-4 } [\sigma_{\rm gas}(t)]^{1.4 \pm 0.15} \,M_{\odot} \, {\rm yr^{-1} kpc^{-2}},
\label{eq:SFRKenn}
\end{equation}
where the efficiency parameter is derived from the fit to the SFR of local star forming galaxies.

\begin{figure}[htbp]
% Use the relevant command to insert your figure file.
  % For example, with the graphicx package use
  \centering
  \includegraphics[width=0.9\textwidth]{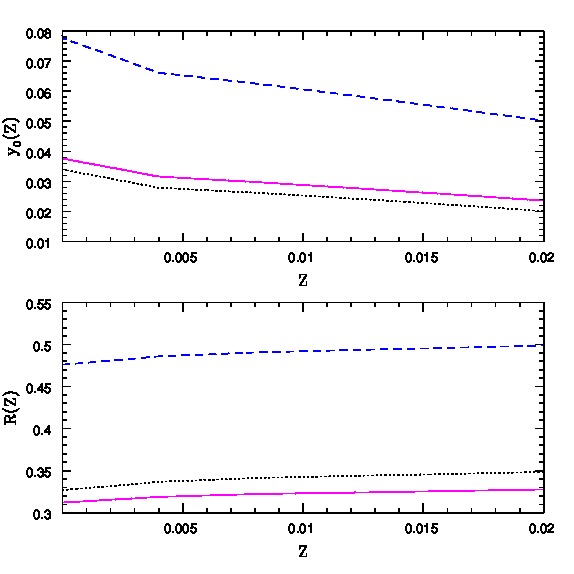}
% figure caption is below the figure
  \caption{{\it Upper panel}: the yield  of oxygen per stellar generation  computed for different metallicities and IMFs. The blue dotted line refers to \cite{Chabrier2003} IMF, the magenta line is the \cite{Salpeter1955} IMF and the black line is the \cite{Kroupa1993} IMF. {\it Lower panel}: returned fraction R as a function of IMF and metallicity. Image reproduced with permission from \cite{Vincenzo2016}, copyright by the authors.}
\label{fig:fiore}       % Give a unique label
\end{figure}

\begin{figure}[htbp]
  \centering
  \includegraphics[width=0.8\textwidth]{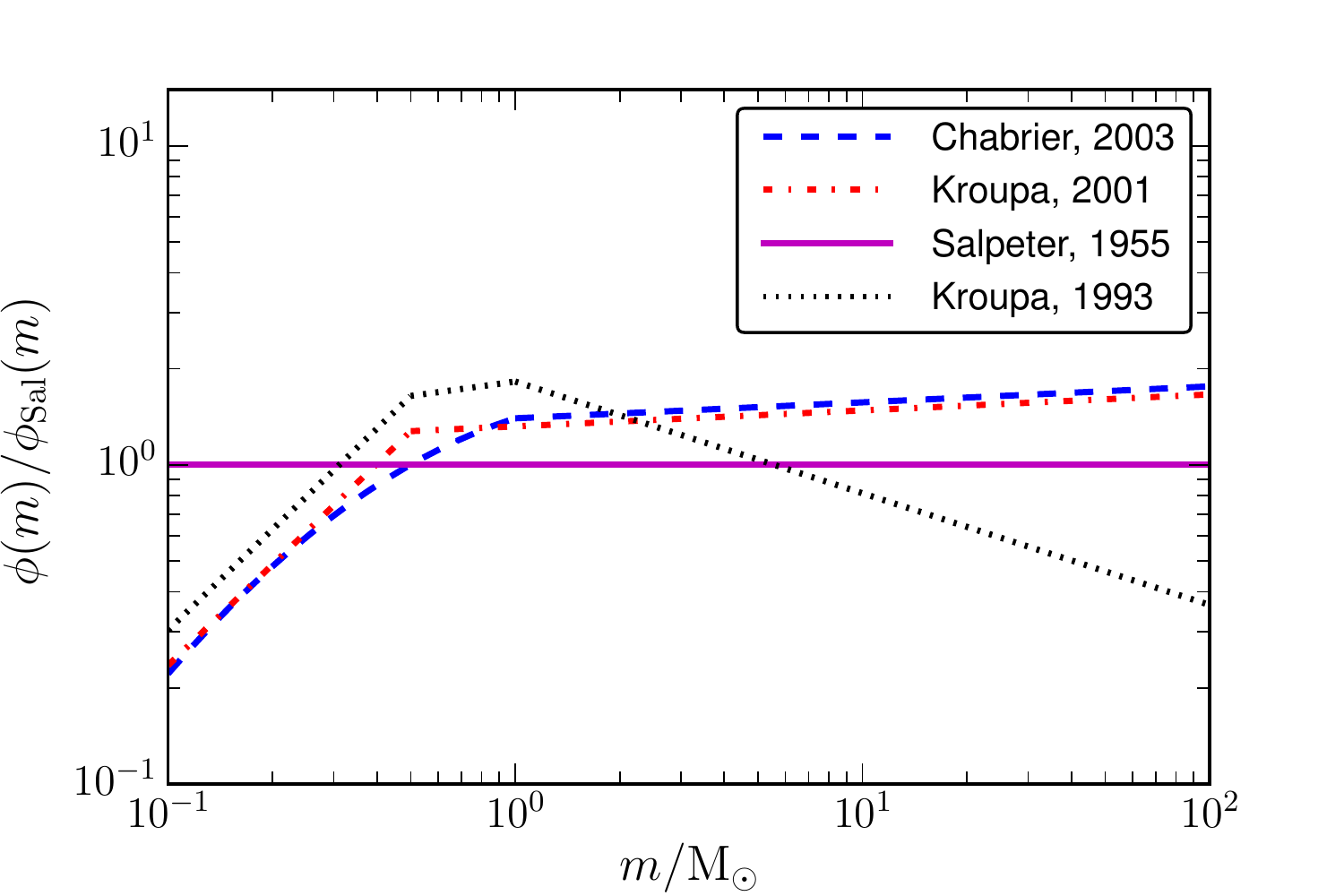}
  \caption{Comparison between the different IMFs adopted in Fig.~\ref{fig:fiore}, normalized to the Salpeter IMF. Image reproduced with permission from \cite{Vincenzo2016}, copyright by the authors.}
\label{fig:imf}       % Give a unique label
\end{figure}

%\begin{figure}[htbp][h]
%\includegraphics[width=14pc]{fiore.jpg}\hspace{2pc}%
%\begin{minipage}[b]{14pc}\caption{Upper panel: the yield  of oxygen per stellar generation  computed for different metallicities and IMFs
%The blue dotted line refers to Chabrier (2003) IMF, the magenta line is the Salpeter (1955%) IMF and the black line is the Kroupa et al. (1993) IMF. Lower panel:returned fraction R as a function of IMF and metallicity. Figure from Vincenzo et al. (2015).}
%\end{minipage}
%\end{figure}

\subsubsection{Parametrization of the IMF}
The IMF, namely the number of stars born in the mass interval, $m-m+dm$, is generally expressed as a power law. It is possible to measure the IMF only in the solar vicinity, since one needs to count the stars as functions of their magnitudes and it is not yet possible to do this in external galaxies. Therefore, the only observational information we have is relative to the solar region of our Galaxy. 

A typical expression of the IMF is:
\begin{equation}
  \phi(m)dm=Cm^{(1+x)}dm,
  \label{eq:imf}
  \end{equation}
where C is the so-called normalization constant obtained from the following condition:
\begin{equation}
  \int^{100}_{0.1}{m \phi(m) dm} =1.
  \label{eq:normimf}
\end{equation}

The well-known \cite{Salpeter1955} IMF has $x=1.35$ over the whole range of stellar masses (0.1--100\,$M_{\odot}$) and is widely used in galactic evolutionary models. {It should be noted that also other mass ranges, including stars up to 120\,$M_{\odot}$ and down to 0.05\,$M_{\odot}$, have been adopted in \cite{Prantzos2018} and \cite{Kobayashi2020}, respectively. However, more realistic IMFs derived for the solar vicinity have two or three slopes in the whole range of masses, such as those in Figs.~\ref{fig:fiore} and \ref{fig:imf} (\citealt{Kroupa1993, Chabrier2003, Kroupa2001}).

\subsection {Stellar nucleosynthesis}

The chemical enrichment in galaxies is due to the masses of chemical elements produced by stars of different initial mass and restored to the interstellar medium (ISM) when they die. These masses represent both the newly formed elements and the elements already present in the star at its formation and restored into the ISM without being reprocessed.

In particular, we can define the mass of a newly formed element in a star of mass $m$ as:
\begin{equation}
  M_{im}=\int^{\tau_m}_0 {\dot M_{\rm lost} \cdot [X(i)-X_o(i)] dt},
  \label{eq:newyield}
  \end{equation}
where $\tau_m$ is the lifetime of a star of mass $m$, $X_o(i)$ is the original abundance of the element $i$, $X(i)$ is the final one and $\dot M_{\rm lost}$ is the mass loss rate of the star.

With this quantity we can define what is called \emph{stellar yield}, namely:
\begin{equation}
  p_{im}=M_{im}/m.
  \label{eq:yield}
  \end{equation}

To obtain the total stellar mass ejected, we  should add to Eq.~\eqref{eq:newyield} the mass ejected without being processed, namely:
\begin{equation}
  M_{imo} = X_o(i)\cdot M_{\rm lost},
\label{eq:oldyield}
\end{equation}
where $M_{imo}$ is the stellar mass in the form of the element $i$, already present in the star at birth. With $M_{\rm lost}$  we intend the total mass ejected by
the star into the ISM during its lifetime.

Each stellar mass can produce and eject different chemical elements and the yields are a function of the stellar mass but also of the original stellar metal content that we will indicate with $Z$. These yields are computed by means of detailed nucleosynthesis calculations taking into account all the main nuclear reactions in stars.

Here we summarize briefly the element production in stars:
i) stars with $M<0.8\,M_{\odot}$ do not contribute to galactic chemical enrichment since they have lifetimes longer than the Hubble time.
ii) Low and intermediate mass stars (0.8--8\,$M_{\odot}$) produce He, N, C and heavy s-process elements (e.g., Ba, Y, Sr). They die as C-O WDs when single, and can die as Type Ia SNe when in binaries. Type Ia SNe are, in fact, believed to originate in WDs in binary systems.
iii) Massive stars ($M>8$--$10\,M_{\odot}$) produce mainly alpha-elements (O, Ne, Mg, S, Si, Ca),  some Fe, light s-process elements and perhaps r-process elements and explode as CC-SNe. However, r-process elements originating in neutron binary mergers seems to represent one of the most promising channel for r-process element production (\citealt{Korobkin2012, Eichler2015}), especially after the GW170817 gravitational-wave event connected to merging neutron stars and the detection of heavy elements (e.g., \citealt{Evans2017, Pian2017, Tanvir2017, Troja2017}). On the other hand, \cite{Cote2019} and \cite{Kobayashi2020} concluded that magneto-rotational supernovae could be the main site of the r-process in the Galaxy. 
iv) Novae are binary systems made of a WD plus a Main Sequence or Red Giant low mass star. These systems suffer outbursts which do not destroy the WD, but create new elements through explosive H-burning. These new elements are: CNO isotopes ($^{13}$C, $^{15}$N, $^{17}$O), perhaps $^{7}$Li, plus some radioactive elements (e.g., $^{22}$Na, $^{26}$Al) (see \citealt{Jose2007}).

To summarize: {CC-SNe do produce the bulk of $\alpha$-elements and only part of Fe on timescales negligible relative to a Hubble time, whereas Type Ia SNe do produce the bulk of Fe plus traces of elements from C to Si on a large range of timescales, going from 35 Myr to a Hubble time. Low and intermediate mass stars dying in the same range of times as Type Ia SNe, do produce the bulk of heavy s-process elements, $^{14}$N and  part of He and $^{12}$C. Novae can contribute to the enrichment of CNO isotopes, $^{7}$Li and radioactive elements on long timescales, whereas merging neutron stars (MNS) can contribute substantially to r-process elements (e.g., Eu) either on short or moderately long merging timescales (see later).}

\subsubsection{The yield per stellar generation}

To understand the chemical enrichment by a simple stellar population, namely stars born at the same time and with the same chemical composition, we define the \emph{yield per stellar generation} of a single chemical element, as in \cite{Tinsley1980}:
\begin{equation}
  y_i={\int^{\infty}_{1}{mp_{im} \phi(m)dm} \over (1-R)},
  \label{eq:yieldgen}
\end{equation}
where $p_{im}$ is the stellar yield of the newly produced and ejected element $i$ by a star of mass $m$, as defined before, and $R$ is the returned fraction (see Eq.~\ref{eq:Rfraction}). Therefore, the yield $y_i$ is the mass fraction of the element $i$ newly produced by a generation of stars, relative to the fraction of mass locked up in remnants (white dwarfs, neutron stars and black holes) and brown dwarfs ($M< 0.1\,M_{\odot}$). 

We define \emph{returned fraction} the fraction of mass ejected into the ISM by an entire stellar generation, namely:
\begin{equation}
  R= {\int^{\infty}_{1}{(m-M_{\rm rem}) \phi(m) dm} \over \int^{\infty}_{0.1}{m \phi(m) dm}},
  \label{eq:Rfraction}
\end{equation}
where $M_{\rm rem}$ is the remnant mass, which can be either a WD or a neutron star or a black hole.

The term fraction originates from the fact that $R$ is divided by the normalization integral of the IMF, which is equal to unity (Eq.~\ref{eq:normimf}). The upper mass limits here is indicated by $\infty$ but normally is assumed to be $100\,M_{\odot}$.
%\begin{equation}
%\int^{100}_{0.1}{m \phi(m)dm}=1.
%\end{equation}

In order to define $y_i$ and $R$,
we have made a very specific assumption: the Instantaneous Recycling Approximation (I.R.A.), stating that \emph{all stars more massive than $1\,M_{\odot}$ die instantaneously, while all stars less massive than $1\,M_{\odot}$ live forever}. This assumption allows us to solve analytically the chemical evolution equations, but it is a very poor approximation for chemical elements produced partly or entirely on long timescales such as C, N and Fe. On the other hand, for oxygen, which is almost entirely produced by short lived CC-SNe, I.R.A.\ can be  an acceptable approximation.
In Fig.~\ref{fig:fiore} we report the yield per stellar generation of oxygen,  $y_O$, as well as the returned fraction, $R$,  computed for different initial stellar metallicities and three different IMFs (\citealt{Chabrier2003, Salpeter1955, Kroupa1993}). As one can see, the variation of $y_O$ and $R$ with metallicity $Z$ is negligible, whereas the dependence on the assumed IMF is strong.  In particular, the \cite{Chabrier2003} IMF predicts the largest differences in these two quantities, and the reason can be found in the larger number of massive stars in this IMF, relative to the other ones (see Fig.~\ref{fig:imf}). In Fig.~\ref{fig:fiore} the \cite{Kroupa2001} universal IMF, suggesting that the IMF in stellar clusters is an universal one, does not appear and the reason is that is very similar to that of \cite{Chabrier2003}, as one can see in Fig.~\ref{fig:imf}. Finally, the IMF of \cite{Kroupa1993}, obtained for the solar vicinity, contains less massive stars than all the other IMFs adopted by \cite{Vincenzo2016}. The adopted stellar yields to compute $y_O$ and $R$ are those of \cite{Romano2010} their model 15.

\subsection {Gas flows: infall, outflow, radial flows}

In order to build a realistic galaxy one has to assume the presence of gas flows both in and out. The gas inflows are considered either as gas accretion or radial gas flows and they are influencing the chemical evolution of galaxies: in the case of accretion, usually assumed to occur at a constant rate or exponentially decreasing in time, the main effect is to dilute the metal content, except if the metallicity of the infalling gas is equal or larger than that of the pre-existing gas, but this is a rather unlikely situation.

The most common parametrization of the gas infall is:
\begin{equation}
  A(t)= \Lambda e^{-t/\tau},
  \label{eq:infall}
  \end{equation}
  where A(t) is the gas accretion rate, namely how many solar masses are accreted per unit time, $\Lambda$ (adimensional) and $\tau$ are two free parameters.
  In particular, $\tau$ is the timescale for gas accretion, namely the time necessary to accumulate half of the mass of the system. In some analytical chemical models (e.g., \citealt{Matteucci1983}), it has been assumed that $A(t) \propto \psi(t)$, but this is a questionable assumption since is indeed the SFR which is affected by $A(t)$ but not viceversa.

  In the case of outflows or galactic winds (when the mass is lost from the galactic potential well), the effect is also that of decreasing the metal concentration by simply decreasing the gas which is available for star formation.

  Galactic outflows are generally assumed to occur at a rate proportional to the SFR, such as:
\begin{equation}
  W(t) = -\lambda \psi(t),
  \label{eq:wind}
 \end{equation}
where $W(t)$ is the wind rate, namely how many solar masses are lost per unit time from the galaxy, $\lambda$ is a free adimensional parameter
and $\psi(t)$ is the SFR.

In the case of radial gas flows, the most common assumption is that they are directed inward, as a dynamical consequence of gas infalling onto the disk. In fact, the infalling gas has a lower angular momentum than the circular motion in the disk, and mixing with the gas in the disk produces a net radial inflow. Such an inflow can favour the formation of abundance gradients as long as its speed is $<2$km/sec (\citealt{Tinsley1980}). Many models for chemical evolution of the Milky Way have assumed inward radial gas flows in the disk (\citealt{Mayor1981, Lacey1985, Goetz1992, Portinari2000, Schonrich2009, Spitoni2011, Grisoni2018, Vincenzo2020}).

\section{Analytical models of chemical evolution}
\label{sec:anal}

First of all, we discuss the so-called \emph{Simple Model} for
the chemical evolution of the solar neighbourhood.
We note that the solar neighbourhood is defined as a region 
centered in the Sun and extending roughly 1 kpc in all directions. 

\subsection{Basic assumptions and solution of the Simple Model}

In this paragraph we follow the definition of the \emph{Simple Model} as
given in \cite{Tinsley1980}; in particular, the \emph{Simple Model} is based 
on the following assumptions:

\begin{itemize}
\item {1} the system is one-zone and closed, namely there are no inflows 
or outflows
\item {2} the initial gas is primordial
\item {3} $\varphi(m)$ is constant in time
\item {4} the gas is well mixed at any time.
\end{itemize}

The well known solution of the Simple Model is:
\begin{equation}
  Z= y_{Z} ln({ 1 \over \mu}),
  \label{eq:SModel}
\end{equation}
where $\mu={M_{\rm gas} \over M_{\rm tot}}$ is the gas mass fraction ($M_{\rm tot}$ is the mass of stars plus gas), and $y_Z$ is the yield per stellar generation of the metals. This solution is obtained after assuming I.R.A.\ and integrating the equation describing the evolution of $Z$ between $M_{\rm gas}(0)=M_{\rm tot}$ and $Z(0)=0$ and $Z(t)$.

The yield which appears in Eq.~\eqref{eq:SModel} is known as \emph{effective yield}, and is simply defined as the yield $y_{Z_{\rm eff}}$ that would be deduced if the system were assumed to be described by the Simple Model.
Therefore:
\begin{equation}
  y_{Z_{\rm eff}}={Z \over ln(1/\mu)}
  \label{eq:SolSM}
\end{equation}

%If $y_{Z_{\rm eff}} > y_{Z}$, then the actual system has attained a higher
%metallicity for a given gas fraction $\mu$.

The effective yield represents the highest degree of chemical enrichment for a given IMF. In fact, systems where the hypothesis 1) is relaxed (i.e. infall and/or outflow) have true yields lower than the effective yield (see later).
The Simple Model for describing the evolution of the solar vicinity was discarded since it predicts too many long living stars (G-dwarfs) at low metallicities, the well known ``G-dwarf problem'', which was solved by assuming gas infall for the formation of the solar vicinity. Moreover, this model and also all the analytical chemical models, cannot follow, because of the I.R.A., the evolution of elements restored into the ISM on long timescales, such as Fe, which is formed mainly in Type Ia SNe and is the main tracer of stellar metallicity. On the other hand, for elements formed on short timescales by massive stars, such as O, the analytical solutions assuming I.R.A.\ can be acceptable.

The solution of Eq.~\eqref{eq:SModel} is valid for a \emph{primary} element, namely an element  formed directly from H and He, as opposed to a \emph{secondary} element which is formed from metals already present in the star at birth.
For the abundance of a secondary element $X_S$, such as $^{14}$N, which is produced during the CNO cycle, although it can have also a primary origin if the C and O out of which is formed have been synthesized in the star (cases of dredge-up in AGB stars and rotation in massive stars), the solution of the Simple Model is (\citealt{Tinsley1980}):
\begin{equation}
  X_S= {1 \over 2}({y_S \over y_Z Z_{\odot}})Z^{2},
  \label{eq:secondary}
\end{equation}
where $y_S$ is the yield per stellar generation for the generic secondary element $S$. As one can see from Eq.~\eqref{eq:secondary}, the ratio between the abundance of a secondary element and the abundance of its primary progenitor evolves  proportionally to the abundance of the progenitor (e.g., $X_S/Z \propto Z$).

\subsection{Analytical solution for gas outflows}

A more realistic situation would involve gas flows in the studied system, in particular outflow and infall.
The situation in which there is only gas outflow can be described by the following solution (\citealt{Matteucci1983}):
\begin{equation}
  Z={y_Z  \over (1+ \lambda)} ln[(1+\lambda) \mu^{-1} -\lambda].
  \label{eq:solwind}
\end{equation}

After assuming a wind rate of the form:
\begin{equation}
  W(t)=-\lambda(1-R) \psi(t),
  \label{eq:W}
\end{equation}
where $\lambda \ne 0$ is the adimensional wind parameter, and integrating between 0 and t and between $M_{\rm tot}=M_{\rm gas}(0)$ and  $M_{\rm gas}(t)$.
 
It is clear that for $\lambda=0$  Eq.~\eqref{eq:solwind} becomes the solution of the Simple Model (Eq.~\ref{eq:SModel}).
The meaning of Eq.~\eqref{eq:solwind} is immediately clear, the true yield is lower than the effective yield in presence of only outflows.

\subsection{Analytical solution for gas infall}

In the case of only gas infall, the solution  for a primordial chemical composition ($Z_{\rm inf}=0$) of the infalling gas is (\citealt{Matteucci1983}):
\begin{equation}
  Z= {y_Z \over \Lambda}[1-(\Lambda-(\Lambda-1)\mu^{-1})^{-\Lambda/(1-\Lambda)}],
  \label{eq:solinfall}
\end{equation}
where the accretion rate has been chosen to be:
\begin{equation}
  A(t)=\Lambda (1-R) \psi(t),
  \label{eq:accretion}
\end{equation}
with  $\Lambda$ a positive constant different from zero and from 1.
Also in this case, the true yield in Eq.~\eqref{eq:solinfall} is lower than the Simple Model effective yield, and by imposing $\Lambda=0$ the Eq.~\eqref{eq:solinfall} becomes the solution of the Simple Model.

If $\Lambda=1$ the solution is:
\begin{equation}
  Z=y_Z[1-e^{-(\mu^{-1}-1)}],
  \label{eq:lambda1}
\end{equation}
which is the well-known solution for the \emph{extreme infall case}
(\citealt{Larson1972, Tinsley1980}), where the amount of gas remains constant in time.

\subsection{Analytical solution for gas infall plus outflow}

If  $A(t)=\Lambda (1-R) \psi(t)$ and $W(t)= - \lambda (1-R) \psi(t)$ are both active, the
analytical solution is (\citealt{Recchi2008}):
\begin{equation}
Z={(\Lambda Z_{\rm inf} +y_Z) \over \Lambda}[1-(\Lambda -\lambda) - (\Lambda- \lambda-1) \mu^{-1}]^{{\Lambda \over \Lambda-\lambda -1}},
\label{eq:infout}
\end{equation}
for $\Lambda \ne 0\, \ne 1$ and $\lambda \ne 0$. This general solution allows us to consider also an enriched infall (i.e., $Z_{\rm inf}\ne 0$).
A similar situation has been studied by \cite{Lilly2013} and called ``bathtube model''.

\subsection{Analytical solution for biased galactic outflow plus infall}

Both theory (e.g., \citealt{Vader1986, Recchi2001, Recchi2008}) and observations (e.g., \citealt{Martin2002}) have suggested that galactic outflow can be metal-enhanced, in the sense that metals produced by supernovae are lost more easily from a galaxy than the total gas made mainly by H and He. \cite{Recchi2008} found an analytical solution for galactic winds carrying out mostly metals.
In such a case, the wind rate is defined as:
\begin{equation}
W(t)Z^{o}=  -\alpha Z \lambda (1-R) \psi(t),
\label{eq:Wrich}
\end{equation}
where $Z^{o}$ is the metallicity of the outflowing gas which, in this case, 
can be different from the metallicity $Z$ present in the galactic gas and can be defined as:
\begin{equation}
  Z^{o}= \alpha_{\rm ef} Z,
  \label{eq:Zout}
\end{equation}
with $\alpha_{\rm ef}> 1$ being the ejection efficiency.

The equation for metals in this case is:
\begin{equation}
{d(ZM_{\rm gas}) \over dt}=(1-R) \psi(t)[\Lambda Z_{\rm inf} + y_Z -(\lambda \alpha_{\rm ef} + 1)Z],
\label{eq:metrich}
\end{equation}
where $Z_{\rm inf}$, $\Lambda$ and $\lambda$ are the same parameters as defined above. The solution of this equation is:
\begin{equation}
Z= {{\Lambda Z_{\rm inf} + y_Z} \over {\Lambda + (\alpha_{\rm ef}-1) \lambda}}\biggl\lbrace 1-[(\Lambda- \lambda) - (\Lambda -\lambda -1) \mu^{-1}]^{{\Lambda + (\alpha-1) \lambda} \over {\Lambda- \lambda -1}}\biggr\rbrace.
\label{eq:solmetrich}
\end{equation}
One can immediately see that if we assume $\alpha_{\rm ef}=1$ we obtain again Eq.~\eqref{eq:infout}.

\subsection{More complex analytical solutions} 
Interesting and more refined analytical solutions are those proposed by \cite{Spitoni2017} which follow the evolution of the gas mass, total mass and metallicity of a given galaxy. In particular, these solutions are obtained for an exponential infall gas law, such as in Eq.~\eqref{eq:infall} ($A(t)= \Lambda e^{-t/\tau}$),
with $\tau$ being the timescale of the infall and $\Lambda$ a normalization constant obtained from the following condition:
\begin{equation}
  \int^{t_G}_0{\Lambda e^{-t/\tau} dt}=M_{\rm infall}
\label{eq:Spito}
\end{equation}
where $t_G$ is the galactic lifetime and $M_{\rm infall}$ is the mass that is accumulated by gas infall over the time $t_G$.

A galactic outflow law is also taken into account with the expression:
\begin{equation}
  W(t)= -\lambda \psi(t),
\label{eq:Spitowind}
\end{equation}
which is similar to that of Eq.~\eqref{eq:W} except for the factor $(1-R)$.

In the \cite{Spitoni2017} model formulation, there is also an explicit consideration of the SFR law, in particular:
\begin{equation}
  \psi(t)= \nu M_{\rm gas}(t),
\label{eq:SFRgas}
\end{equation}
where $\nu$ is the efficiency of star formation, as previously defined in
Eq.~\eqref{eq:SFR} and $M_{\rm gas}$ is the gas mass at the time $t$.

The basic equations of this model are:
\begin{equation}
  {dM_{\rm tot} \over dt}= \Lambda e^{-t/ \tau} -\lambda \psi(t)
\label{eq:Spitoeq1}
\end{equation}

\begin{equation}
  {dM_{\rm gas} \over dt}= -(1-R) \psi(t) +\Lambda e^{-t/ \tau} -\lambda \psi(t)
\label{eq:Spitoeq2}
\end{equation}

\begin{equation}
  {dM_Z \over  dt} = [-Z(t) + y_Z](1-R) \psi(t) -\lambda Z(t) \psi(t) + Z_{\rm inf} \Lambda e^{-t/ \tau}.
\label{eq:Spitoeq3}
\end{equation}
The quantity $Z_{\rm inf}$ represents the metallicity of the infalling gas.

The equation for metals can be written as:
\begin{equation}
  \dot Z(t) =y_Z(1-R) \nu + {\Lambda [Z_{\rm inf} - Z(t)]e^{-t/ \tau} \over M_{\rm gas}(t)}.
  \label{eq:Spitoeq4}
\end{equation}
The integration of the above equations is performed with the following initial conditions: at $t=0$ $M_{\rm tot}(0)= M_{\rm gas}(0)$.
%and they are extremely small quantities.
The initial metallicity $Z(0)=0$ and the 
chemical composition of the infalling gas is primordial ($Z_{\rm inf}=0$).
The solution for the gas mass is:
\begin{equation}
  M_{\rm gas}(t) = e^{-\alpha t}\left[{\Lambda (e^{-t/ \tau +\alpha t} -1) \tau \over \alpha \tau -1} + M_{\rm gas}(0)\right],
\label{eq:Spitoeq5}
\end{equation}
and  for the global gas metallicity $Z$:
\begin{equation}
  Z(t)= {y_Z \nu (1-R) \over \alpha \tau -1} \cdot{M_{\rm gas}(0)t(\alpha \tau -1)^2 +\Lambda \tau[t - \tau(1+ \alpha t) + \tau e^{-\alpha t -t/ \tau}] \over \Lambda \tau (e^{\alpha t-t/ \tau} -1) +M_{\rm gas}(0) (\alpha \tau-1)}
\label{eq:Spitoeq6}
\end{equation}
In the two equations above it has been assumed  $\alpha= (1 + \lambda -R) \nu$.

It is interesting to compute also the average metallicity of stars ($\langle Z_*(t) \rangle$) which can be expressed as:
\begin{equation}
  \langle Z_{*}(t) \rangle = \int_0^t  {dt^{'}Z(t^{'}) \psi(t^{'})} / \int_0^t{dt^{'} \psi(t^{'})}.
  \label{eq:Spitoeq7}
\end{equation}

The above analytical solutions (Eq. \ref{eq:Spitoeq5}  and Eq. \ref{eq:Spitoeq6}) can be very useful to study the Mass-Metallicity relation in either star forming or passive galaxies with the necessary condition of considering the global metallicity $Z$, which is dominated by the abundance of oxygen for which the I.R.A.\ is a good approximation. Clearly, these solutions cannot be applied to the study of the evolution of elements produced on long timescales, such as iron.

In Fig.~\ref{fig:Spito17} we show the effects of galactic gas outflows with wind parameter $\lambda=0.5$ and gas infall as in Eq.~\eqref{eq:infall} with infall timescale $\tau=2$ Gyr. It is evident the effect of galactic winds taking out of the galaxy a large fraction of gas and therefore metals.

\begin{figure}[htbp]
  \centering
  \includegraphics[width=0.9\textwidth]{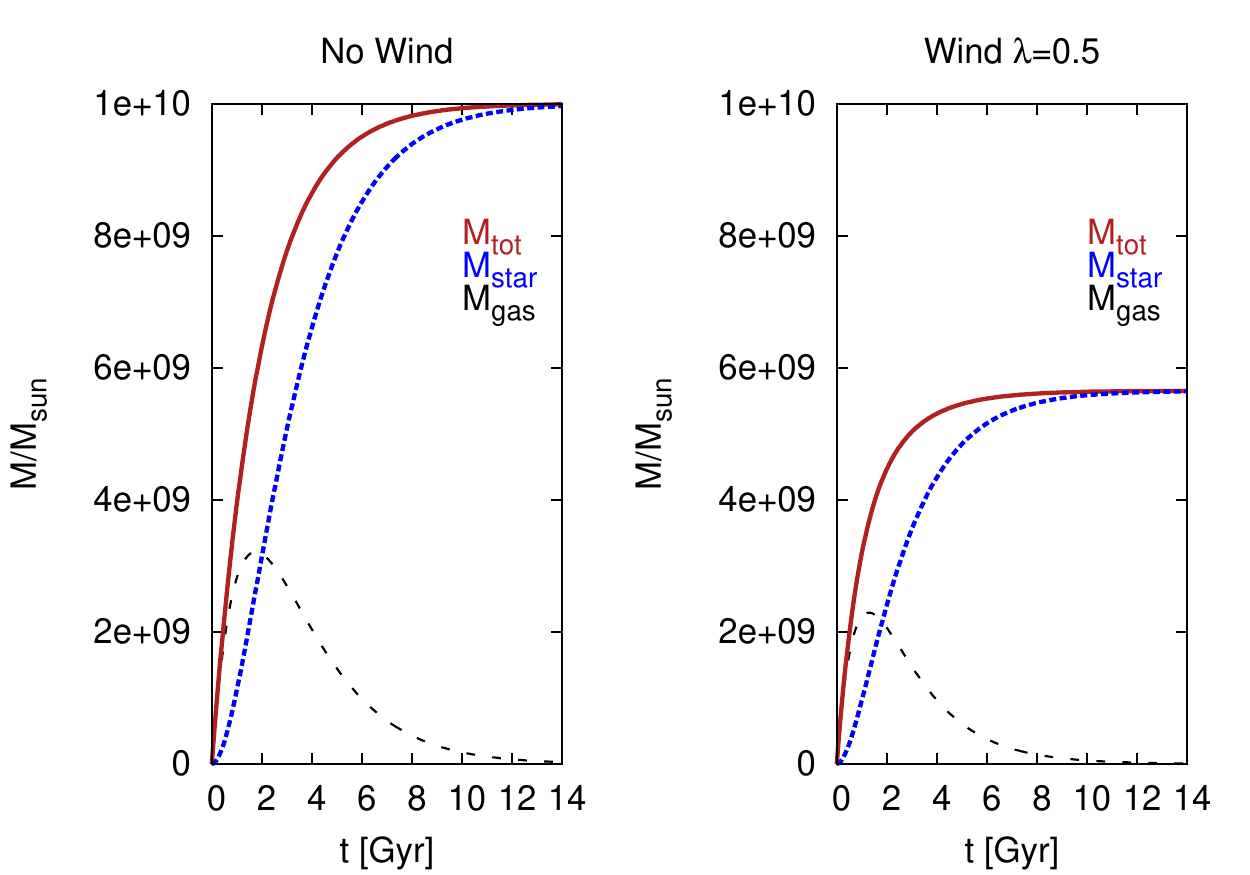}
  \caption{The effects of galactic winds on the predicted time evolution of the galaxy gas mass $M_{\rm gas}$ (gray dashed lines), stellar mass $M_{\star}$ (blue dotted lines), and total mass $M_{\rm tot} = M_{\rm gas} + M_{\star}$ (red solid lines). 
 The model assumes an infall timescale $\tau=2$ Gyr and infall mass  $M_{\rm infall}=10^{10}\,{M}_{\odot}$, and a star formation efficiency $\nu=1\,{\rm Gyr}^{-1}$. 
 {\it Left panel}: model without galactic winds  ($\lambda=0$); {\it Right panel}: model with wind parameter $\lambda=0.5$. Image reproduced with permission from \cite{Spitoni2017}, copyright by ESO.}
\label{fig:Spito17}       % Give a unique label
\end{figure}

\section{Numerical chemical evolution models of the Milky Way}
\label{sec:numer}

We have seen in the previous section, that if we assume the I.R.A.\ the equations for gas and metals can be solved analytically. However, the I.R.A.\ is a good approximation only for the chemical elements which are produced on timescales negligible relative to the age of the Universe, but is a very bad approximation for elements produced on long timecales. The former are basically the $\alpha$-elements which are mainly produced in CC-SNe (i.e., O and Mg), although some $\alpha$-elements (e.g., Si and Ca) are also produced in a non-negligible quantity by long living stars (Type Ia SNe). The latter are elements, such as Fe and N, which are mainly produced by low and intermediate mass stars either by single stars (e.g., N) or binary systems (e.g., Fe from SNe Ia).

If we assume any of the analytical solutions derived before and we apply them to two single chemical elements, instead of $Z$, then we can state that the ratio of the abundances of such elements is a constant and equal to the ratio of the corresponding yields per stellar generation (see Fig.~\ref{fig:fiore}). Therefore, if we take for example O and Fe, then we conclude that the ratio $X_{\rm O}/X_{\rm Fe}$ is constant during the entire Galactic lifetime. We know from studying the abundance ratios in the main Galactic components that this is not true. In fact, the [O/Fe] ratio in halo stars is larger than solar and it decreases towards the solar value ([O/Fe]=0 dex) in disk stars (see later). To study in detail and in the correct way the evolution of the abundances of single elements we must relax I.R.A.\ and solve the chemical evolution equations numerically.

\subsection{Basic equations of chemical evolution}
A suitable compact form for the complete chemical evolution equations is the following, where we write one equation for each chemical element, such as:
\begin{equation}
  \dot \sigma_i(t) = -\psi(t)X_i(t)+ \dot R_i(t) + \dot \sigma_{i,inf}(t) -\dot \sigma_{i,w}(t),
\label{eq:basiceq}
\end{equation}
where $\sigma_i(t)= \sigma_{\rm gas}(t)X_i(t)$ is the fractional surface mass density of the element $i$ in the ISM at the time $t$. The quantity $X_i(t) = \sigma_i(t)/\sigma_{\rm gas}(t)$ is the abundance by mass of the element $i$ and $\sigma_{\rm gas}(t)$ is the surface mass density of the ISM.
The quantity $\psi(t)$ is the SFR, the quantity $\dot R_i(t)$ is the rate of variation of the returned mass in form of new and old chemical elements. The quantities $\dot \sigma_{i,inf}(t)$ and $\dot \sigma_{i,w}(t)$ are the infall and outflow rate, respectively. All the previous quantities are expressed in terms of \,$M_{\odot} \, {\rm pc}^{-2} \, {\rm Gyr}^{-1}$.

The term $\dot R_i(t)$ can be written as:
\begin{equation}
  \dot R_i(t)= \int_{m(t)}^{m_{\max}} {\psi(t-\tau_m)Qm_i(t)(t-\tau_m) \phi(m)dm},
\label{eq:recycl}
\end{equation}
where $m(t)$ is the minimum mass dying at the time $t$, $\phi(m)$ is the IMF, $\psi(t-\tau_m)$ is the SFR at the time at which the star of mass $m$ was born ($\tau_m$). The quantity $Qm_i(t)$ contains all the stellar nucleosynthesis (old and newly produced element $i$ by a star of mass $m$). The dependence of $Qm_i$ upon $(t-\tau_m)$ is relative to the abundance of the element $i$ already present at the star birth (the fraction of $i$ restored without being processed). This integral contains the products of all stars such as single stars dying as WDs, SNeIa and CC-SNe.

Concerning the SFR, a common parametrization is the Schmidt--Kennicutt law expressed in Eq.~\eqref{eq:SFR}.
%\begin{equation}
%  \psi(t)= \nu \sigma_{\rm gas}(t)^{k},
%  \end{equation}
%where $\sigma_{\rm gas}(t)$ is the gas surface density, $\nu$ is the efficiency of star formation and is expressed in units of $t^{-1}$ and $k= 1.4$ (Kennicutt 1998).

The gas infall rate is generally described by an exponential law, as expressed by Eq.~\eqref{eq:infall}. Such a law should be written for each chemical element in the following way:
\begin{equation}
  \dot \sigma_{i, \rm inf}= \Lambda e^{-t/ \tau}X_{i, \rm inf}(t),
  \label{infnum}
\end{equation}
with $\sigma_{i, \rm inf}$ being the surface gas density of the element $i$ present in the infalling gas and $X_{i,inf}$ its  abundance. Normally, the chemical composition of the infalling gas is assumed to be primordial (no metals). In reality, the infall metallicity is likely not primordial, but models have shown that it must be quite low (e.g., \citealt{Tosi1988b}) and consistent with the metallicity observed in the Magellanic Stream, for example.

The parameter $\Lambda$ is derived by imposing the condition of Eq.~\eqref{eq:Spito} which gives:
\begin{equation}
  \Lambda = {\sigma_{\rm inf}(t_G) \over \tau(1-e^{-t_G/ \tau})},
\label{eq:paraminfall}
\end{equation}
where $t_G \sim 13.7$ Gyr, and $\sigma(t_G)_{\rm inf}$ is the present time total surface mass density accumulated by infall (infall mass if the chemical equation is written in terms of masses and not surface densities) of the studied system. If we study the solar vicinity the total present time surface mass density is $\sim 48 \pm 8 \,M_{\odot}\, {\rm pc}^{-2}$ (\citealt{Kuijken1989}). A more recent estimate from \cite{McKee2015} suggests  $\sim 41 \pm 3.4 \,M_{\odot}\, {\rm pc}^{-2}$.

Finally, the rate of outflow has generally the form:
\begin{equation}
  \dot \sigma_{i,w}(t)= -w_i \psi(t) X_i(t),
\label{eq:windnum}
\end{equation}
where $w_i$ is the so-called mass loading factor or wind efficiency parameter relative to the specific element $i$. It can be different for different elements (i.e., differential galactic winds).

In summary, Eq.~\eqref{eq:basiceq} can be solved only numerically and the most complex computation is due to the term $\dot R_i(t)$. This quantity appears as an integral in Eq.~\eqref{eq:basiceq} involving the SFR and the IMF, and it grows with the integration time, since more and more stars die at any successive timestep. It contains the detailed nucleosynthesis occurring in each star contributing to the chemical enrichment.

\subsection{The supernova rates}

The galactic chemical enrichment (the term $\dot R_i(t)$ in Eq.~\eqref{eq:basiceq} occurs mainly by means of supernovae of all Types (Ia, Ib, Ic and II), C-O WDs, novae and MNS. Massive stars end their lives as SNe II, Ib, Ic and SNe Ib,c are originating from the upper mass range: some of these SNe can have a very high explosion energy ($\sim 10^{52}$ erg), while normally is $\sim 10^{51}$ erg and are called \emph{hypernovae}. Therefore, it is necessary to compute in detail the SN, nova and MNS  rates in order to estimate their chemical pollution. In the following we summarize how to compute all these rates.

\subsubsection{Rate of CC-SNe}

The rate of CC-SNe (Type II, Ib, Ic) can be computed by assuming either that they originate from single massive stars or massive binaries. In particular, the SNeII are exploding single massive stars in the mass range
($8 \le M/\,M_{\odot} \le M_{\rm WR}$), although the upper limit, namely the limiting mass for the formation of a Wolf--Rayet (WR) star,  is very uncertain depending upon mass loss, rotation and other physical assumptions of the stellar models.

The SNe Ib and Ic are either the explosions of single stars with $M>M_{\rm WR}$ or stars
in binary systems in the mass range 12--20\,$M_{\odot}$  (\citealt{Baron1992}) or 14.8--45\,$M_{\odot}$ (\citealt{Yoon2010}). Finally, a fraction od CC-SNe are called ``hypernovae''. These SNe originate from masses $>M_{\rm WR}$ but with explosion energy ($\sim 10^{52}$ erg) a factor of 10 higher than the other SNe.

The rate of Type II SNe can be written as:
\begin{equation}
  R_{\rm SNII}(t)= \int^{M_{\rm WR}}_8{\psi(t-\tau_m) \phi(m)dm},
\label{eq:SNIIrate}
\end{equation}
while the rate for Type Ib, Ic is given by the previous equation, where the integration goes from $M_{\rm WR}$ to $M_{\max}$, which can be any mass from 70 to 120\,$M_{\odot}$, plus the rate of explosion of massive binaries in a given range:
\begin{equation}
  R_{\rm SNIb,c}(t) = (1-\gamma)\int^{M_{\max}}_{M_{\rm WR}}{\psi(t-\tau_m) \phi(m)dm} +  \gamma \int^{45}_{14.8} {\psi(t-\tau_m) \phi(m)dm},
\label{eq:SNIbcrate}
\end{equation}
where $\gamma$ is a free parameter indicating how many massive binary systems are present in the range 14.8--45\,$M_{\odot}$, and is normalized to obtain the present time rate of Type Ib,c SNe in galaxies ($\gamma=0.15$--$0.30$, \citealt{Grieco2012a}).

\subsubsection{Supernovae Ia}

The rate of SNe Ia is very important since these SNe are believed to be the major producers of Fe in the Universe, which is the main tracer of stellar metallicity. Two basic scenarios for Type Ia SNe have been proposed in the last years: i) the single degenerate (SD), where a WD in a binary system explodes after accretion of matter from a companion, which fills its Roche lobe when becoming a red giant (but it could be also a slightly evolved Main Sequence star), and reaches and overcomes the Chandrasekhar mass (\citealt{Whelan1973}), and ii) the double degenerate (DD), where two WDs merge after losing angular momentum because of  gravitational wave emission, and explode after reaching the Chandrasekhar mass (\citealt{Iben1984, Han2004}). The Chandrasekhar mass in absence of H is $M_{\rm Ch} \sim 1.44\,M_{\odot}$ and represents the limiting mass for the stability of a C-O WD. In the last years, many variations to these two basic scenarios have been suggested (see \citealt{Hillebrandt2013, Ruiter2020} for reviews and \citealt{KobaNomoto2020}), including sub-Chandrasekhar masses for the exploding WD.
The alternative scenarios have been proposed to explain the so-called peculiar SNeIa. Concerning the SD scenario, a point of concern has always been the requested precise rate of mass accretion from the donor. To overcome this problem, \cite{Hachisu1999} suggested a scenario where a wind from the WD stabilizes the accretion from the donor and provides a wider channel to the occurrence of SNeIa.

\cite{Greggio2005} proposed a general formulation for the rate of Type Ia SNe, which can include any possible progenitor model for such supernovae, if expressed in analytical form.
In this way, the SNIa rate is given by the product of the SFR and the function describing the distribution of the explosion times (delay time distribution, hereafter DTD).

In particular, we can write:
\begin{equation}
  R_{\rm SNIa}(t) = \kappa_{\alpha} \int_{\tau_i}^{\min(t, \tau_x)}{A(t-\tau) \psi(t-\tau) {\rm DTD}(\tau) d\tau},
\label{eq:SNIarate}
\end{equation}
where $\psi(t-\tau)$ is the SFR, $\tau$ is the total delay time, namely the nuclear stellar
lifetime of the secondary star plus a possible delay due for example to the gravitational time delay in the DD model; this time is defined in the range $(\tau_i, \tau_x)$ so that:
\begin{equation}
  \int_{\tau_i}^{\tau_x}{{\rm DTD}(\tau) d\tau} =1.
\label{eq:normSNIa}
\end{equation}
The function ${\rm DTD}(t)$ can be any function able to describe the sequence of SNIa explosions, the time $\tau_i$ is the minimum time requested for the explosion of SNeIa. In the SD scenario, $\tau_i$ is the lifetime of a
$\sim 8 \,M_{\odot}$ star, which is considered the maximum mass for the formation of a C-O WD, although the precise value of this mass depends on the stellar evolution model prescriptions.  In the DD scenario, $\tau$ is the lifetime of a $\sim 8 \,M_{\odot}$ star plus the minimum gravitational time delay ($\sim 1$ Myr, \citealt{Greggio2005}). The time $\tau_x$ is the maximum time for the explosion of a SNIa and in the SD model is the Hubble time corresponding to the lifetime of a $\sim 0.8 \,M_{\odot}$, whereas in the DD model this value can be several Hubble times, depending on the initial separation of the two WDs which determines the timescale for merging. 

The parameter $A(t-\tau)$ is the fraction of binary systems, in the IMF, possessing the right characteristics to produce a SN Ia. It is normally considered as a constant but in principle it could vary with time. Generally, $A$ is derived by reproducing the present time observed Type Ia SN rate of the object under study.

The quantity $\kappa_{\alpha}$ contains the IMF, and is given by:
\begin{equation}
  \kappa_{\alpha} = \int_{m_L}^{m_U}\phi(m)dm,
\label{eq:kalfa}
\end{equation}
where $m_L=0.1\,M_{\odot}$ and $m_U=100\,M_{\odot}$. The normalization of the IMF is the usual one (see Eq.~\ref{eq:normimf}).

The advantage of the DTD formulation is that we can test both theoretical and empirical Type Ia SN rates. These empirical rates can be approximated by analytical expressions such as that of \cite{Totani2008} which goes like $DTD(t) \propto t^{-1}$. Another empirical Type Ia SN rate is that suggested by \cite{Mannucci2006}: here, the Type Ia rate is bimodal and predicts that 50\% of SNe Ia explode before 100 Myr and they are called \emph{prompt SNeIa}, while the other 50\% explode on timescales $>100$ Myr and they are called \emph{tardy SNeIa}. It should be noted that the DTD of the empirical rate of \cite{Totani2008} well follows the theoretical DD rate of \cite{Greggio2005}.

In Fig.~\ref{fig:DTD} we show several DTDs, both theoretical and empirical. For an extensive review about the empirically derived DTD functions for SNIa progenitors, we address the reader to \cite{Maoz2012}; in this paper, they conclude that a variety of methods to derive the DTD converges on a DTD such as that of \cite{Totani2008}, at least in the range of $1< \tau < 10$ Gyr, thus suggesting the DD scenario as the preferred one. However, we cannot exclude that the SD can also work, if the problems related to the accretion rate from the donors are solved, as discussed above, and it can also be consistent with the empirical one (see \citealt{KobaNomoto2020}).
It has been shown that in order to reproduce the abundance patterns in the solar vicinity the best scenarios are the SD and the DD ones, which predict a low number of prompt SNeIa, as opposed for example to the \cite{Mannucci2006} model. In particular, in the SD and DD model the DTD function contains  $< 20\%$ of prompt SNeIa. If the prompt SNeIa represent a fraction larger than 30\%, the agreement with the observed abundance patterns in the Milky Way is lost (\citealt{Matteucci2006, Matteucci2009}). In Fig.~\ref{fig:DTD} one can see that the DTD for the DD scenario does not differ much from the one for the SD scenario of \cite{MatteucciRecchi2001}.

\begin{figure}[htbp]
  \centering
  \includegraphics[width=0.6\textwidth]{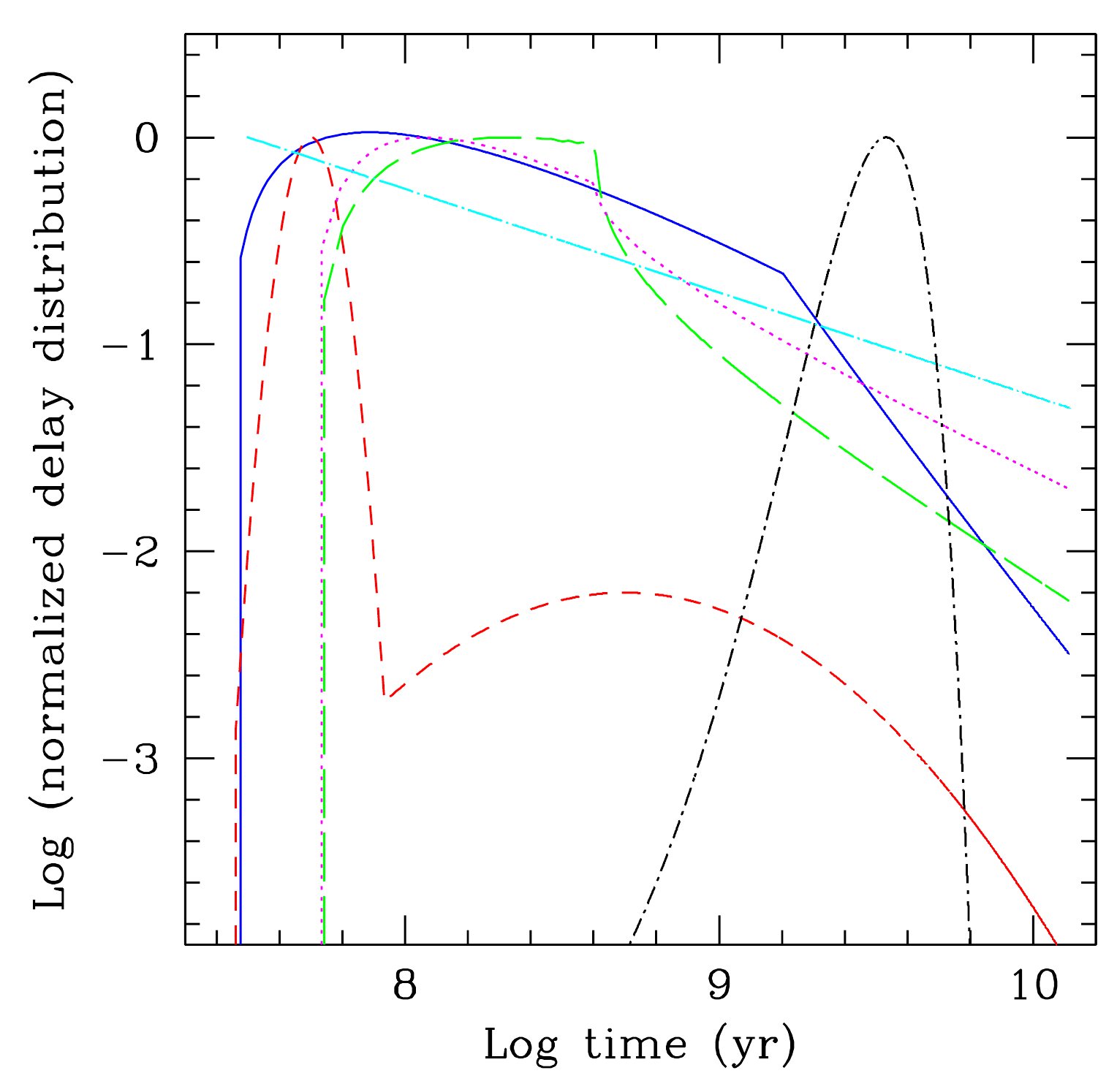}
\caption{Various DTD functions normalized to their own maximum value: the continuous blue line is the DTD for the SD scenario (\citealt{MatteucciRecchi2001}); the long dashed green line is the DTD of \cite{Greggio2005} for the DD wide channel; the dotted magenta line is the DTD for the DD close channel of \cite{Greggio2005}; the dashed red line is the bimodal DTD of \cite{Mannucci2006}; the short dashed-dotted black line is the DTD of \cite{Strolger2004} derived from the observed cosmic Type Ia SN rate and the cyan long dashed-dotted line is the DTD of \cite{Pritchet2008} with DTD $ \propto t^{-0.5 \pm 0.2}$. Image reproduced with permission from \cite{Matteucci2009}, copyright by ESO.}
\label{fig:DTD}       % Give a unique label
\end{figure}

\subsubsection{Novae}
Not only supernovae can enrich the ISM but also nova outbursts can eject newly formed elements, as discussed in Sect.~\ref{sec:ingr}. One way to compute the nova rate theoretically is to assume that it is proportional to the rate of formation of C-O WDs (\citealt{Dantona1991}):
\begin{equation}
  R_{\rm novae}(t) =  \delta \int^8_{0.8}{\psi(t-\tau_{m_2}-\Delta t) \phi(m)dm},
\label{eq:novae}
\end{equation}
where $\Delta t \sim 1$ Gyr (but it can be longer) is the delay time between the formation of the WD and the first nova outburst, and $\delta$ is a free parameter representing the fraction of WDs in the IMF belonging to binary systems giving rise to novae, and is normalized to reproduce the present time nova rate, after assuming that each nova system suffers roughly $10^{4}$ outbursts during its lifetime. The time $\tau_{m_2}$ is the lifetime of the secondary star (the less massive one) which determines the start of the mass accretion on the WD. In our Galaxy, the present time nova rate is estimated to be $\sim$20--25 nova yr$^{-1}$ (see \citealt{dellavalle2020}). The value of $\delta$  derived by  (\citealt{Romano2003a}) for the Milky Way is $\sim 0.01$.

\begin{figure}[htbp]
  \centering
  \includegraphics[width=0.8\textwidth]{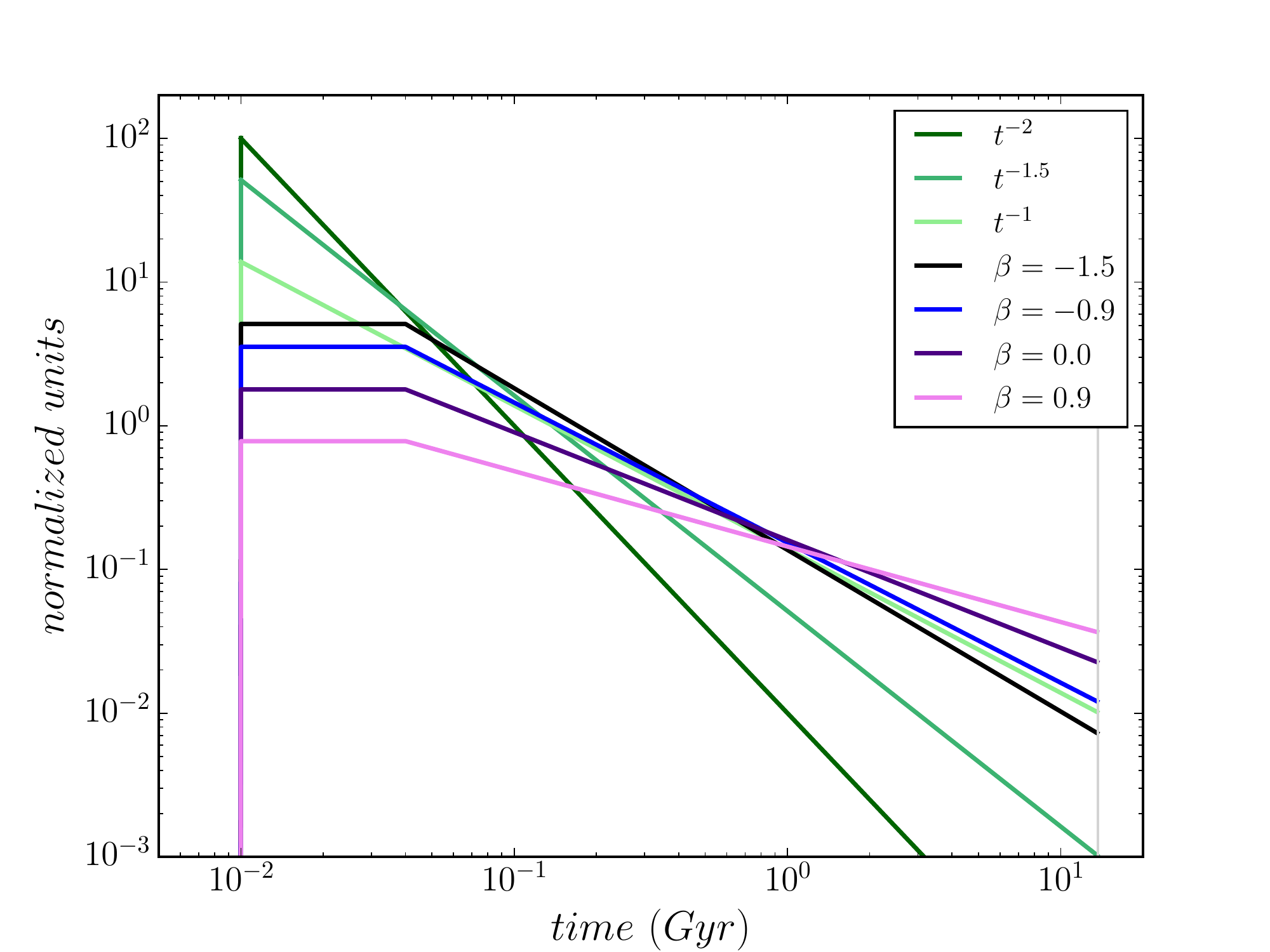}
  \caption{Seven different DTD functions for describing the MNS. Three are described by simple power law functions (${\rm DTD} \propto t^{-a}$, with $a=-1.0, -1.5, -2.0$). The other four DTDs are the functions derived by \citet{Simonetti2019} for different values of the parameter $\beta$ which is the exponent of the function describing the initial separation of the two  neutron stars ($g(S)\propto S^{\beta}$ with $S$ being the separation). In such cases, three different phases are visible in the DTD functions: an initial null plateau, a plateau representing the close binary sistems which promply merge and the tail for wide binary systems. The area under each curve is the same and equal to 1. Image reproduced with permission from \citet{Simonetti2019}, copyright by the authors.}
\label{fig:DTDMNS}       % Give a unique label
\end{figure}

\subsubsection{Rate of MNS (kilonovae)}
The gravitational-wave event GW170817 (\citealt{Abbott2017}), identified also as kilonova AT2017gfo, has confirmed that MNS, or kilonovae, can be important producers of r-process elements (see Sect.~\ref{sec:ingr}). For this reason, it is important to compute the MNS rate in galaxies.

In \cite{Matteucci2014} the MNS rate has been derived by simply assuming that
is proportional to the rate of formation of neutron stars.

In particular,
\begin{equation}
  R_{\rm MNS}(t) = \alpha_{\rm MNS} \int^{50}_{8}{\psi(t- \tau_m- \Delta_{\rm MNS}) \phi(m)dm},
  \label{eq:kilonovae}
\end{equation}
with $\alpha_{\rm MNS}$ being a free parameter indicating how many MNS systems are present in the mass range 9--50\,$M_{\odot}$, assumed to be the range of progenitors of neutron stars. I remind here that the upper mass in this range is largely uncertain depending upon mass loss, rotation, convection and other stellar physical parameters. The quantity $\Delta_{\rm MNS}$ represents the time necessary for the neutron stars to merge, after emission of gravitational waves, and is very important in order to reproduce the abundance pattern of Eu/Fe in the Milky Way, as we will see in the following sections. \cite{Matteucci2014} assumed a constant time delay of 1\,Myr, 10\,Myr and 100\,Myr, and normalized the value of $\alpha_{\rm MNS}$ to reproduce the present time observed rate of MNS in our Galaxy ($\sim 80 \,{\rm events}\, {\rm Myr}^{-1}$, as in \citealt{Kalogera2004}), and in such a case is derived to be $\alpha_{\rm MNS}=0.018$.

\citet{Simonetti2019}, where we address the reader for details, proposed a more refined calculation of the MNS rate based on a DTD function similar to what discussed for Type Ia SNe.

In particular, the suggested rate is:
\begin{equation}
  R_{\rm MNS}(t) = k_{\alpha} \int_{\tau_i}^{\min(t, \tau_x)}{\alpha_{\rm MNS}\psi(t-\tau){\rm DTD}_{\rm MNS}(\tau) d \tau},
  \label{eq:MNS}
\end{equation}
where $\kappa_{\alpha}$ is defined by Eq.~\eqref{eq:kalfa} and the ${\rm DTD}_{\rm MNS}$ is normalized as in Eq.~\eqref{eq:normSNIa}. The gravitational time delay $\Delta_{\rm MNS}$ has been substituted by the ${\rm DTD}_{\rm MNS}$ function. This is a more detailed representation of the MNS rate and the ${\rm DTD}_{\rm MNS}$ function depends upon the gravitational time delay and the distribution of the separations of the neutron stars, which is the most important parameter.

The distribution of separations is expressed as:
\begin{equation}
  g(S) \propto S^{\beta},
\label{eq:beta}
\end{equation}
where $S$ is the initial separation between the two neutron stars and $\beta$ is a free parameter. Figure~\ref{fig:DTDMNS} shows the DTDs for four different values of $\beta= -1.5; -0.9; 0.0; 0.9$, and also DTDs $\propto t^{a}$ with $a=-2; -1.5; -1.0$. \cite{Cote2018} had explored first the two DTD functions with $a=-1.0$ and $-1.5$.
The choice of the ${\rm DTD}_{\rm MNS}$ is very important in trying to reproduce the evolution of typical r-process elements, such as Eu, as we will see in the next Sections.
  
\section{Historical model approaches}
\label{sec:history}

In the past years, different approaches to the study of the chemical evolution of the Galaxy have been developed. We summarize them here:
\begin{itemize}

\item {\it The serial approach}

  The formation of the Galaxy is modeled by a continuous accretion of gas during which the halo, thick and thin disks are formed in a temporal sequence (one-infall model), as proposed by \cite{Chiosi1980, MatteucciGreggio1986, Matteucci1989, Boissier1999} among others, or by two different  infall episodes forming the halo plus thick disk and the thin disk, respectively (the two-infall model, \citealt{Chiappini1997}), occurring in a temporal sequence. \cite{Micali2013} proposed a three-infall model where the thick disk phase is treated separately from the halo and thin disk ones.

\item{\it The parallel approach}

  The formation of the Galaxy is modeled by different episodes of gas accretion occurring in parallel but at different rates (e.g., \citealt{Pardi1995, Chiappini2009, Grisoni2017}).
  
\item{\it The stochastic approach}

  This approach is relative only to the formation of the stellar halo, when the gas mixing was probably not efficient and a large spread in the chemical abundances in stars is expected, reflecting the pollution by single supernovae. Among the papers of this kind I remind here as examples, those of \cite{Argast2000} and \cite{Cescutti2008}.

\item{\it The stellar accretion approach}

  Also this approach applies only to the stellar halo and assumes that it totally or partly formed by accretion of stars belonging to dwarf galaxies satellites of the Milky Way (e.g., \citealt{Prantzos2008b, delucia2008}). This approach follows from the original suggestion of \cite{Searle1978}, who suggested that the outer Galactic halo formed by merging of sub-galactic fragments over a long timescale. This assumption has mainly been explored by semi-analytical models of galaxy formation, and we will discuss them in more detail in the following sections.
  
\end{itemize}

\subsection{Serial approach and time delay model}
The serial approach was the first proposed: it assumes that one-infall episode has given rise first to the halo and then to the disk(s). In other words, the stars formed with ISM metallicity up to ${\rm [Fe/H]}=-1.0$\,dex belong to the halo, while the stars formed when ${\rm [Fe/H]}>-1.0$\,dex belong to the disk(s). These models predict a unique line as a function of time describing the average evolution of chemical abundances. An example of this approach is shown in Fig.~\ref{fig:MG86}, where the predicted [O/Fe] vs.\ [Fe/H] relations are compared with the data available at that time. The best model in Fig.~\ref{fig:MG86} (model a) was obtained by assuming that 2/3 of the Fe production comes from Type Ia SNe and 1/3 from CC-SNe (\citealt{MatteucciGreggio1986}), in excellent agreement with the recent nucleosynthesis yields by \cite{Kobayashi2020}. In fact, CC-SNe do produce some Fe and this fraction is important, because if one assumes that only Type Ia SNe do produce Fe, the agreement with data is lost (line b in Fig.~\ref{fig:MG86}). This figure is a very nice representation of the so-called \emph{time-delay model}, namely the interpretation of the [O/Fe] vs.\ [Fe/H] diagram: at low metallicity only CC-SNe contribute to Fe enrichment and the [O/Fe] ratio is roughly flat, since it reproduces the (O/Fe) production ratio in massive stars, while for ${\rm [Fe/H]}>-1.0$\,dex the SNe Ia start to be important in the Fe production and continue to be so up to the solar metallicity. From the best model in Fig.~\ref{fig:MG86}, one can extract the timescale for the formation of the Galactic halo: this timescale is simply the time at which the ISM reached the abundance ${\rm [Fe/H]}=-1.0$\,dex, which roughly marks the maximum halo star metallicity. \cite{MatteucciGreggio1986} derived  this timescale to be $\sim$ 1.0--1.5 Gyr.

\begin{figure}[htbp]
  \includegraphics[width=\textwidth]{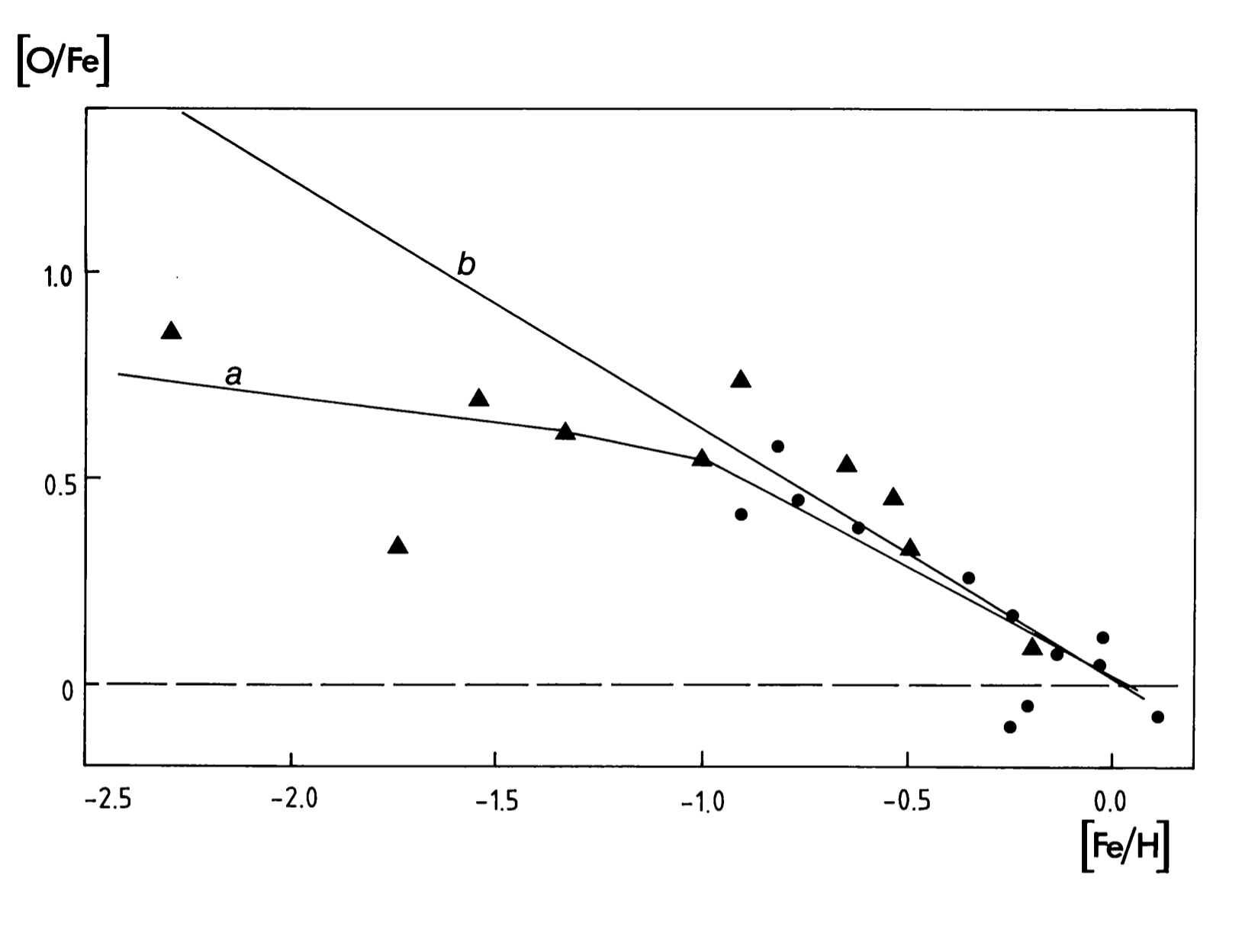}
  \caption{The plot of [O/Fe] vs.\ [Fe/H] for solar neighbourhood stars compared with predictions from a one-infall model (\citealt{MatteucciGreggio1986}). The line labelled a) represents a model where Fe is produced in both CC-SNe  and SNe Ia, while the line labelled b) shows the predictions of a model with Fe contributed only by SNe Ia. The data points are from \cite{Clegg1981} (circles) and \cite{Sneden1979} (triangles). Image reproduced with permission from \cite{MatteucciGreggio1986}, copyright by ESO.}
\label{fig:MG86}       % Give a unique label
\end{figure}

\subsection{Parallel approach}
A typical example of parallel approach can be found in \cite{Pardi1995} where the model predicts three separate curves describing the evolution of the abundances and abundance ratios as functions of time and metallicity. The formation of the three stellar components (halo, thick and thin disk) starts at the same time, but the bulk of star formation occurs at different times. Moreover, there is a connection in the formation of halo-thick disk and thick-thin disk: such connections are represented by the gas lost from the halo which forms the thick disk and the gas shed from the thick which goes to form the thin disk. In Fig.~\ref{fig:pardi} we show the predictions of such a model in comparison with observations. The SFR during the halo and thick disk phases is assumed to be high and short, whereas in the thin disk is lower and more extended in time. It is worth noting that all the predicted curves cover the entire range of metallicities, but this is the evolution of the [O/Fe] ratio in the ISM, which should be compared with the stellar abundance ratios and therefore it is important to know also how many stars have formed at each metallicity. This is described by the stellar MDF. For the halo and thick disk, the numbers of stars formed at metallicities larger than ${\rm [Fe/H]}=-1.0$\,dex for the halo, and $-0.6$\,dex for the thick disk, are negligible. Therefore, Fig.~\ref{fig:pardi} simply suggests that there could be a negligible numberof halo and thick disk stars at high metallicity. However, this parallel model has a weak point: the connection in the formation of halo-thick disk and thick-thin disk is not realistic since the thin disk is much more massive than the halo and thick disk and is hard to explain how it could have been formed from gas shed by the thick disk. Moreover, the parallel approach of \cite{Pardi1995} is at variance with the distribution of the stellar angular momentum per unit mass, indicating that the thick disk did not
form out of gas shed by the halo (\citealt{Wyse1992}).

In the modern view, the Galactic halo and thick disk  might have formed by accretion of extragalactic gas and/or stellar systems and the thin disk out of extragalactic gas. As we will discuss later, in \cite{Chiappini1997},  the two-infall model assumes two main independent gas infall episodes: during the first, the halo and the thick disk formed, while during the second the thin disk assembled on a longer timescale than the other two components.

\begin{figure}[htbp]
  \includegraphics[width=\textwidth]{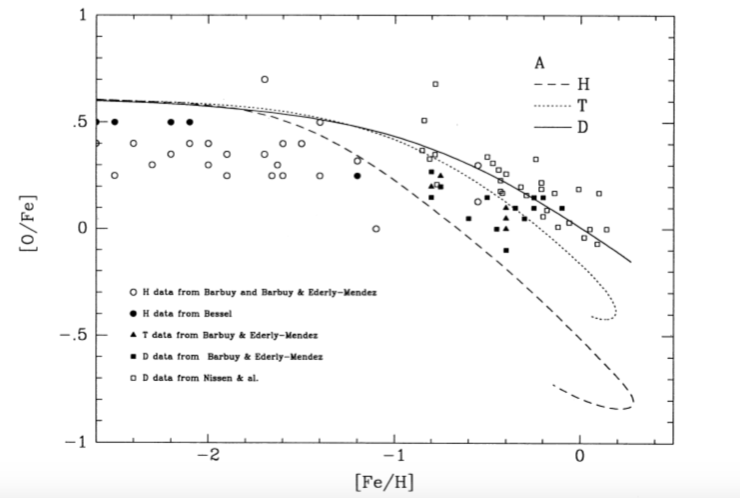}
  \caption{Comparison between predicted [O/Fe] vs.\ [Fe/H] in the solar vicinity for the three Galactic components: halo (H), thick disk (D) and thin disk (T).
  One curve is predicted for each component and the three curves can represent the observed spread in the data. Data from \cite{Barbuy1988, Barbuy1989} and \cite{Nissen1991} as indicated by the legenda inside the figure. Image reproduced with permission from \cite{Pardi1995}, copyright by AAS.}
\label{fig:pardi}       % Give a unique label
\end{figure}

In \cite{Grisoni2017}, it was proposed a new version of the parallel model applied to the thick and thin disks. In particular, it was assumed that the formation by different gas infall episodes of the two disks started at the same time but it proceeded at different rates: the thick disk evolved faster with a high SFR and a short timescale of gas infall, whereas the thin disk assembled on a much longer timescale and moderate SFR. This approach could explain the double sequence of [Mg/Fe] vs.\ [Fe/H] found for the stars in the two disks, as it is evident in AMBRE data (\citealt{delaverny2013}), and shown in Fig.~\ref{fig:G171}.
This ``bimodality'' in the [$\alpha$/Fe] ratios has been found also in APOGEE data (\citealt{Nidever2014, Hayden2015}) and it will be extensively discussed later.

\begin{figure}[htbp]
  \includegraphics[width=\textwidth]{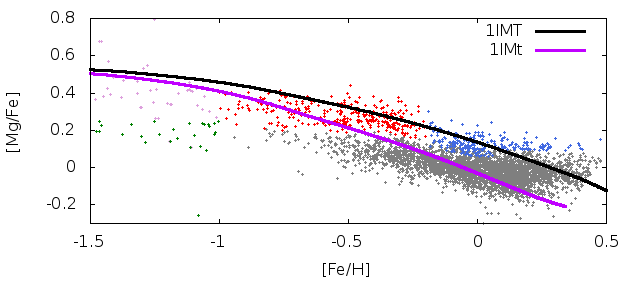}
  \caption{Comparison between predicted [Mg/Fe] vs.\ [Fe/H] in the solar vicinity for the thick and thin disks.The black line represents the prediction for the thick disk while the magenta line for the thin disk. Here the timescale for the formation of the thick disk is $\sim 1$ Gyr whereas that for the thin disk is $\sim 8$\,Gyr. The red points are data relative to thick disk stars, while the grey points are thin disk stars (AMBRE data). The blue points represent stars whose origin is not well understood, they could either be high metallicity thick disk stars or stars migrated from the inner Galactic regions. Image reproduced with permission from \cite{Grisoni2017}, copyright by the authors.}
\label{fig:G171}       % Give a unique label
\end{figure}

\subsection{Stochastic approach}
In the framework of the stochastic approach (\citealt{Travaglio2001, Suzuki2001, Argast2004, Cescutti2008, Cescutti2015, Haynes2019, Voort2015}), a large spread is predicted for all the abundance ratios in halo stars, at least for ${\rm [Fe/H]}< -3.0$\,dex. From an observational point of view, such a large spread is present only for s- and r-process elements, while it is much lower for other elements, such as $\alpha$-elements. As an example of that, we show in Fig.~\ref{fig:Cesc08} the predictions of the inhomogeneous stochastic chemical model by \cite{Cescutti2008}. In particular, in Fig.~\ref{fig:Cesc08} the [Ba/Fe] vs.\ [Fe/H] and the observational data (red triangles) show a large spread, which is well reproduced by the model. Also in this figure, we show the same model applied to [Mg/Fe] vs.\ [Fe/H] and in this case the observed and predicted spreads are both smaller, indicating that if the observed spread is real there are some elements which are more affected by inhomogeneous mixing than others. The different spread observed in s-and r-process elements versus $\alpha$-elements, has been interpreted by \cite{Cescutti2008} as due to the fact that these two groups of elements are produced in different stellar mass ranges, coupled with a random birth of the stars. In particular, the mass range for production of neutron capture elements is resctricted to $\sim$12--30\,$M_{\odot}$, whereas the mass range for production of $\alpha$-elements (e.g. O, Mg) is the whole range of massive stars. However, in subsequent works, the possibility of producing n-capture elements (such as Ba) in the mass range 12--20\,$M_{\odot}$ has been rejected (\citealt{Kobayashi2020}). In any case, the observed spread in s- and r-process elements at low metallicity is still a matter of debate (see also later).

\begin{figure}[htbp]
  \includegraphics[width=15pc]{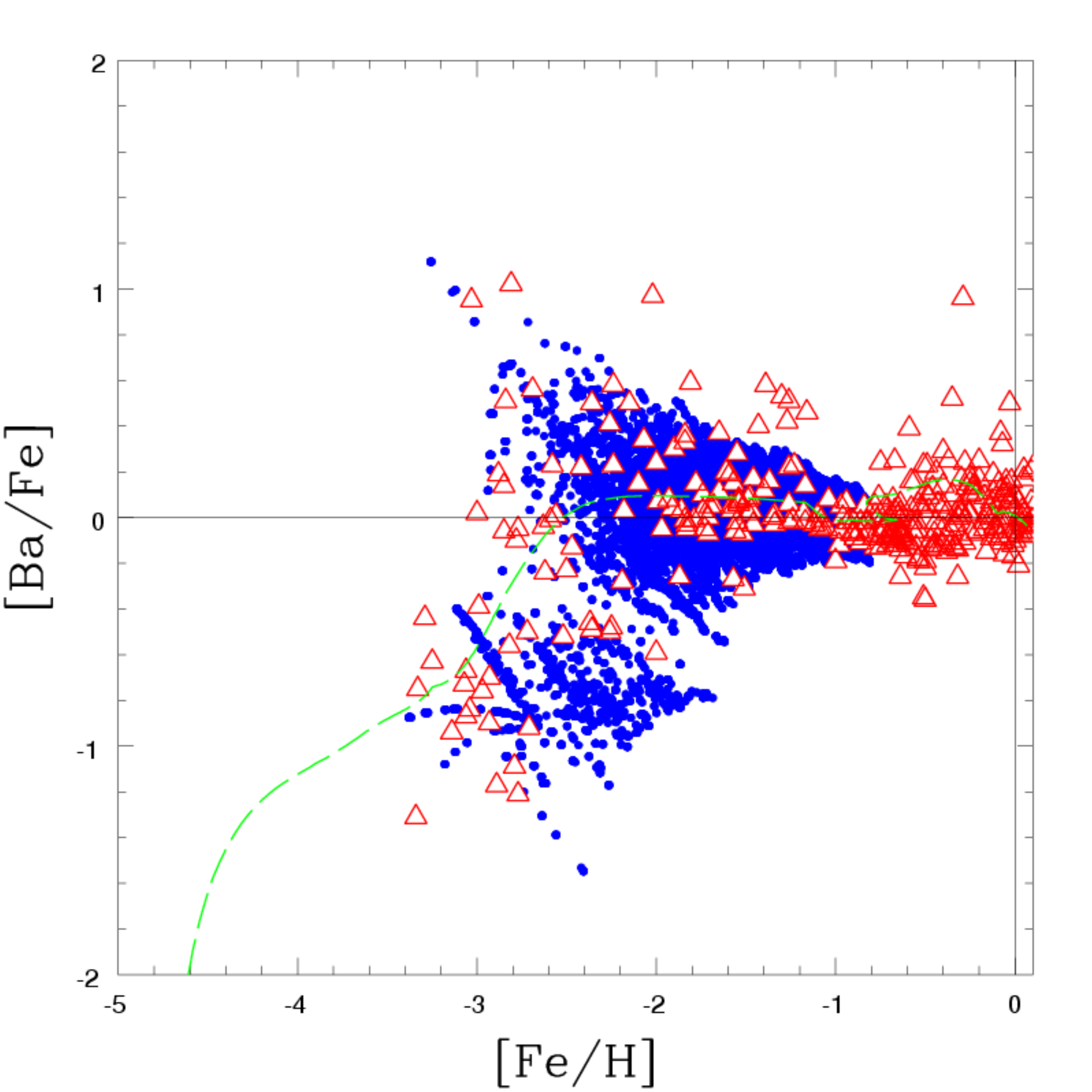}
   \includegraphics[width=15pc]{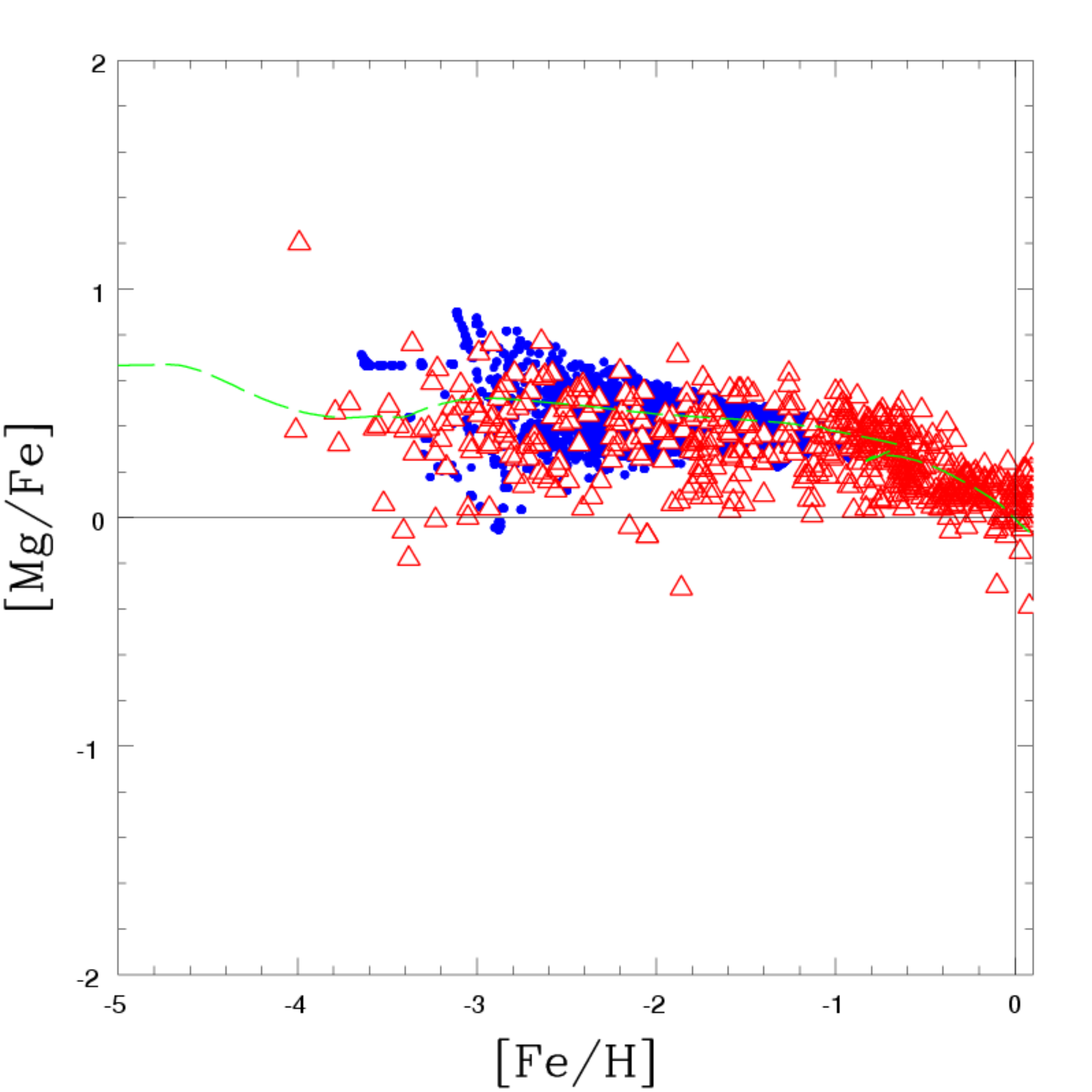}
   \caption{Comparison between predicted [Ba/Fe] vs.\ [Fe/H] in the solar vicinity and observational data (left panel) and [Mg/Fe] vs.\ [Fe/H] (right panel). The blue points represent the predictions of an inhomogeneous stochastic chemical evolution model for the Galactic halo, while the red triangles are the observational points. The dashed green lines are the predicted average behaviours for the [Ba/Fe] and [Mg/Fe] ratios, as  predicted by a homogeneous model (\citealt{Cescutti2006}). Image reproduced with permission from \cite{Cescutti2008}, copyright by ESO, where details on the model and references to the data can be found.}
    \label{fig:Cesc08}       % Give a unique label
\end{figure}

\subsection{The classical two-infall model}
\cite{Chiappini1997} proposed the so-called \emph{two-infall model} assuming two main independent but sequential in time infall episodes giving rise to the halo-thick disk and thin disk, respectively.
The infall law proposed in that paper is:
\begin{equation}
  A(R,t)= a(R)e^{-t/\tau_H} +b(R)e^{-(t-t_{\max})/\tau_D(R)},
\label{eq:twoinfall}
\end{equation}
where $a(R)$ and $b(R)$ are parameters, similar to $\Lambda$ in Eq.~\eqref{eq:infall}, and are derived by imposing that they reproduce the total surface mass density at the present time of the halo-thick disk and thin disk, respectively. The quantities $\tau_H$ and $\tau_D(R)$ are the timescale for the formation of the halo-thick disk and the thin disk, respectively. The quantity $t_{\max}$ is the time for the maximum infall onto the thin disk. In the best model of \cite{Chiappini1997} and \cite{Chiappini2001}, $t_{\max}=1$ Gyr, whereas $\tau_H= 0.8$ Gyr and $\tau_D(R)$ is assumed to vary with Galactocentric distance in such a way that the thin disk forms inside-out:
\begin{equation}
  \tau_D(R)= 1.033R({\rm kpc}) -1.267 \, {\rm Gyr},
\label{eq:tauD}
\end{equation}
where $\tau_D(8) =7\, {\rm Gyr}$ is the timescale for the formation of the solar vicinity.
The SFR was assumed to be a Schmidt--Kennicutt law with a dependence also on the total surface mass density:
\begin{equation}
  \psi(R,t) = \nu(t)\,\left( \frac{\sigma(R, t)}{\sigma(R_\odot, t)} 
\right) ^{2\,(k - 1)}\,\left( \frac{\sigma(R, t_{\rm G})}{\sigma(R, t)} \right)
^{k - 1}\,\sigma_{\rm gas}^{k}(R, t),
\label{eq:SFRChiap}
\end{equation}
where $\nu(t)$ is the efficiency of the star formation process, $\sigma(R,t)$ 
is the total surface mass density at a given radius $R$ and given time $t$, 
$\sigma(R_\odot, t)$ is the total surface mass density at the solar position 
and $\sigma_{\rm gas}(R, t)$ is the gas surface density.
%The quantity $t_{\rm gal}$ is the Galactic lifetime.
The gas surface 
density exponent, $k=1.5$, in agreement with \cite{Kennicutt1998} ($k=1.4 \pm 0.15$).
The efficiency of star formation was assumed to be $\nu_H=2\,{\rm Gyr}^{-1}$ and $\nu_D=1\,{\rm Gyr}^{-1}$ for the halo-thick and thin disk, respectively. The adopted IMF was that of \cite{Scalo1986}.
They assumed also a gas density threshold in the star formation, namely that star formation stops when $\sigma_{\rm gas} \le 7\,M_{\odot}\, {\rm pc}^{-2}$ (\citealt{Kennicutt1989}). Such a threshold has the effect of creating a gap of $<1\,{\rm Gyr}$ in the SFR between the end of halo-thick disk  and thin disk formation. Such a gap in the SFR should be visible in the abundance ratios versus metallicity, as pointed out by \cite{Gratton2000} and \cite{Fuhrmann1998}. In particular, in Fig.~\ref{fig:Chiap01} from \cite{Chiappini2001} we can see that the existence of a gap in star formation creates a loop in the curve describing the [Fe/O] vs.\ [O/H] relation in the solar vicinity, in particular, at [O/H]$\sim -0.5$\,dex, corresponding roughly to ${\rm [Fe/H]} =-1.0$\,dex, This gap is visible also in the data, as discussed in \cite{Gratton2000}. The behaviour of the [Fe/O] ratio in Fig.~\ref{fig:Chiap01} is due to the fact that when star formation stops, O is no more produced while Fe continues to be produced by Type Ia SNe. As a consequence of that, the [Fe/O] ratio increases while [O/H] remains roughly constant, or sligthly diminishes, because of the infall related dilution, then when star formation starts again the [Fe/O] ratio increases for increasing [O/H]. In \cite{Chiappini1997} and \cite{Chiappini2001}, this effect is entirely due to the existence of a gas threshold which is not yet confirmed by observations; however, any event that could stop star formation at the end of the thick disk phase would predict the loop in the abundance ratios (e.g., $\alpha$-elements/Fe).  In \cite{Fuhrmann1998}, a similar effect is observed for Mg, as shown in Fig.~\ref{fig:Fuhr}, where [Fe/Mg] is plotted versus [Mg/H]; here, the abrupt change of [Fe/Mg] ratio at ${\rm [Mg/H]} \sim -0.4$\,dex shows a paucity of old stars and is indicative of a time when the enrichment of Mg from CC-SNe had stopped while Fe enrichment continued, due to Type Ia SNe, even in absence of star formation.

\begin{figure}[htbp]
  \centering
\includegraphics[width=0.8\textwidth]{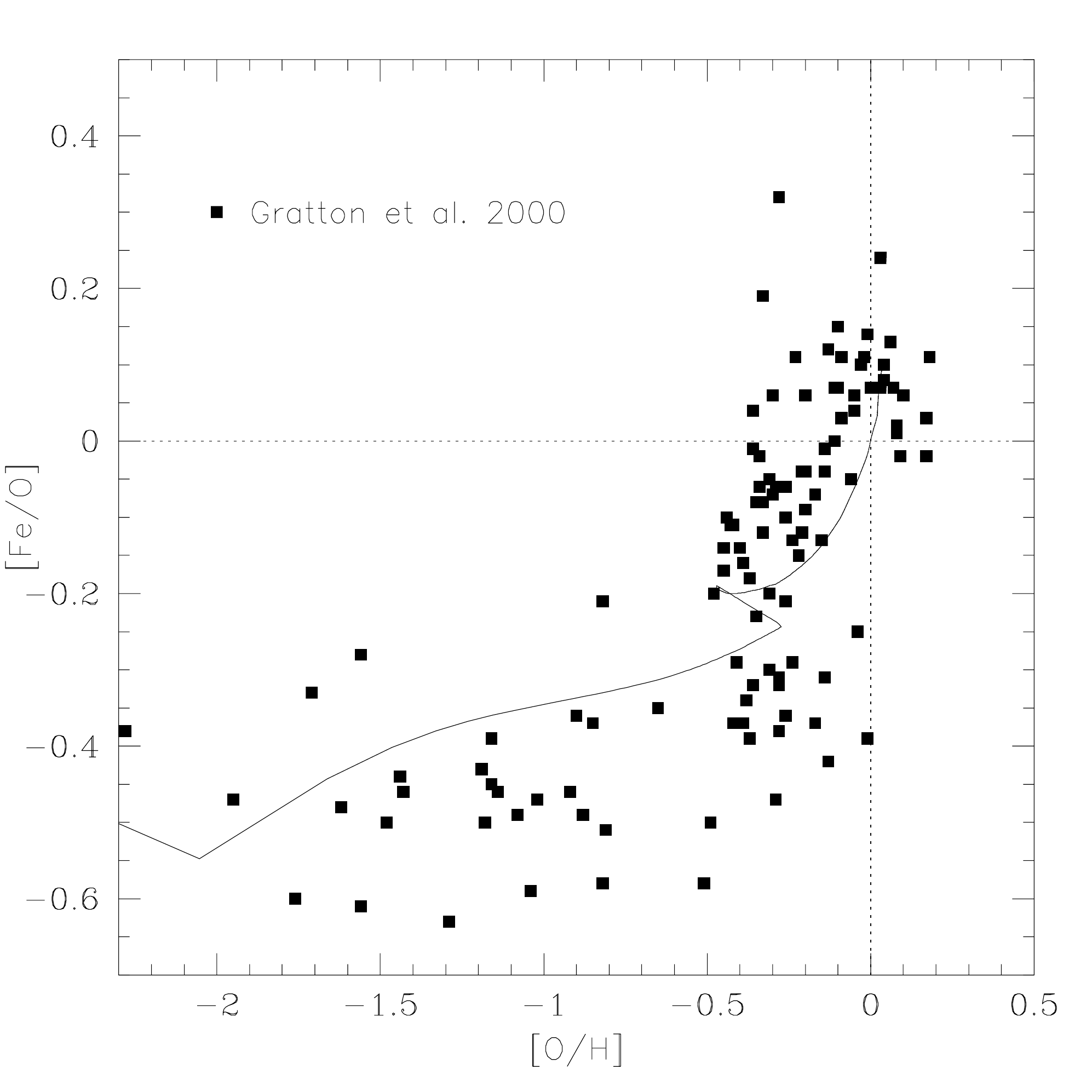}
  \caption{Comparison between predicted and observed [Fe/O] vs.\ [O/H] in the solar vicinity. The points represent the observations of \cite{Gratton2000}. As one can see, the data seem to show a lack of stars corresponding to ${\rm [O/H]} \sim -0.5$\,dex. The model prediction shows a loop in correspondance of the same abundance, and this is due to the stop in the star formation produced by the assumption of a gas density threshold for star formation, as discussed in the text. Image reproduced with permission from \cite{Chiappini2001}, copyright by AAS.}
\label{fig:Chiap01}       % Give a unique label
\end{figure}

\begin{figure}[htbp]
  \centering
\includegraphics[width=0.9\textwidth]{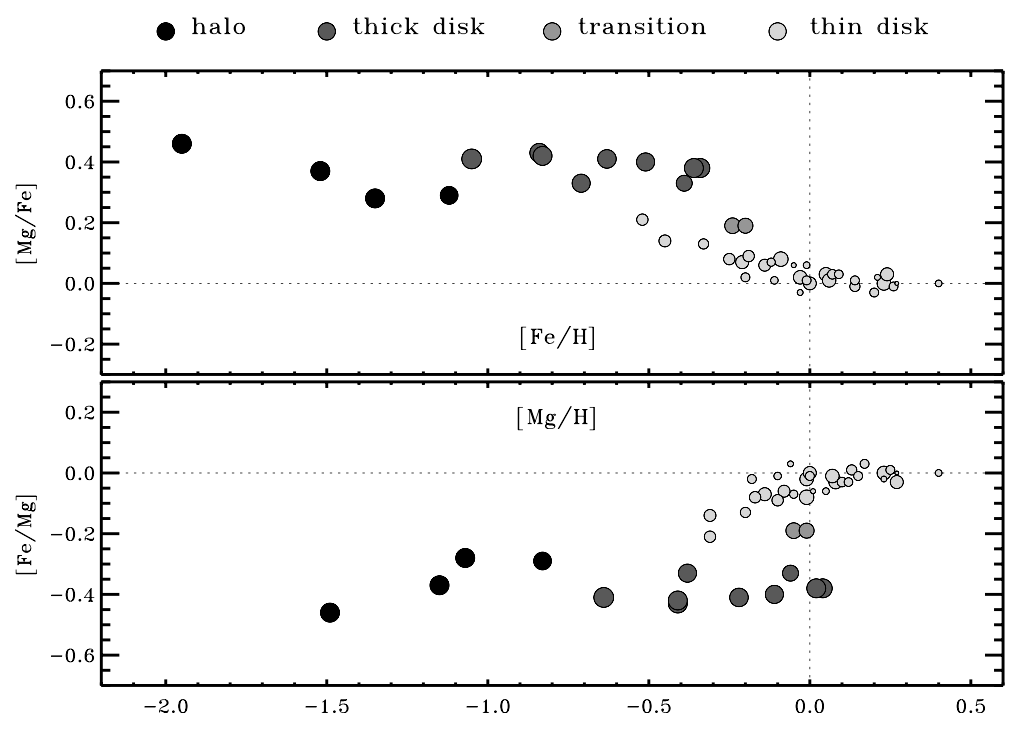}
\caption{Plot of the [Mg/Fe] vs.\ [Fe/H] for solar vicinity stars (upper panel) and [Fe/Mg] vs.\ [Mg/H] (lower panel). The stars belonging to halo, thick, thin disk and transition are indicated. Image reproduced with permission from \cite{Fuhrmann1998}, copyright by ESO.}
\label{fig:Fuhr}       % Give a unique label
\end{figure}

%\begin{figure}[htbp]
%  \centering
%\includegraphics[width=0.9\textwidth]{snaith_2014.png}
%\caption{Estimated SFR vs.\ time for the solar vicinity by \cite{Snaith2014}. It is evident a dip in the SFR 8--9 Gyr ago. This SFR history has been derived by reproducing the [Si/Fe] vs.\ age. The red region represents the $1\sigma$ error and the red horizontal error bar shows the estimated errors in the age (\citealt{Haywood2013}). In the Figure for comparison is shown the black curve obtained by selecting Milky Way-like galaxies via halo abundance matching (\citealt{vanDokkum2013}). Image reproduced with permission from \cite{Snaith2014}.}
%\label{fig:Snaith}       % Give a unique label
%\end{figure}

This effect is present, although less clear, also in the plot [Mg/Fe] vs.\ [Fe/H] in Fig.~\ref{fig:Fuhr}.  More recently, the gap in the SFR after the formation of the thick disk was suggested by \cite{Snaith2014}, who analysed the [Si/Fe] vs.\ age plot based on a sample of F, G  and K dwarfs in the solar vicinity from \cite{Adibekyan2012}. According to their model, the Galaxy underwent an intense phase of star formation between 9 and 13 Gyr ago and this burst formed the thick disk. The estimated SFR shows a dip at 8--9\,Gyr ago and this corresponds to the end of the thick disk phase. Therefore,  there are hints that a dip or a gap in the SFR occurred before the formation of the thin disk.

\begin{figure}[htbp]
  \centering
  \includegraphics[width=0.9\textwidth]{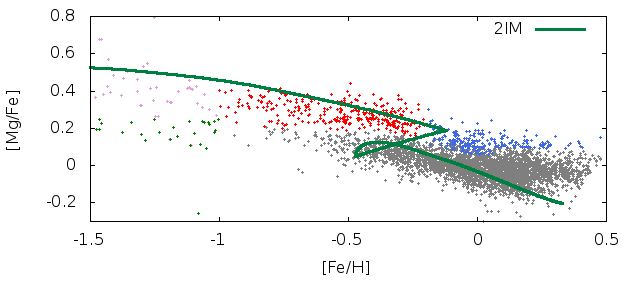}
  \caption{The plot of [Mg/Fe] vs.\ [Fe/H]: model versus AMBRE data (as in Fig.~\ref{fig:G171}). The model prediction (green line) is from the two-infall model of \cite{Grisoni2017}. Image reproduced with permission from \cite{Grisoni2017}, copyright by the authors.}
\label{fig:G172}       % Give a unique label
\end{figure}

\begin{figure}[htbp]
  \centering
  \includegraphics[width=0.9\textwidth]{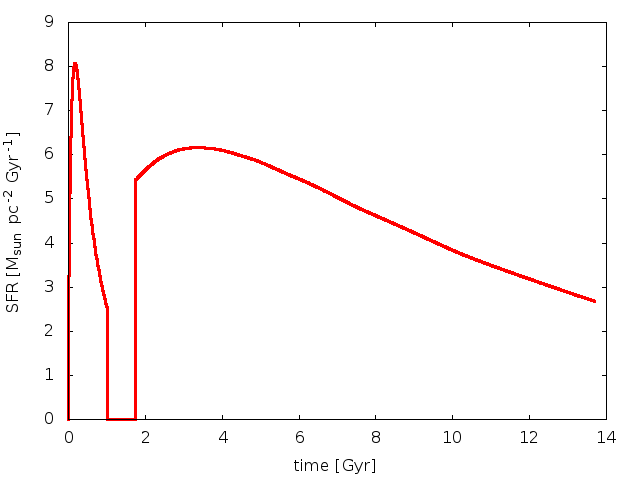}
  \caption{The SFR versus time as predicted by the two-infall model. The gap in the SFR is evident and occurred over 12\,Gyr ago  and lasted $\sim 1$ Gyr, as in the original model of \cite{Chiappini1997}. Image reproduced with permission from \cite{Grisoni2017}, copyright by the authors.}
\label{fig:SFR}       % Give a unique label
\end{figure}

In summary, the two-infall model main results are: i) the thin disk in the solar neighbourhood formed by infall of primordial gas on a long timescale (7--8 Gyr) and ii) this timescale increases with Galactocentric distance giving rise to an inside-out formation of the disk, as already suggested by \cite{Matteucci1989}. iii) The existence of a gas threshold in the SFR in the thin disk creates a gap in the star formation process between the formation of the two disks. This model can still reproduce the most recent data in the solar vicinity; in Fig.~\ref{fig:G172}, we show the results of \cite{Grisoni2017} obtained with the two-infall model applied to the thick and thin disks. As one can see, the loop due to the stop in the star formation is visible, and is due to a gas density treshold of $7 \,M_{\odot}\,{\rm pc}^{-2}$. The timescale for accretion of the thick disk is $\tau_{\rm thick}=0.1\,{\rm Gyr}$ and for the thin disk is $\tau_{\rm thin}=7\,{\rm Gyr}$; the efficiency of star formation is $2\,{\rm Gyr}^{-1}$ for the thick and $1\,{\rm Gyr}^{-1}$ for the thin disk. The adopted IMF is that of \cite{Kroupa1993} for the solar vicinity. The model can reproduce the abundance patterns of thick and thin disk stars, but not the high [$\alpha$/Fe] at high metallicity blue stars (the same shown in Fig.~\ref{fig:G171}). These stars, in the framework of the two-infall model can be explained only as stars migrated from the inner Galactic regions. Stellar migration represents a different and auxiliary paradigm for interpreting the abundance patterns in the solar vicinity, as we will see in the \ref{sec:migr}. Figure~\ref{fig:SFR} displays a plot of the SFR vs.\ time, as predicted by the two-infall model of \cite{Grisoni2017}, and it shows a clear gap in the star formation history, just before the formation of the thin disk.

\subsubsection{A revised two-infall model and the bimodality in the [$\alpha$/Fe] ratios}
A revised two infall model, with the second infall occurring with a delay of $\sim$ 4.3 Gyr relative to the previous one has been suggested by \cite{Spitoni2019}. By means of this model, they were able to reproduce the bimodality in the [$\alpha$/Fe] vs.\ [Fe/H] diagram  ($\alpha$= Mg+Si) for thick and thin disk stars, as observed by APOKASC (APOGEE + Kepler Asteroseismology Science Consortium), as well as the stellar ages measured by asteroseismology, suggesting a large gap in age between thick and thin disk stars (\citealt{Silva2018}). A similar conclusion has been reached by \cite{Noguchi2018}, who suggested two main episodes of cold gas infall, with a hiatus in the star formation 6--7 Gyr ago, in order to reproduce the bimodality. In Fig.~\ref{fig:Spito19} we show the comparison between the predictions by \cite{Spitoni2019} and the APOKASC data.
%The model predicts a gap in the SFR of $\sim 4.3$ Gyr, which can be identified by the parameter $t_{\max}$, and  produces a larger loop in the [$\alpha$/Fe] ratio than the classical two-infall model where the gap in the SFR lasts $< 1$ Gyr, in the [$\alpha$/Fe] ratio and allows us to better explain  the data for the thin disk.
%% In Fig.~\ref{fig:Spito19},
Also displayed is the effect of varying the duration of the gap in star formation (indicated with $t_{\max}$): as one can see, a decrease in $t_{\max}$ produces a smaller loop starting at lower metallicities.
%It is worth noting, that this strong ``bimodality'' between the abundance ratios in thick disk and thin disk stars  is not so evident in the AMBRE sample, shown in Figs.~8 and 13.
Therefore, the flat [$\alpha$/Fe] ratio observed in APOGEE thin disk stars can indeed be due to the strong infall episode which forms the thin disk and occurs at the end of the thick disk phase, a suggestion also made by \cite{Calura2009}, by means of a semi-analytical model of galaxy evolution. A different explanation for the bimodality in the [$\alpha$/Fe] distribution is provided by \cite{Buck2020} and \cite{Sharma2020}, suggesting that it can be explained by stellar migration (see later).

In conclusion, from all the previous discussion it seems that most of the data require separate gas accretion episodes to explain the abundances in the thick and thin disk stars, although the real delay between the two episodes, as well as the physical reason for such a delay, are still not well established.

\begin{figure}[htbp]
  \centering
  \includegraphics[width=0.66\textwidth]{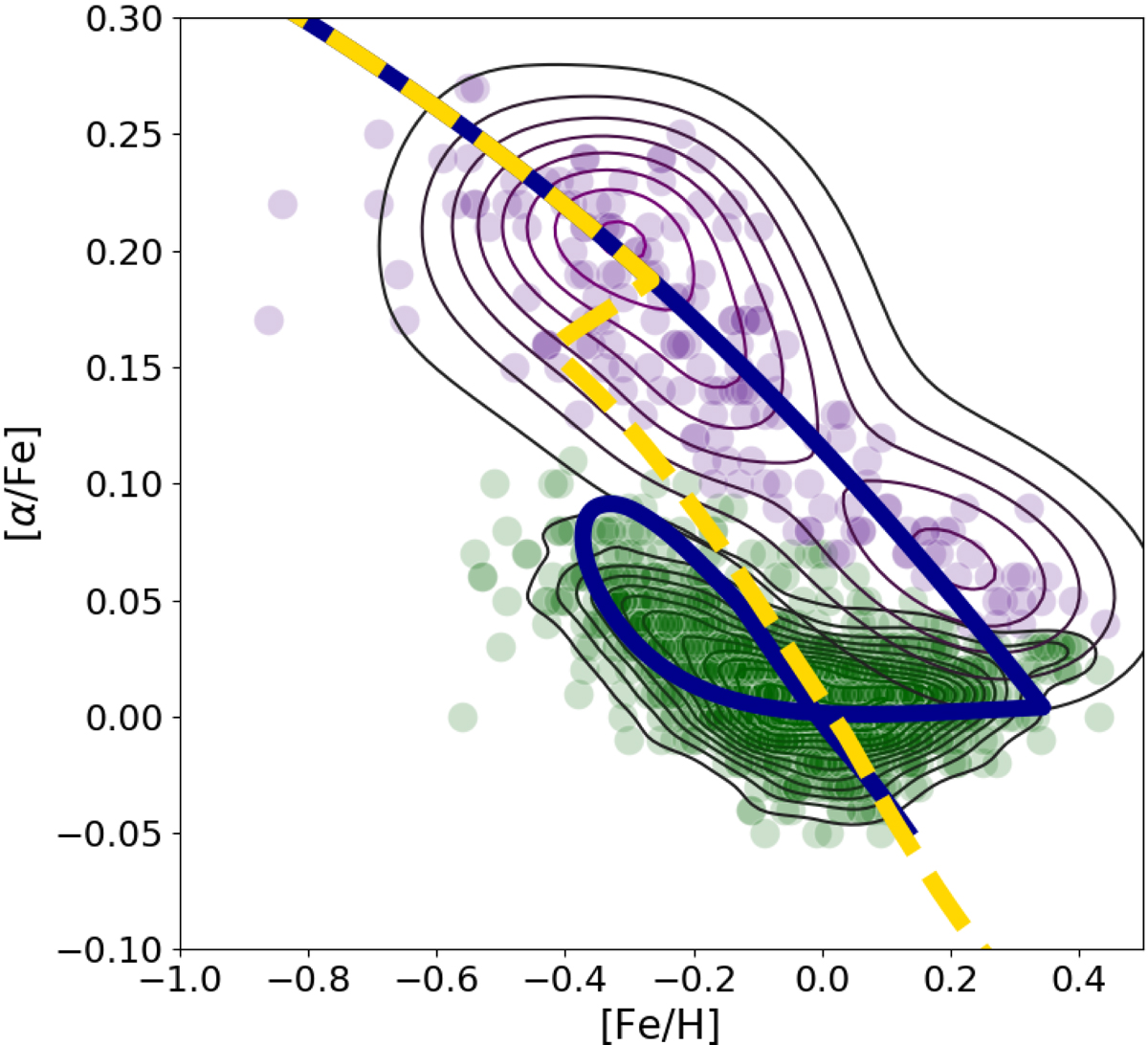}
  \includegraphics[width=0.8\textwidth]{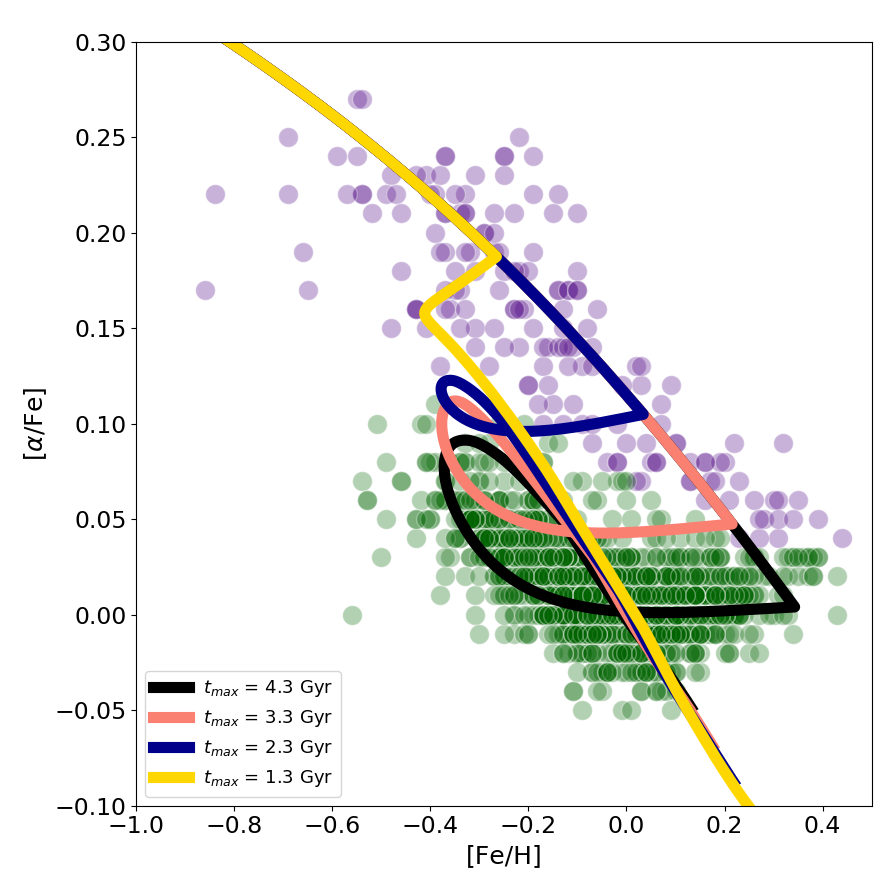}
  \caption{{\it Upper panel}: comparison between predicted [$\alpha$/Fe] vs.\ [Fe/H] in the solar vicinity and observational data by APOKASC (the purple filled circles are the observed ``high-$\alpha$ stars'' and green filled circles are the ``low-$\alpha$ stars''. The black-purple and black-green contour lines enclose the observed high- $\alpha$ and low-$\alpha$ stars, respectively. The blue line is the prediction of \cite{Spitoni2019}, while the yellow dashed line represents the prediction of the classical two-infall model. {\it Lower panel}: the effect of changing the timescale of the gap in the star formation ($t_{\max}$) between the thick and thin disk. Image reproduced with permission from \cite{Spitoni2019}, copyright by ESO.}
    \label{fig:Spito19}       % Give a unique label
\end{figure}

\subsection{A three-infall model}
It is worth mentioning another approach to the formation and evolution of the Milky Way: the ``three-infall model'' by \cite{Micali2013}. In this model, halo, thick and thin disk were formed by means of three different infall episodes separated by periods of quiescent star formation. Therefore, a continuous and not parallel evolution, similar to the two-infall model but with the thick disk well separated from the halo and thin disk. Thresholds in the gas density for star formation are also present in this model (4, 5 and 7\,$M_{\odot}\,{\rm pc}^{-2}$ for halo, thick and thin disk, respectively). By fitting of the solar vicinity [X/Fe] vs.\ [Fe/H] diagrams, they suggested that the halo formed on a timescale of 0.2\,Gyr, the thick disk of 1.25\,Gyr and the thin disk, at the solar ring, of 6\,Gyr. This model was also able to reproduce simultaneously the different observed stellar MDFs in the thick and thin disk, a constraint not always considered in chemical model.

\begin{figure}[htbp]
  \centering
  \includegraphics[width=0.8\textwidth]{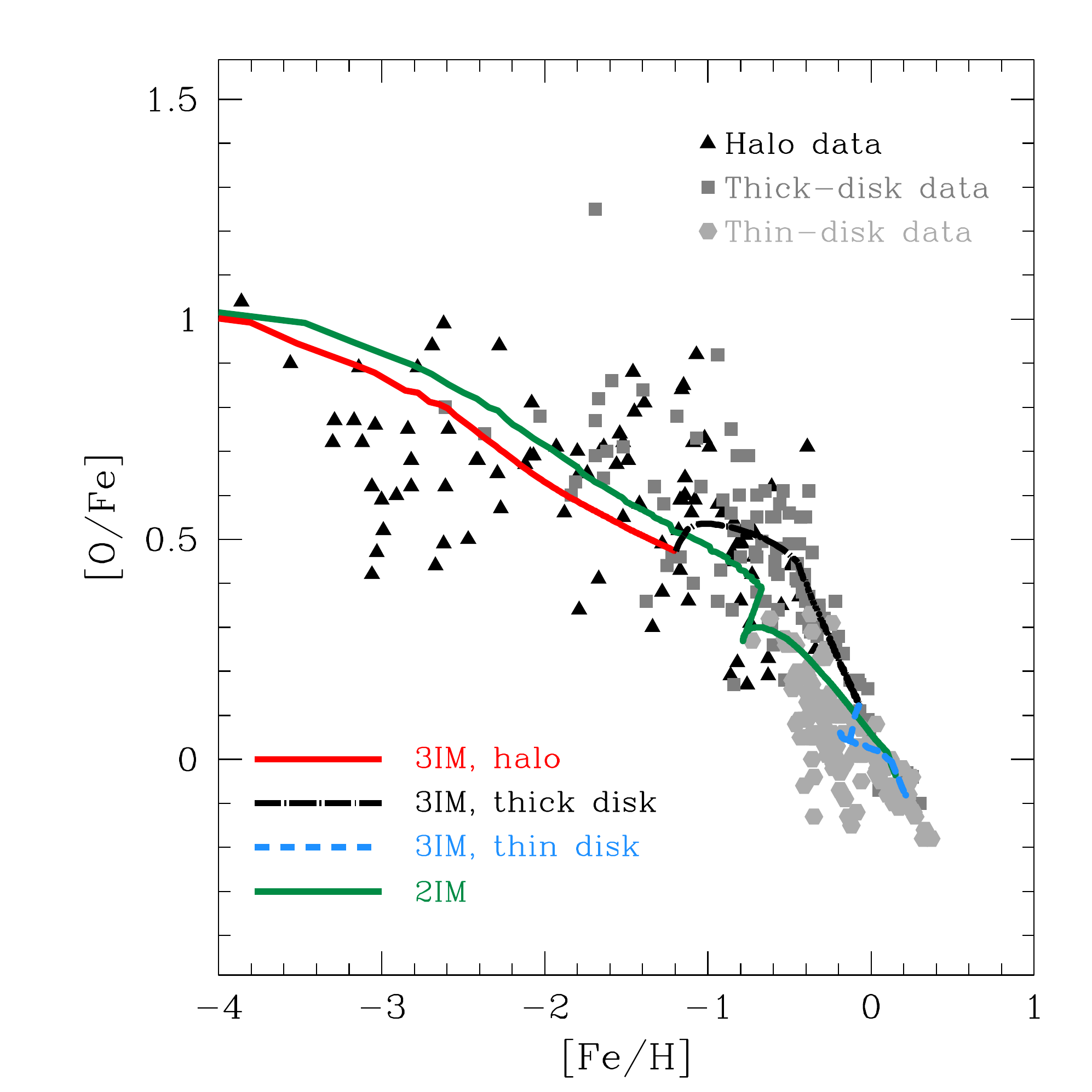}
  \caption{The plot of [O/Fe] vs. [Fe/H]: model predictions from the three-infall model (labelled 3IM, red, black and blue lines) and from the two-infall model (labelled 2IM, green line) compared to observations. References to the data can be found in \cite{Micali2013}. Image reproduced with permission from \cite{Micali2013}, copyright by the authors.}
\label{fig:3inf}       % Give a unique label
\end{figure}

As one can see from Fig.~\ref{fig:3inf}, the three-infall model predicts a more pronounced thick disk phase (black line) and shorter periods of quenched star formation between halo and thick and thick and thin disks, relative to the two-infall model.

\section{The evolution of heavy elements in halo and disks}
\label{sec:heavyhd}

Numerical chemical evolution models are able to follow in detail the evolution of the abundances of a large number of chemical elements, both light (He, D, Li, Be and B) and heavy (C, N, O, $\alpha$-elements, Fe-peak elements, s- and r- process elements).

Here, we will summarize  the results obtained in the last 10 years about the evolution of heavy elements. In Fig.~\ref{fig:RKTM} we report the results of \cite{Romano2010}; these authors showed the abundances predicted by means of the two-infall model and two different sets of stellar yields, which are the most important parameters in galactic chemical evolution (see \citealt{Cote2018}). From the comparison between model predictions and observations, it is evident that one set of yields should be preferred in order to reproduce the observed abundance patterns, as well as the solar abundances. However, some elements (in particular K, V, Ti and Sc) cannot be reproduced by any set of yields. The best yields suggested by \cite{Romano2010} include the results of \cite{Kobayashi2006} for nucleosynthesis in massive stars including hypernovae, but for C, N, O the yields with mass loss and rotation from the Geneva group are preferred. The yields of \cite{Karakas2010} are suggested for low and intermediate mass stars and those of \cite{Iwamoto1999} for Type Ia SNe. They also suggested that the lack of agreement with some chemical species can be improved by including and improving the treatment of processes such as hot bottom burning in intermediate mass stars and stellar rotation in nucleosynthesis studies.

\begin{figure}[htbp]
  \includegraphics[width=\textwidth]{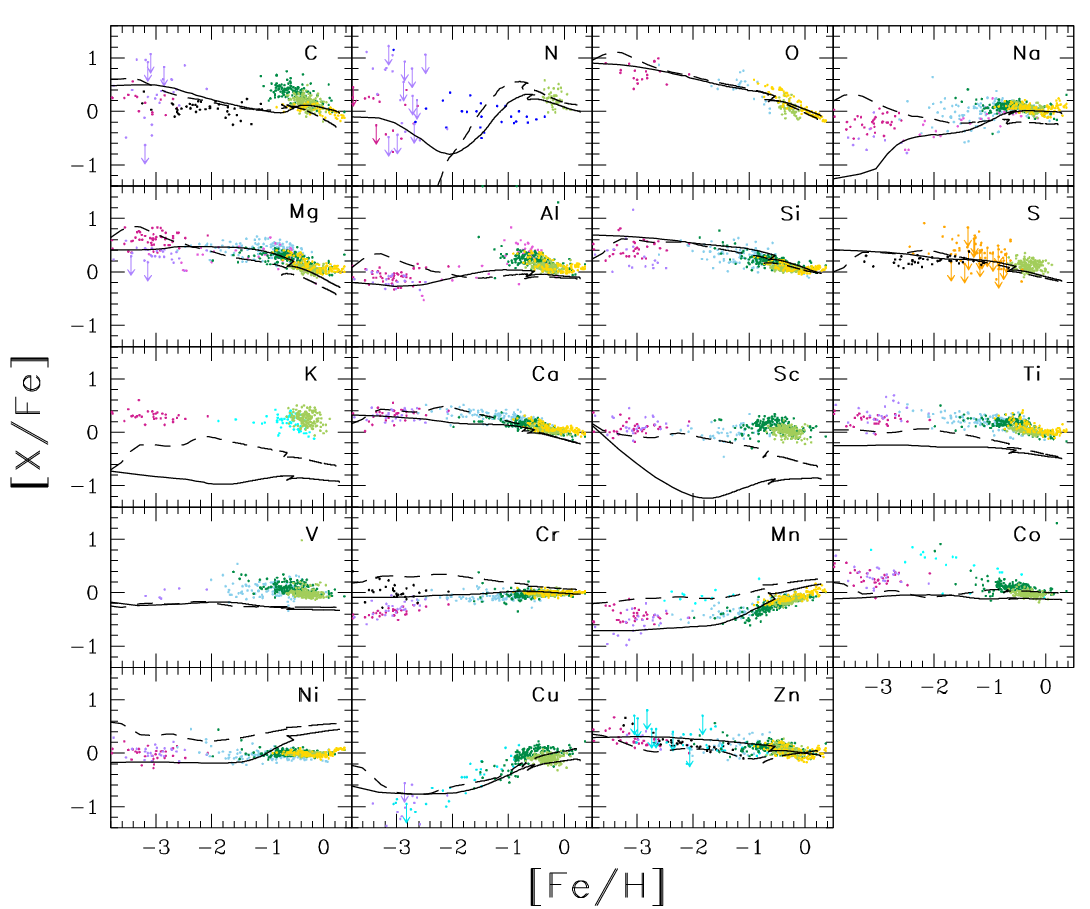}
  \caption{Plot of the [X/Fe] vs.\ [Fe/H] relations for elements from C to Zn. The solid curves represent the predictions of Model~15 of \cite{Romano2010}, assuming the set of yields described in the text and considered as the best one (continuous lines). We note that this set of yields includes hypernovae. The dashed lines refer to a set of yields including those of \cite{Woosley1995} for massive stars, without mass loss and rotation, and those by \cite{vandenHoek1997} for low and intermediate mass stars. Some elements, such as V, K, Ti and Sc cannot be reproduced by any set of yields. The data are from a large compilation made by \cite{Romano2010}, where the reader can find the references. Image reproduced with permission from \cite{Romano2010}, copyright by ESO.}
    \label{fig:RKTM}       % Give a unique label
\end{figure}

\cite{Prantzos2018} presented a study similar to that of \cite{Romano2010}, with a one-infall Galactic model, where recent yields for low and intermediate mass stars  depending on metallicity, as well as yields of massive stars including the combined effect of metallicity, mass loss and rotation, are adopted. In particular, for low and intermediate mass stars they adopted yields from the FRUITY database\footnote{\url{http://fruity.oa-abruzzo.inaf.it/}}, and the yields for massive stars of \cite{Limongi2018} including rotation and mass loss. In Fig.~\ref{fig:Prantzos} we show the results of \cite{Prantzos2018} for a large number of heavy elements: two sets of theoretical curves are shown and they are representative of models with rotating and non-rotating massive stars, respectively. The evolution of most of the $\alpha$-elements is fairly well reproduced, as in \cite{Romano2010}. The elements Ne and Ar, for which no data are available, show the same typical behaviour of $\alpha$-elements. The inclusion of stellar rotation does not change the results for $\alpha$-elements (apart from a small increase of C and O at low metallicities). For what concerns Mg, the adopted yields underproduce Mg, a problem common to previous sets of yields (e.g., \citealt{Woosley1995}), but not to the yields of \cite{Kobayashi2006} and \cite{Nomoto2013}. Again, the elements K, Sc, Ti and V are not well reproduced by the adopted yields. However, the inclusion of rotation improves sligthly the agreement  with the abundances of K and Sc at low metallicities. The yields fail in reproducing the evolution of Zn, a feature shared by the yields of \cite{Woosley1995} and \cite{Kobayashi2006}. The Fe-peak elements (Cr, Mn, and Ni) are largely produced by Type Ia SNe, except for Co which is mainly produced in hypernovae.
The increase of the [Mn/Fe] here is therefore due only to the metallicity dependence of Mn yields in massive stars, otherwise Mn should evolve in lockstep with Fe.  \cite{CescuttiMcWilliam2008} proposed instead that the secondary-like behaviour of Mn could be explained if the yields from SNe Ia have a dependence on metallicity, although such a feature is not present in the yields commonly adopted for Type Ia SNe (\citealt{Iwamoto1999, Kobayashi2020}). The case of Co is not yet clear: no yields are able to reproduce the increase of [Co/Fe] for [Fe/H]$<-2.0$\,dex. On the other hand, the increase of Ni with metallicity in \cite{Prantzos2018} models is due to the  Ni overproduction in the typical nucleosynthesis model adopted for Type Ia SNe (model W7 and W70 of \citealt{Iwamoto1999}). However, more recent nucleosynthesis studies (e.g., \cite{Kobayashi2020}) do not overproduce Ni.

A particular discussion should be reserved to $^{14}$N, a typical secondary element that instead reveals a primary behavior at low metallicities; see Figs.~\ref{fig:RKTM} and \ref{fig:Prantzos}, where the data show a roughly constant [N/Fe] ratio. Actually, $^{14}$N shows some primary contribution also at solar metallicities, as already indicated by \citet{Matteucci1985} and
  \citet{Diaz1986}, due to the N produced during the hot bottom burning phase of intermediate mass stars. If N were produced as a typical secondary elements in massive stars, the [N/Fe] ratio should increase with [Fe/H]. These facts suggest that N is produced as a primary element in massive stars, as first suggested by \cite{Matteucci1986}, and rotating massive stars can indeed produce primary N (\citealt{Meynet2002, Chiappini2006, KobayashiK2011, Limongi2018}), as shown clearly in Fig.~\ref{fig:Prantzos}.

The evolution of the $^{12}$C abundance is still a matter of debate. For many years it was believed that $^{12}$C originates from low and intermediate mass stars, but with the consideration of mass loss and rotation in massive stars, it appeared that massive stars can be the dominant source of carbon. The most recent observational data, showing an overabundance of C relative to Fe, seem to confirm the massive star origin of $^{12}$C, as discussed by \cite{Romano2020}.

\begin{figure}[htbp]
  \includegraphics[angle=270,width=\textwidth]{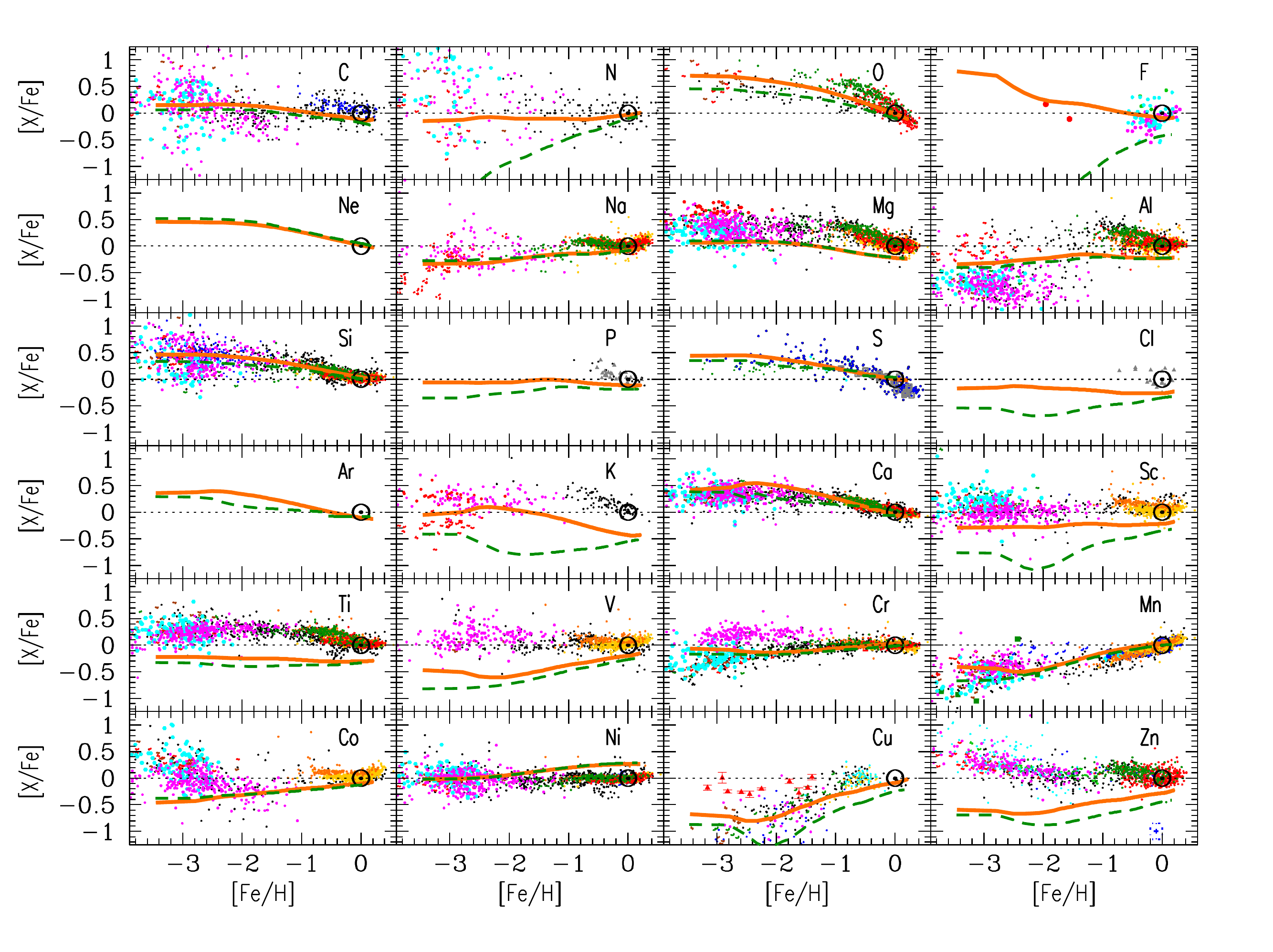}
  \caption{Comparison between observed and predicted [X/Fe] vs.\ [Fe/H] relations by \cite{Prantzos2018} who adopt two sets of stellar yields, one with rotating massive stars (continuous lines) and the other with non-rotating massive stars (dashed lines). The adopted yields are described in the text and do not include hypernovae. Image reproduced with permission from \cite{Prantzos2018}, copyright by the authors.}
    \label{fig:Prantzos}       % Give a unique label
\end{figure}

\subsection{Very heavy elements: s- and r-processes}
The s- and r-process elements deserve a separate treatment, first of all because unlike all the other heavy elements their abundances show a very large spread in halo stars, and because most of them can be produced partly as s- and partly as r-process elements.

\subsubsection{s-process elements}
We remind that s-process elements are formed by slow neutron capture on heavy seeds (e.g., Fe), relative to the timescale of the $\beta$-decay in nuclei. The r-process elements instead derive from rapid neutron capture. The s-process elements, such as Sr, Y, Zr (first peak),  Ba, La, Ce, Nd, Sm (second peak) and Pb (third peak), are mainly produced during He-shell burning in low mass stars (1--3\,$M_{\odot}$, see \citealt{Busso1999, Cristallo2009, Cristallo2011, Karakas2016}), but they are also produced during He-core burning in massive stars ($M\ge 10\,M_{\odot}$) (\citealt{Baraffe1992}).
Various authors have computed the chemical evolution of s-process elements (e.g., \citealt{Travaglio1999, Cescutti2006, Cescutti2014, Cescutti2013, Cescutti2015, Rizzuti2019}). In particular,  \cite{Travaglio1999} and \cite{Cescutti2006} suggested that s-process elements, such as Ba, are mainly produced by AGB stars but partly also by massive stars. \cite{Cescutti2014} suggested that magneto-rotationally driven SNe (\citealt{Winteler2012}) can represent a promising source of s-process elements in the early Galaxy and that part of the production of Sr, Ba and Y most likely came from spinstars (e.g., \citealt{Pignatari2008, Frish2012, Limongi2018}). On the other hand, \citet{Kobayashi2020} concluded that the first-peak elements come from electron-capture SNe, while the magneto-rotationally driven SNe are responsible for the production of r-process elements. The inhomogeneous model of \cite{Cescutti2014} was also able to reproduce the observed large spread at low metallicity observed for these elements. We have already shown the typical behaviour of a s-process element in Fig.~\ref{fig:Cesc08}; where the plot of [Ba/Fe] vs.\ [Fe/H,] shows an initial increase at low metallicity followed by an almost constant behaviour for ${\rm [Fe/H]}> -2.0$\,dex. The behaviour at low metallicity is due to the fact that Ba is assumed to be produced partly as a r-process element from stars in the range 10--30\,$M_{\odot}$, and therefore it appears already at low metallicities.

\subsubsection{r-process elements}
For what concerns r-process elements (Eu, Gd, Dy, Er, Yb) they can be produced during explosive nucleosynthesis in CC-SNe (\citealt{Cowan1991, Woosley1994, Wanajo2001}), but still many uncertainties are present in the production of r-process elements by CC-SNe since there are too few neutrons during explosive nucleosynthesis to reproduce the solar chemical composition (\citealt{Arcones2007}). An alternative and more succesfull source of r- process elements is the merging of compact objects, such as neutron stars and black holes (\citealt{Lattimer1977, Meyer1989, Freiburghaus1999, Rosswog1999, Korobkin2012, Thielemann2018}).

\begin{figure}[htbp]
  \centering
  \includegraphics[width=0.8\textwidth]{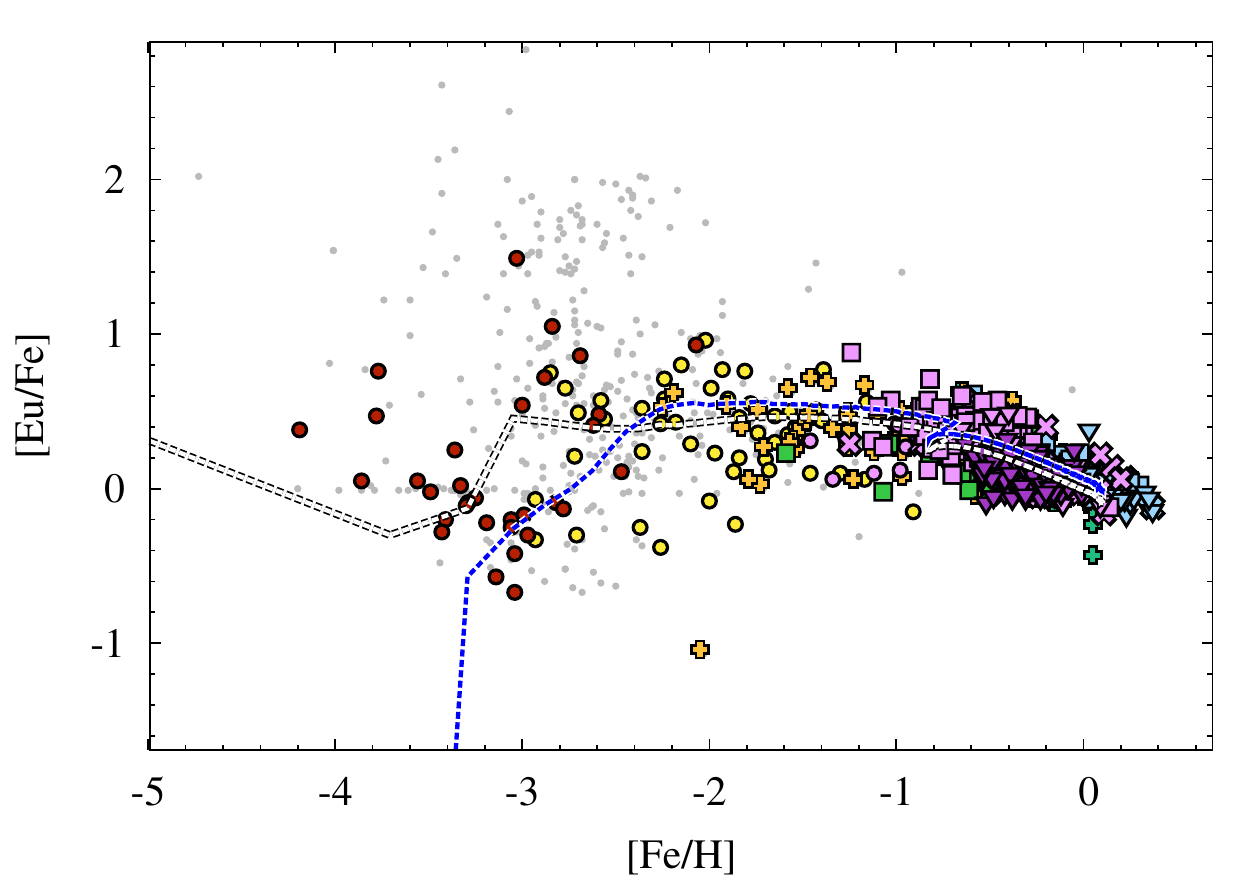}
  \includegraphics[width=0.8\textwidth]{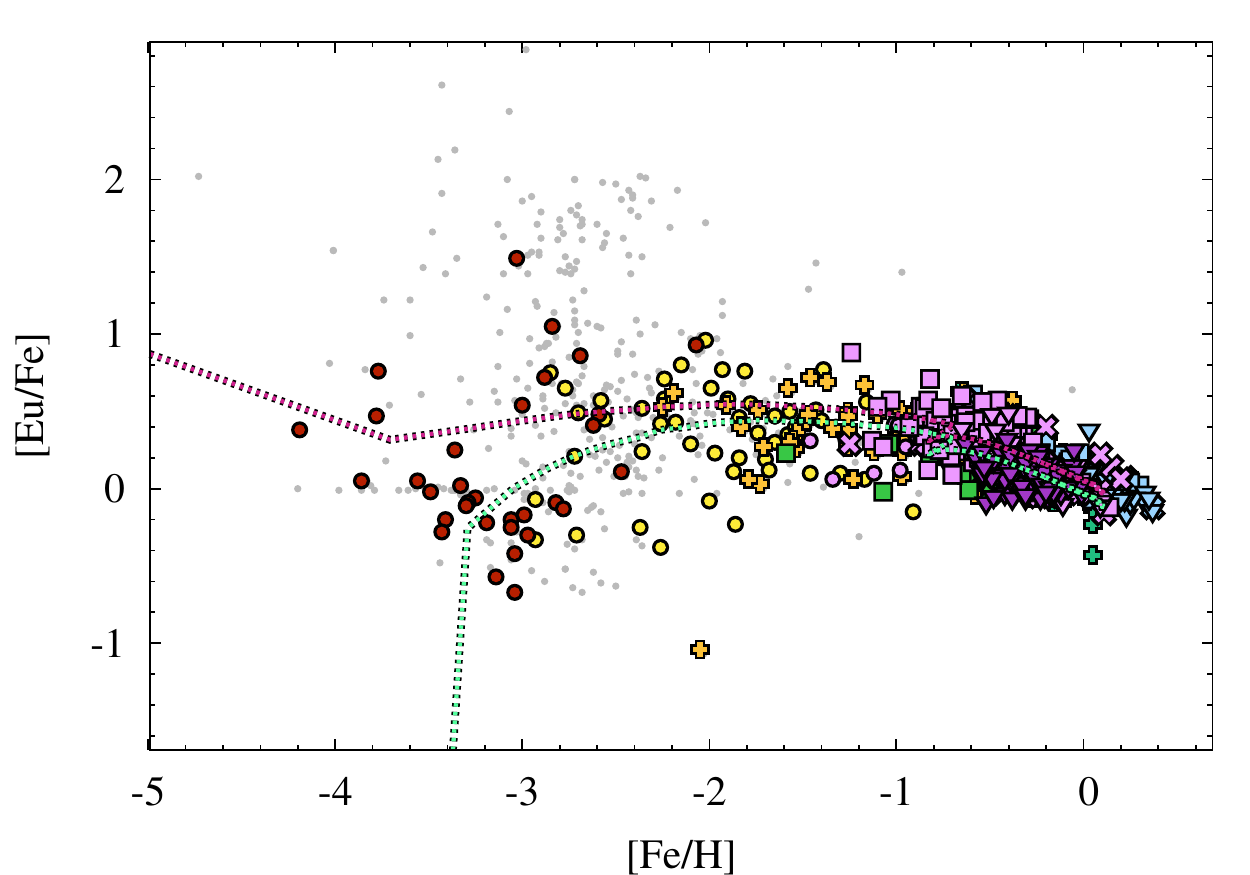} 
\caption{Plot of the [Eu/Fe] vs.\ [Fe/H] for solar vicinity stars. The upper panel shows two models including both MNS and CC-SNe as Eu producers: they differ by the assumed delay time for MNS, 1\,Myr for the white curve and 10\,Myr for the blue curve and for the yields of Eu from massive stars; in both models the range of progenitors of neutron stars is 9--30\,$M_{\odot}$. The lower  panel shows models where only MNS produce Eu with a delay constant and equal to 1\,Myr. The difference between the two curves are the mass range of neutron star progenitors. One model assumes a range 9--50\,$M_{\odot}$ (purple line), and the other a range of 9--30\,$M_{\odot}$ (green line). The white line in the upper panel and the purple one in the lower panel are two examples of how we can fit the data either with CC-SNe plus MNS or only with MNS. Image reproduced with permission from \cite{Matteucci2014}, copyright by the authors.}
\label{fig:Eu}       % Give a unique label
\end{figure}

This has been confirmed by the gravitational event GW170817 (\citealt{Abbott2017}, due to MNS,  where r-process elements have been detected (\citealt{Evans2017, Pian2017, Tanvir2017, Troja2017}). The Eu is easily observable in stellar atmospheres and is a pure r-process element, therefore most chemical evolution models refer to this element. \cite{Matteucci2014} included nucleosynthesis from merging of compact objects in the two-infall model and followed in detail the evolution of the abundances of Eu and Fe in the Milky Way. Concerning Eu, both CC-SNe and MNS were taken into account as Eu producers. The main conclusions in that paper can be summarized as follows: i) if each merging event produces $\sim 3\cdot 10^{-6}\,M_{\odot}$ of Eu and the time delay for merging, due to gravitational-wave emission, is constant and short ($\sim$ 1 Myr) for all systems, and the progenitors of neutron stars (or black holes) giving rise to the merging lie in the range 9--50\,$M_{\odot}$, then MNS can be responsible of the entire Eu production in the solar vicinity; ii) If, instead, the time delays for merging are longer, then also CC-SNe need to be considered as Eu producers to fit the observed [Eu/Fe] vs.\ [Fe/H] plot at low metallicities. In this case, the amount of Eu produced per merging event should be lower ($\sim 2\cdot 10^{-6}\,M_{\odot}$), but always in the range suggested by nucleosynthesis calculations ($10^{-7}$--$10^{-5}\,M_{\odot}$, \citealt{Korobkin2012}) and at the lower extreme of the range suggested for GW170817. This conclusion was confirmed by other papers (\citealt{Cescutti2015, Cote2018}; \citet{Simonetti2019, Molero2021}. In Fig.~\ref{fig:Eu} we show the results of \cite{Matteucci2014} for [Eu/Fe] vs.\ [Fe/H] in the cases of only MNS as Eu producers and of both CC-SNe and MNS.

\begin{figure}[htbp]
  \centering
\includegraphics[width=0.8\textwidth]{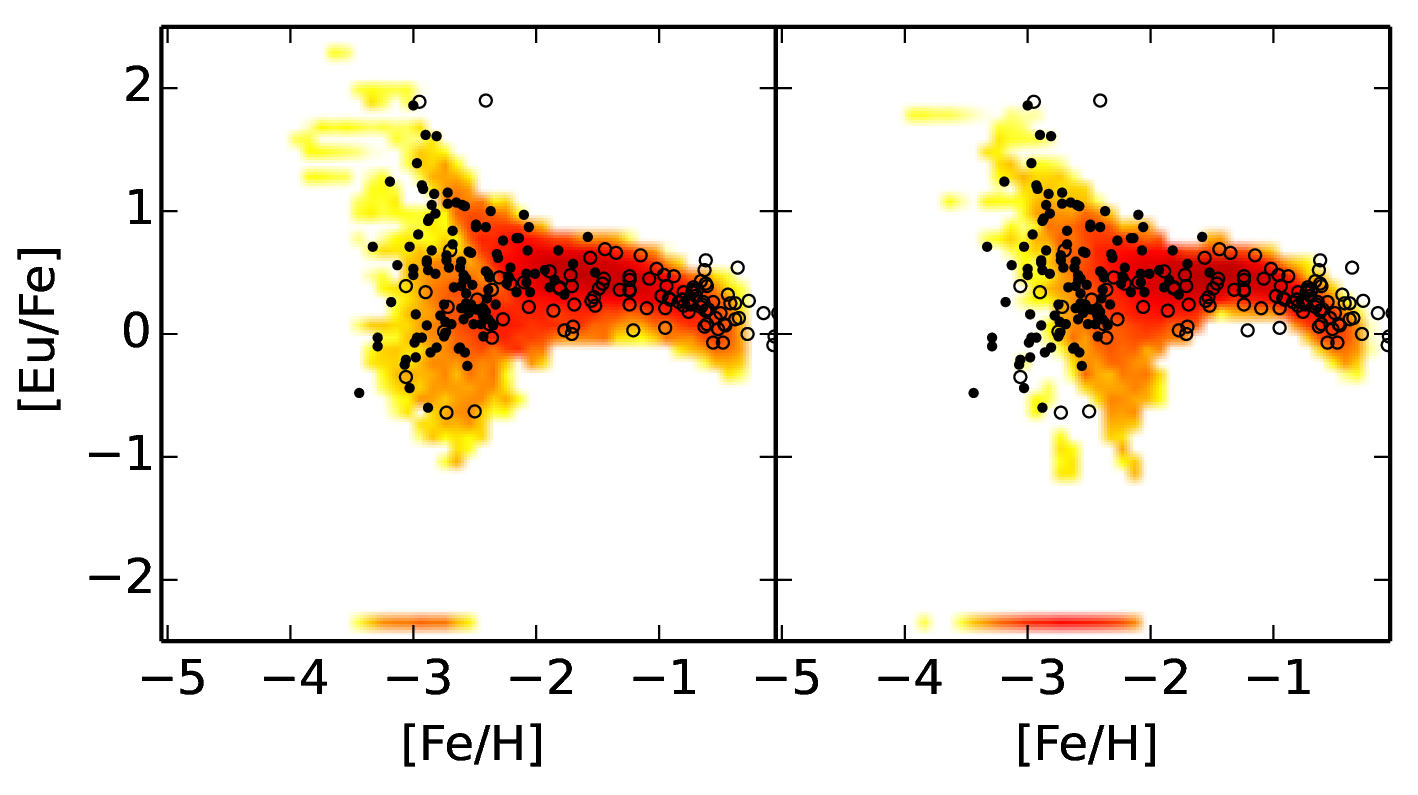}
\caption{Plot of the [Eu/Fe] vs.\ [Fe/H] for Galactic halo stars. The model predictions are indicated by the red and yellow areas, while the data are the black circles. The left panel shows a model with only MNS as Eu producers with a constant delay of 1\,Myr, as in \cite{Matteucci2014}. The right panel shows the model including MNS with fixed delay time of 100 Myr plus the contribution of magneto-rotationally driven SNe acting only for metallicities $Z< 10^{-3}$. See \cite{Cescutti2015} for the references to the data.
Image reproduced with permission from  \cite{Cescutti2015}, copyright by ESO.}
\label{fig:Cesc15}       % Give a unique label
\end{figure}

As we have said before, the observational data concerning Eu show a very large spread and the model of \cite{Matteucci2014} cannot reproduce such a spread but just the average behaviour of the [Eu/Fe] ratio, since this model assumes instantaneous mixing approximation. In order to explain the spread, if real, one has to adopt an inhomogeneous model for the Galactic halo, as shown in several recent works (\citealt{Cescutti2015, Voort2015, Shen2015, Komiya2016, Wehmeyer2019, Cavallo2020}). An example of inhomogeneous model results for the Galactic halo are shown in Fig.~\ref{fig:Cesc15}, where the model of \cite{Cescutti2015} assumes both MNS and magneto-rotationally driven SNe as Eu producers. As one can see, in both cases the observed spread is rather well reproduced. The need for introducing CC-SNe as Eu producers arises from the fact that, in this way, one can adopt MNS delay times much longer than 1\,Myr, in agreement with the GW170817 which occurred in an early type galaxy, where star formation had stopped several Gyr ago. Longer delay times are also required to fit the observed cosmic rate of short gamma-ray bursts (\citealt{Ghirlanda2016}), as shown by several papers (\citealt{Simonetti2019, Molero2021}).
A  way to reconcile the observed [Eu/Fe] ratios in the Milky Way, which requires a fast Eu production at early times, together with long coalescence timescales,  could be that the fraction of systems giving rise to MNS decreases with time, as suggested by \citet{Simonetti2019}.
On the other hand, in some cosmological simulations the need of a short coalescence timescale to reproduce the [Eu/Fe] at low metallicity seems not to be present, as shown by \cite{Shen2015}, who adopted a minimum delay time of 100 Myr and claimed to be able to predict Eu abundances different from zero also in stars of very low metallicity as well as the observed spread. This is clearly an effect of the stellar enrichment in cosmological simulations, but it is not clear from the paper how many stars with zero Eu abundance are predicted at very low metallicity and the predicted [Eu/Fe] along the disk looks too flat compared to the data. Finally, \cite{Schonrich2019} showed that with a DTD for MNS with a typical timescale of 150 Myr they could reproduce the [Eu/Fe] pattern in the Galaxy, if they allow for a 2-phase interstellar medium (hot and cold).

In conclusion, we still do not know whether merging of compact objects can be entirely responsible for the production of r-process elements.

\section{Abundance gradients along the thin disk}
\label{sec:abgrad}

It is well known that the abundance of metals decreases with Galactocentric distance along the thin disk. Negative abundance gradients are present also in external spirals and they represent an important tool to understand the formation mechanism of galactic disks. Negative abundance gradients of metals can be obtained if the star formation has been more efficient in the inner than in the outer regions of the disk. In the Milky Way, the SFR is indeed higher towards the Galactic center at the present time (\citealt{Rana1991, Stahler2004}). There are several ways to reproduce abundance gradients by means of chemical evolution models. One way is to assume an inside-out formation of the disk, as suggested originally by \cite{Larson1976} and then by \cite{Matteucci1989, Boissier1999} and \cite{Pilkington2012}: in this scenario, the disk forms by gas accretion occurring much faster in the inner than in the outer disk regions, thus creating a gradient in the SFR. This inside-out mechanism can be easily achieved by assuming a time scale of gas accretion increasing with Galactocentric distance in an exponential infall law of the type of Eq.~\eqref{eq:tauD}. Other ways of obtaining and/or steepening a metal gradient are the presence of a threshold in the gas density for star formation, a star formation efficiency ($\nu$ in Eqs.~\ref{eq:SFR} and \ref{eq:SFRChiap}) decreasing with Galactocentric distance, and inwards radial gas flows. All these situations have been studied (e.g., \citealt{Chiappini2001, Colavitti2009, Portinari2000, Schonrich2009, Spitoni2011, Mott2013, Cescutti2007, Bilitewski2012, Cavichia2014, Grisoni2018, Palla2020}). Moreover, also an equidense infall rate has been suggested (e.g., \citealt{Tosi1982}, \citealt{Tosi1988b}).
In Fig.~\ref{fig:Palla}, we report an example of predicted and observed abundance gradients: the models shown are all based on an inside-out formation of the thin disk plus other mechanisms such as variable star formation efficiency and radial gas flows.
In particular, radial gas flows should have a speed not exceeding 2--4\,km/s (0.7 km/s is the speed found in the cosmological simulation of \cite{Vincenzo2020}), otherwise the inward flows erase the gradient (\citealt{Tinsley1980}), but this speed can be constant or variable (\citealt{Spitoni2011}) or even vary with time (\citealt{Palla2020}). Radial gas flows were first studied by \cite{Mayor1981} and \cite{Lacey1985} and are a consequence of gas infall onto the disk.

\begin{figure}[htbp]
  \includegraphics[width=\textwidth]{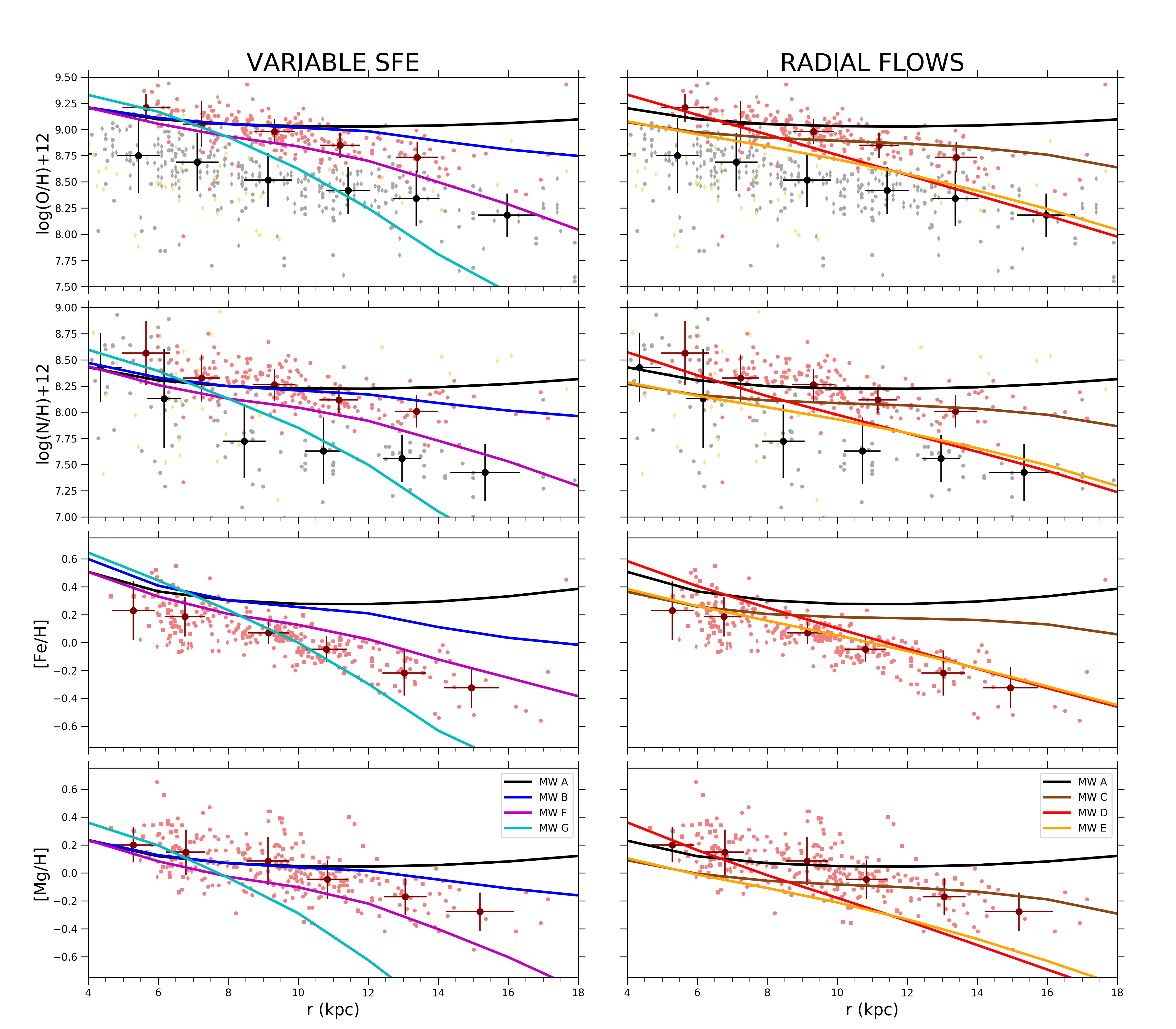}
  \caption{Observed and predicted abundance gradients along the thin disk. Black points with error bars represents the HII regions and PNe, while the red points with error bars refer to stars (Cepheids and young open clusters). Model A includes the inside-out mechanism but no radial flows and constant efficiency of star formation along the thin disk. Clearly it predicts too flat gradients. The models labeled B, F and G in the left panel include different prescriptions for a variable star formation efficiency, while models labeled C, D and E in the right panel contain radial gas flows with different prescriptions for the flow speed. All the models include the inside-out formation mechanism. For details and references see \cite{Palla2020}. Image reproduced with permission from \cite{Palla2020}, copyright by the authors.}
    \label{fig:Palla}       % Give a unique label
\end{figure}

As one can see from Fig.~\ref{fig:Palla}, the gradients derived from nebular data (HII regions and Planetary Nebulae (PNe), black points with error bars) are different from those derived from Cepheids and young open clusters (red points with error bars). This is probably due to an observational bias in at least one of the two observational techniques. The models are all based on the inside-out mechanism for the formation of the thin disk: it is clear from the figure that also other mechanisms, such as variable star formation efficiency and radial gas flows, should be included to properly reproduce the gradients.

\subsection{The [$\alpha$/Fe] ratio bimodality at different Galactocentric distances}
Following the study of \cite{Hayden2015}, in \cite{Queiroz2020} a complete map of the [$\alpha$/Fe] ratios vs.\ [Fe/H], extending from the bulge to the outer Galactic regions ($R_G > 20$\,kpc), has been presented. This map is based on high resolution spectroscopic data from APOGEE-2 survey Data Release 16 (DR16). These results show that the [$\alpha$/Fe] bimodality extends from the very inner Galactic regions to the outermost ones, suggesting that the bimodality is a clear property of the chemical enrichment of the Milky Way. These data are displayed in Figs.~\ref{fig:Q1} and \ref{fig:Q2}, where the diagram [$\alpha$/Fe] vs.\ [Fe/H] is shown in 2\,kpc wide bins of the Galactocentric radius $R_G$. In these figures, also a subdivision in $z$ is present (the height above the Galactic plane).
The interpretation of these data is very important to understand how the thick and thin disks formed.

\begin{figure}[htbp]
  \includegraphics[width=\textwidth]{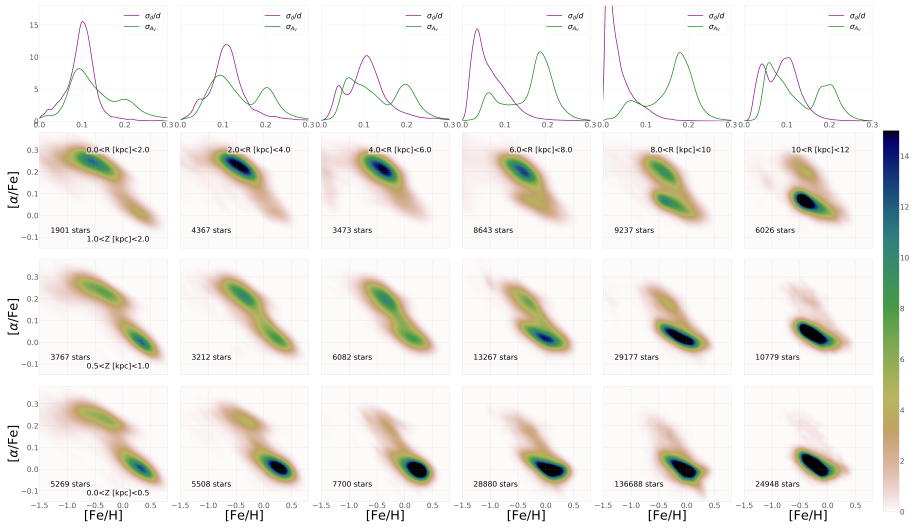}
  \caption{Data from APOGEE DR16 showing [$\alpha$/Fe] vs.\ [Fe/H] diagrams at different Galactocentric distances (bins of 2\,kpc) and different heights above the Galactic plane ($0<Z<2$\,kpc). The range of distances is quite large and starts from the very inner Galactic regions and arrives up to 12 kpc.
    The upper panels show the kernel-density estimates of the uncertainty distributions in StarHorse extinction and distances for each Galactocentic distance bin. Image reproduced with permission from \cite{Queiroz2020}, copyright by ESO.}
    \label{fig:Q1}       % Give a unique label
\end{figure}

\begin{figure}[htbp]
  \includegraphics[width=\textwidth]{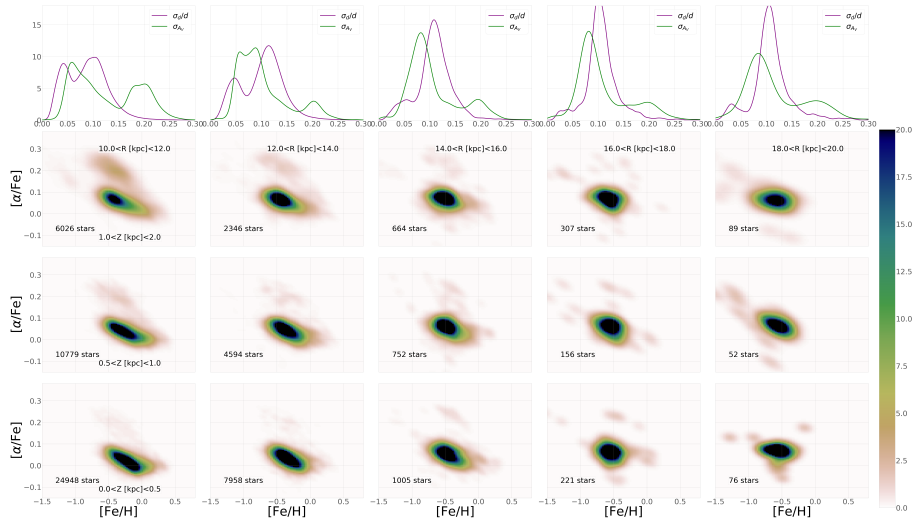}
  \caption{The same as Fig.~\ref{fig:Q1} but extended to the outermost Galactic regions
    ($R_G=20$\,kpc). Image reproduced with permission from \cite{Queiroz2020}, copyright by ESO.}
    \label{fig:Q2}       % Give a unique label
\end{figure}

As we have already discussed, in \cite{Spitoni2019} it has been suggested that a rather long period of quiescent star formation between the formation of the thick and thin disks can explain the observed [$\alpha$/Fe] bimodality in the solar vicinity data. In Figs.~\ref{fig:Q1} and \ref{fig:Q2}, it is evident that the low [$\alpha$/Fe] sequence moves towards lower [Fe/H] values with Galactocentric distance. This is a sign of an inside-out formation  of the thin disk: in fact, a longer accretion timescale for the thin disk means lower SFR and lower [$\alpha$/Fe] ratios, according to the time-delay model, as discussed in the previous Sections. Concerning the innermost Galactic regions, the two sequences are less distinct than in the outer parts and probably an enriched infall to form the thin disk is required to reproduce it, as shown by \cite{Palla2020} and \cite{Spitoni2021}(see Fig.~\ref{fig:Spito21}). Finally, a trend is present also with the height above the Galactic plane, and indicates that the high-$\alpha$ sequence is more pronounced at larger heights.

\begin{figure}[htbp]
  \includegraphics[width=\textwidth]{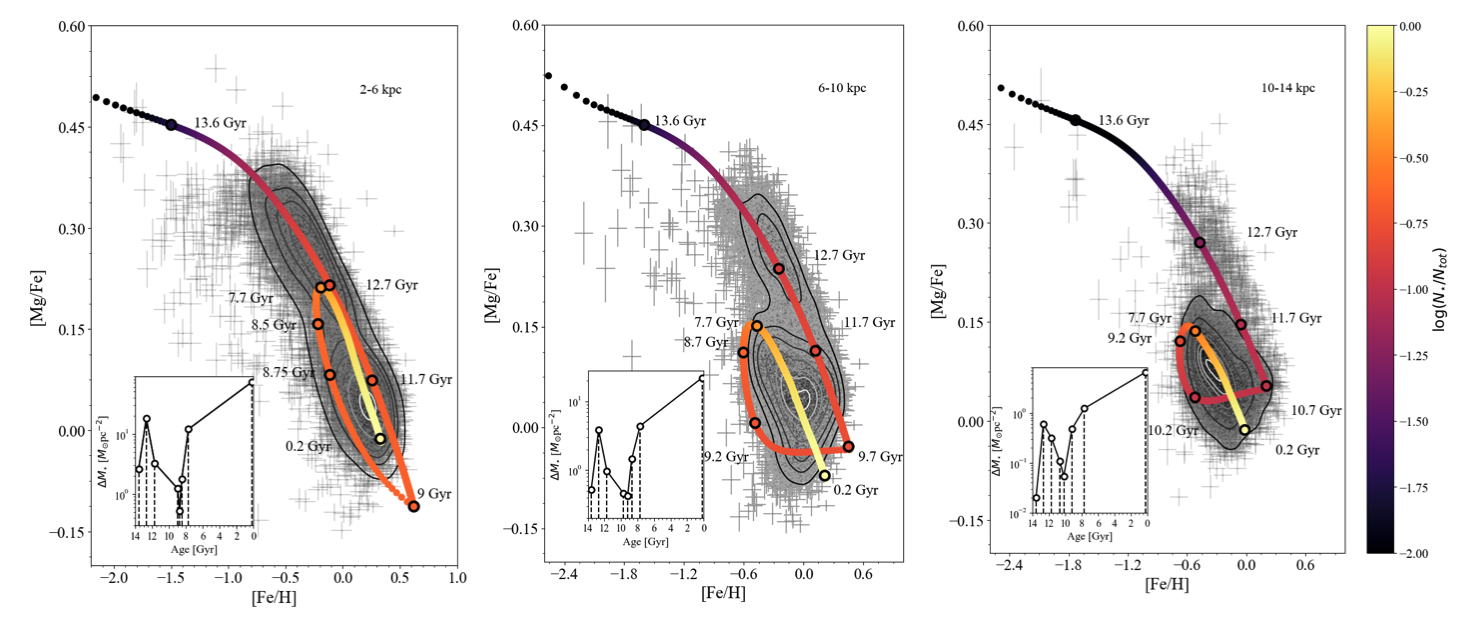}
  \caption{Observed [Mg/Fe] vs.\ [Fe/H] abundance ratios from APOGEE DR16 (grey points with associated errors) in the Galactocentric regions enclosed between 2--6 kpc (left panel), 6--10 kpc (middle panel), and 10--14 kpc (right panel)  compared with  the best-fit chemical evolution models (\citealt{Spitoni2021}), thick curves) in the different regions. The contour lines enclose fractions of 0.90, 0.75, 0.60, 0.45, 0.30, 0.20, 0.05 of the total number of observed stars. The color coding represents the cumulative number of stars formed during the Galactic evolution normalized to the total number $N_{\rm tot}$. The open circles mark the model abundance ratios of stellar populations with different ages. In the insets, we show the surface stellar-mass densities $\Delta M_{\star}$ formed in different age bins at different Galactocentric distances as a function of age, where the bin sizes are delimited by the vertical dashed lines and correspond to the same age values as indicated in the [Mg/Fe] vs.\ [Fe/H] plots. The difference between the data adopted here and in \cite{Queiroz2020} resides in the derived stellar distances. Here are derived from \cite{Leung2019}, while in \cite{Queiroz2020} are obtained by means of the StarHorse tool. Image reproduced with permission from \cite{Spitoni2021}, copyright by ESO.}
    \label{fig:Spito21}       % Give a unique label
\end{figure}

\section{A different paradigm: stellar migration}
\label{sec:migr}

Generally, in chemical evolution models it has been assumed that gas can move along the disk (radial gas flows) but not stars. \cite{Roskar2008}, by means of a cosmological simulation for galaxy formation which included gas and stars,  showed that stars tend to move outwards while the gas move inwards, and this is caused by resonant scattering at corotation. This result confirmed an idea of \cite{Binney2000} according to which the gas would participate in resonant scattering together with stars.
The stellar migration can significantly alter our interpretation of the observational data, simply because the stars we observe at the solar ring might have formed at inner radii and then been scattered at the solar radius. It clearly depends on the entity of this phenomenon quantified by the percentage of stars which have not been formed in situ. A great deal of work on the stellar migration followed the paper of \cite{Roskar2008}: in particular, \cite{Schonrich2009} presented a chemical model including radial gas flows and stellar migration. The physical causes for stellar migration that they considered are: the change of angular momentum by either i) scattering at an orbital resonance or ii) a non-resonant scattering caused by a molecular cloud. As a consequence, the star moves inward or outwards depending on whether angular momentum is lost or gained. They referred to the process of change of angular momentum as ``churning''. This churning process should preserve the star circular orbit. However, if the scattering changes the epyciclic amplitude but not the angular momentum the process is called ``blurring''. The chemical evolution model of \cite{Schonrich2009} including churning and blurring described the evolution of thick and thin disks, with the stars of both disks formed in situ. The predicted radial mixing of stellar populations with high vertical velocity dispersion from inner regions of the disk to the solar vicinity, provided an explanation of why the vertical velocity dispersion increases with age with a steeper slope than predicted by theory. The abundance gradients in the ISM were also reproduced thanks to the inward radial gas flows. Finally, they also suggested that the thick disk could have been formed by migrated stars.

In principle, stellar migration can explain the observed spread in the age-[Fe/H]  and [X/Fe] vs.\ [Fe/H] relations in the solar vicinity, as first pointed out by \cite{Francois1993}, who considered the different stellar birthplaces of stars now observed in the solar neighbourhood, as derived by \cite{Grenon1989}. Several chemical evolution models included stellar migration (\citealt{Kubryk2015, Minchev2011, Minchev2013, Spitoni2015, Buck2020, Vincenzo2020, Sharma2020, Khoperskov2020}) and reached different conclusions. \cite{Kubryk2015} considered both blurring and churning in a detailed chemical evolution model with the thin disk forming inside-out and concluded that stellar migration of old and metal poor stars from the inner regions to the solar vicinity helps in reproducing the features of both thin and thick disk, and in particular the spread in the age-metallicity relation which is larger than the spread observed in the [$\alpha$/Fe] vs.\ [Fe/H] relations. \cite{Francois1993} suggested that the spread in the [$\alpha$/Fe] relations should be less than the spread in [Fe/H], since their differences at various Galactocentric distances are less than the differences in the  age-[Fe/H] relations. \cite{Minchev2011} and \cite{Minchev2013} presented chemo-dynamical models including stellar migration, triggered by mergers at high redshift and central bar at later times, and suggested that a sizeable fraction of old metal-poor high [$\alpha$/Fe] stars can reach the solar vicinity from the inner regions. They also claimed to reproduce the majority of observational features in the thick and thin disks. \cite{Spitoni2015}, included in a simple way stellar migration in the two-infall model, following the prescriptions of \cite{Minchev2013}, and assumed velocities of 1 km/sec for the migrating stars. They were able to reproduce very well the G-dwarf metallicity distribution in the solar vicinity, accounting for those high metallicity stars that were missing in previous works (e.g., \citealt{Chiappini1997}). In fact, they found that a fraction not larger than 20\% of the stars in the solar vicinity might have migrated from the inner regions. These results are shown in Fig.~\ref{fig:Gdwarf}. As one can see, the effect of stellar migration improves the agreement with the observed G-dwarf metallicity distribution at high metallicity, but it does not have a dramatic effect on the predictions, even if all the stars observed in the solar neighbourhood were originating from somewhere else. This is an important consideration because it tells us that stellar migration can improve the agreement with observations but is not crucial to explain them. \cite{Vincenzo2020}, by means of cosmological simulations for the formation and evolution of the Milky Way, concluded that stellar migration involves old and metal rich stars and that it occurs more outwards than inwards (see Fig.~\ref{fig:migrVinc}). This means that stellar migration is not so important to interpret data in the solar neighbourhood. They also suggested that stellar migration acts in flattening the stellar metallicity gradient and this could explain the different slopes of the observed nebular and stellar gradients. \cite{Buck2020}, as mentioned before, explained the bimodality in the [$\alpha$/Fe] vs.\ [Fe/H] plane for the thick and thin disk by means of stellar migration. He suggested that low-$\alpha$ stars originated from both inner and outer disk, while most of the high-$\alpha$ stars originated from the inner disk. Moreover, \cite{Sharma2020} by means of a chemo-dynamical model, suggested that the [$\alpha$/Fe] bimodality cannot be reproduced without stellar migration. On the other hand, \cite{Khoperskov2020}, by means of N-body simulations, concluded that stellar migration has a negligible effect on the [$\alpha$/Fe] vs.\ [Fe/H] relation. In conclusion, there is not yet agreement about the importance and the entity of stellar migration along the thin disk of the Milky Way.

\begin{figure}[htbp]
 \includegraphics[width=0.5\textwidth]{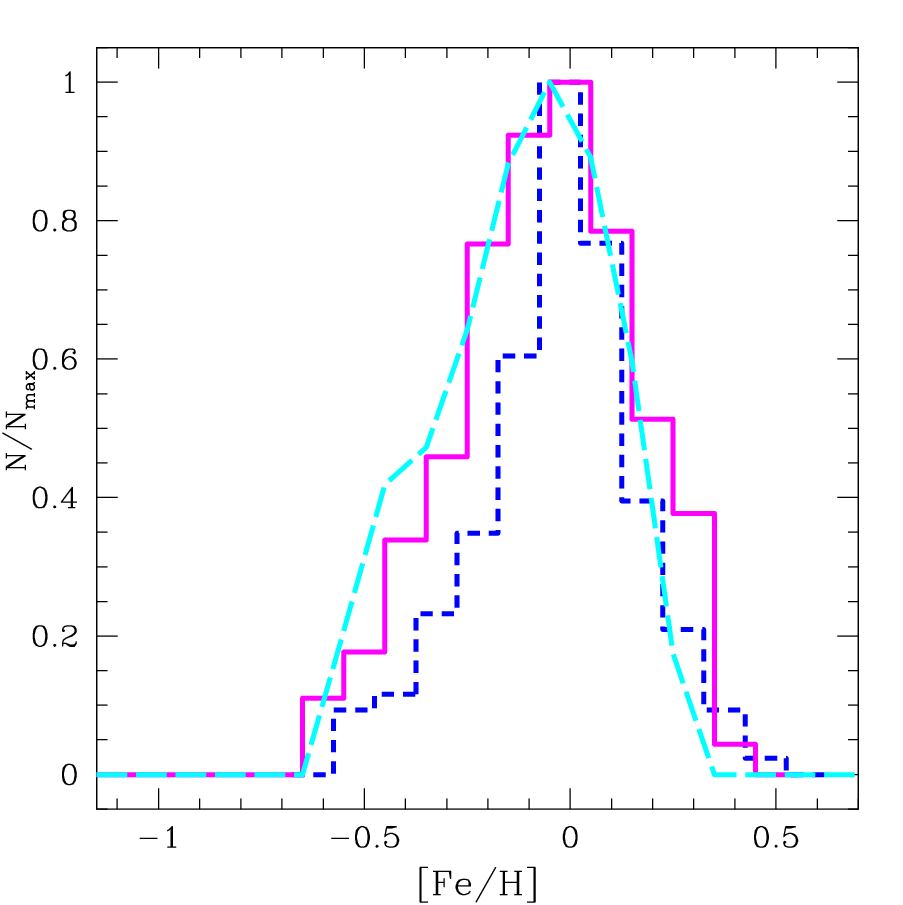}
  \includegraphics[width=0.5\textwidth]{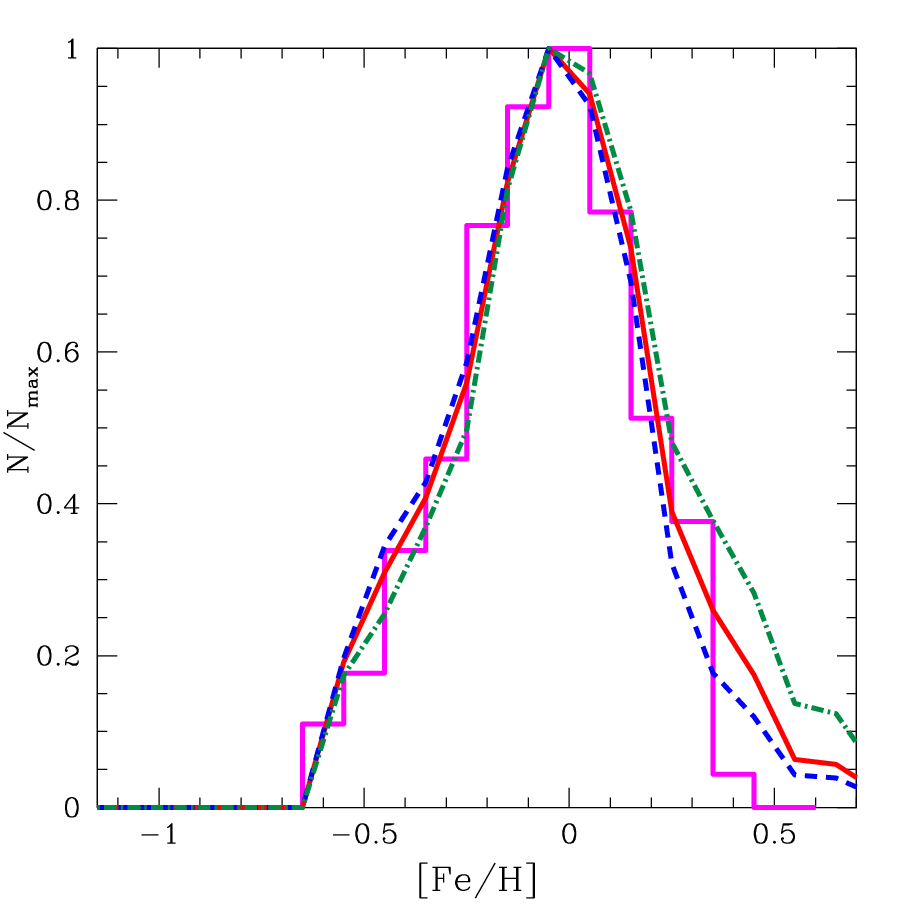}
\caption{Plot of the G-dwarf metallicity distribution in the solar vicinity. {\it Left panel}: comparison between data and  model results without stellar migration (cyan line). The data are from \cite{Adibekyan2013} (solid magenta histogram) and \cite{Fuhrmann2011} (short-dashed blue histogram). {\it Right panel}: comparison between data and models with stellar migration. The dashed blue line refers to the predictions of a case with  migration (10\% and 20\% of the stars born at 4 and 6 kpc, respectively, which end up at 8 kpc). Solid red line is migration (20\% and 40\% of stars born at 4 and 6 kpc, respectively, which end up at 8 kpc), and the green dashed-dotted line is extreme migration (all stars born at 4 and 6 kpc which end up at 8 kpc). Image reproduced with permission from \cite{Spitoni2015}, copyright by AAS.}
 \label{fig:Gdwarf}       % Give a unique label
\end{figure}

\begin{figure}[htbp]
  \centering
 \includegraphics[width=0.8\textwidth]{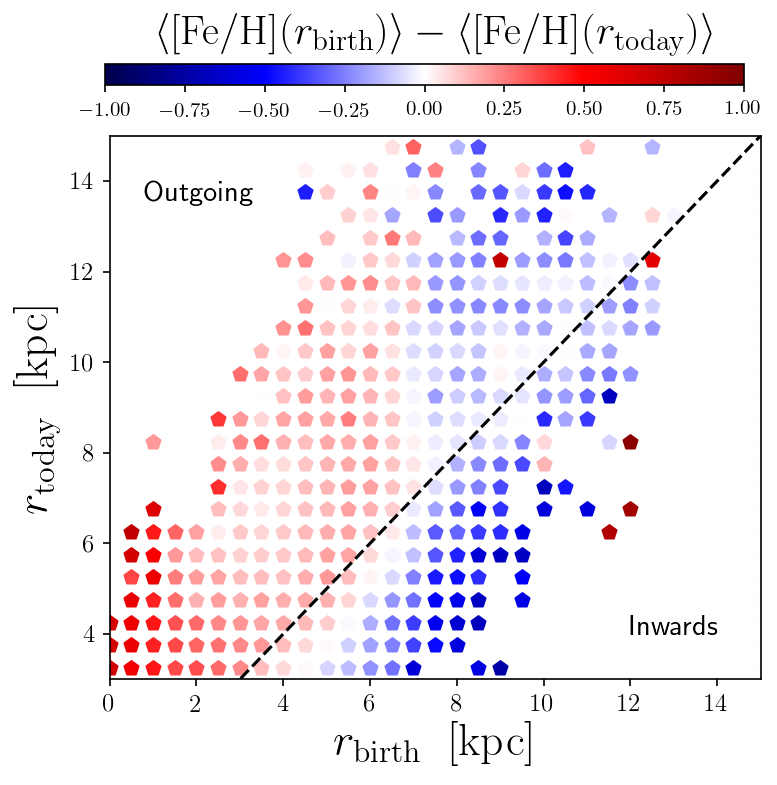}
 \caption{The difference between the average [Fe/H] at the birth-radius, $\langle [Fe/H](r_{\rm birth}) \rangle$, and the average [Fe/H] at the present-day radius, $\langle [Fe/H](r_{\rm today}) \rangle$, in the $r_{\rm today}$--$r_{\rm birth}$ diagram for the thin-disk stars. The dashed line indicates no migration. Image reproduced with permission from \cite{Vincenzo2020}, copyright by the authors.} 
    \label{fig:migrVinc}       % Give a unique label
\end{figure}

\section{The evolution of light elements in halo and disks}
\label{sec:light}

Light elements (H, D, He and $^{7}$Li) have been produced during the Big Bang but some of them are produced also in stars, such as $^{4}$He, $^{3}$He and $^{7}$Li. Deuterium instead is only destroyed inside stars. The primordial abundances of light elements, and in particular that of D, are quite important since they are related to the baryon density of the Universe ($\Omega_B$), or equivalently to the baryon-to photon ratio ($\eta$). Data on cosmic microwave background (CMB) can provide directly the value of $\eta$ and therefore of $\Omega_B$, and thus the primordial abundances of D, He and Li. Results from the Wilkinson Microwave Anisotropy Probe (WMAP, \citealt{Spergel2003}) first, and then Planck Collaboration (\citealt{Coc2014}) were used to derive the primordial abundances. All the primordial abundances derived from the CMB agree with observations, except for $^{7}$Li. Primordial D is measured in high redshift unevolved objects, such as Damped Lyman-$\alpha$ Systems (DLAs), obtaining a lower limit to the primordial value (e.g., \citealt{Omeara2006, Pettini2008}). The primordial abundance of $^{4}$He is measured in unevolved galaxies, such as blue compact ones (e.g., \citealt{Aver2015, Peimbert2017, Izotov2014}) and intergalactic medium (see \citealt{Cooke2018}). Concerning $^{7}$Li, its primordial abundance was believed to be that shown by old metal poor halo stars ($-2.4 \le {\rm [Fe/H]} \le -1.4$), which show a rather flat Li abundance (the Spite-plateau, \citealt{Spite1982}), and in particular 12+log(Li/H) = A(Li) = 2.05--2.2\,dex (\citealt{Spite1986, Bonifacio1997}). On the other hand, WMAP results, confirmed later by Planck, suggested a primordial Li abundance of $\sim$ 2.6\,dex. This discrepancy is known as ``cosmological Li problem''. Moreover, more recent work on very metal poor halo stars  (\citealt{Sbordone2010, Melendez2010, Hansen2014, Bonifacio2015}) challenged  also the existence of the Spite plateau: in fact, for metallicities ${\rm [Fe/H]}< -2.8$\,dex, the Li abundance was found to decrease, but  \cite{Aguado2019} found a halo star with ${\rm [Fe/H]}< -6.1$\,dex and Li abundance close to the Spite plateau value.

\subsection{The chemical evolution of Li}
Here we will focus on the Galactic evolution of the abundance of $^{7}$Li: in fact, the abundance of $^{7}$Li seems to increase from metal poor ([Fe/H]$< -1.0$\,dex) to metal rich stars. This growth is seen by looking at the abundance of $^{7}$Li in the stars lying on the upper envelope of the points in the diagram A(Li) vs.\ [Fe/H] (\citealt{Rebolo1988}), namely the youngest stars and meteorites. This is due to the fact that $^{7}$Li is very easily destroyed in stars during their evolution and therefore the $^{7}$Li diagram shows a quite large spread, according to the different degree of destruction of this element in stars in different evolutionary phases. This can be clearly seen in Fig.~\ref{fig:Izzo}, where the most recent Li data are compared to predictions of chemical evolution models. The increase by roughly a factor of ten in the Li abundance during the Galactic lifetime is here reproduced by assuming that $^{7}$Li is produced in stars and ejected into the ISM.
There is a main channel for producing $^{7}$Li in stars: there should be a site in which the reaction $^{3}{\rm He}(\alpha, \gamma)^{7}{\rm Be}$ can occur, but $^{7}$Be should be rapidly transported by convection in regions of lower temperature where it decays into $^{7}$Li by k-capture. This is known as \emph{Cameron--Fowler mechanism}. Another channel proposed in the literature is the \emph{$\nu$-process} occurring in massive stars (\citealt{Woosley1990}): as the core of a massive stars collapses to form a neutron star, the flux of neutrinos originating from the neutronization process is so intense that it can interact with the overlying shells of heavy elements. Neutrinos excite heavy element atoms and even He to particle unbound levels.

\begin{figure}[htbp]
  \includegraphics[width=\textwidth]{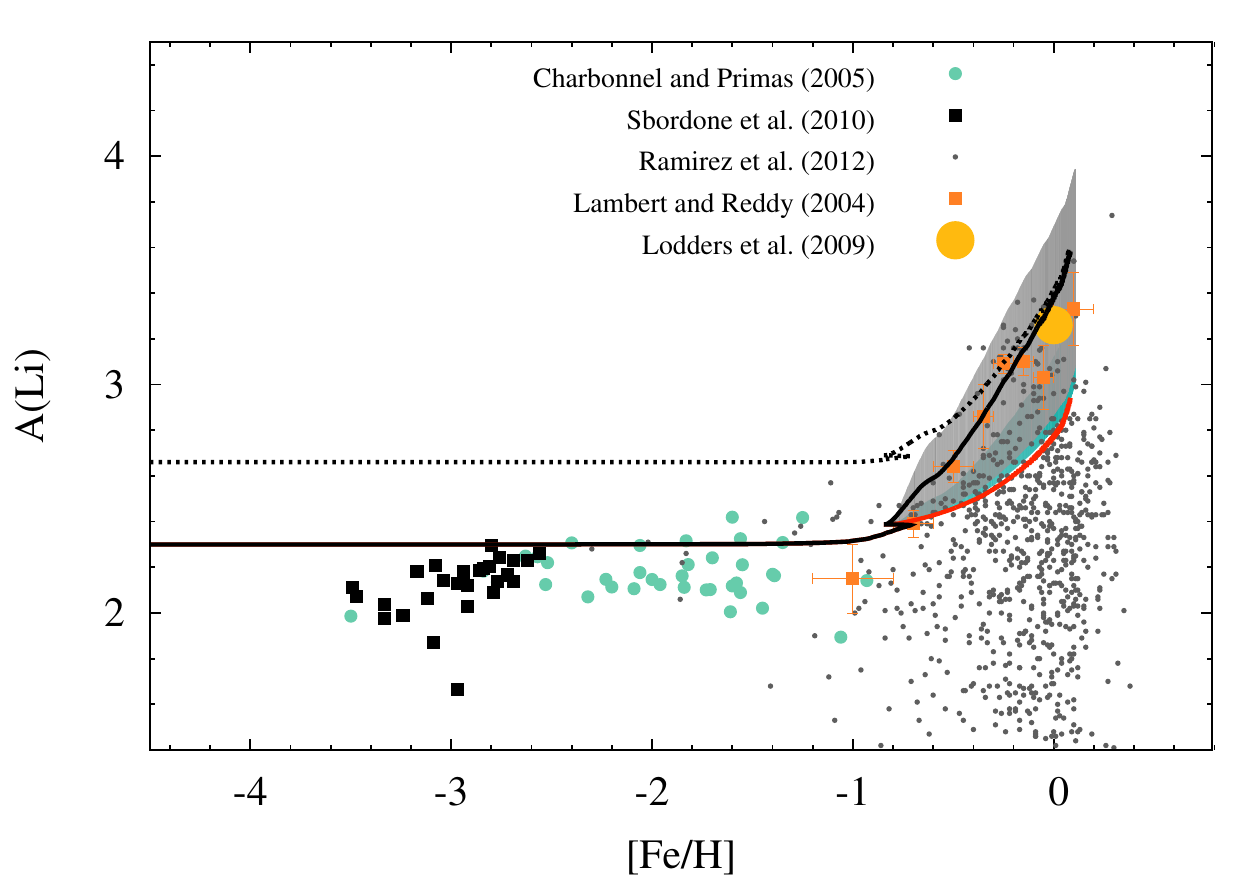}
  \caption{The relation A(Li) vs.\ [Fe/H] for the solar neighbourhood stars and meteorites, compared to the predictions of chemical evolution models (lines and coloured areas). The dotted line refers to a model starting from the primordial Li abundance deduced from WMAP and Planck, while the continuos line refers to a primordial abundance of 2.2\,dex, and corresponding to the Spite plateau.Image reproduced with permission from \cite{Izzo2015}, copyright by AAS.}
  \label{fig:Izzo}
\end{figure}

Several stellar $^{7}$Li producers have been proposed so far and they are: CC-SNe, giant stars, AGB-stars, novae and cosmic rays. \cite{Dantona1991} tested the contribution of novae to the chemical enrichment in $^{7}$Li, and concluded that this contribution cannot be neglected and that is probably responsible for the steep rise of A(Li) for [Fe/H]$> -1.0$\,dex. None of the other suggested sources could explain that, because CC-SNe and AGB stars have too short lifetimes and red giant stars produce too little Li. On the other hand, novae are systems contributing to chemical enrichment on long timescales and in principle, although many uncertainties are still present in Li nucleosynthesis in novae (see \citealt{Jose2003, Jose2007}),  they could produce the right amount of $^{7}$Li (see \cite{Starrfield2020}). The first and only claimed detection of $^{7}$Li in novae is from \cite{Izzo2015}, while detection of $^{7}$Be, which decades into $^{7}$Li, seems to be more common (\citealt{Tajitsu2015, Tajitsu2016, Molaro2016, Izzo2018}, but see also \citealt{Shore2020}).  Figure~\ref{fig:Izzo} demonstrates two predictions for Li evolution: one for the case in which the primordial Li is that shown by the Spite-plateau (continuous line) and the other for the case in which we assume the Planck primordial Li abundance (dotted line). In this latter case, we should assume that $^{7}$Li has been destroyed in halo stars to the level of the Spite plateau.  This difference between the value of Li in the Spite plateau and that derived from Big Bang nucleosynthesis based on Planck results, is called ``cosmological Lithium problem'' (see \citealt{Matteucci2021}). In both cases of primordial Li, the steep rise of its abundance in disk stars is well reproduced by the contribution of novae. The nova rate was computed as in Eq.~\eqref{eq:novae} and the parameter $\delta$  constrained to reproduce an observed present time nova rate in the Galaxy of 20 events yr$^{-1}$, in agreement with recent estimates from \cite{dellavalle2020}.

\begin{figure}[htbp]
  \includegraphics[width=0.5\textwidth]{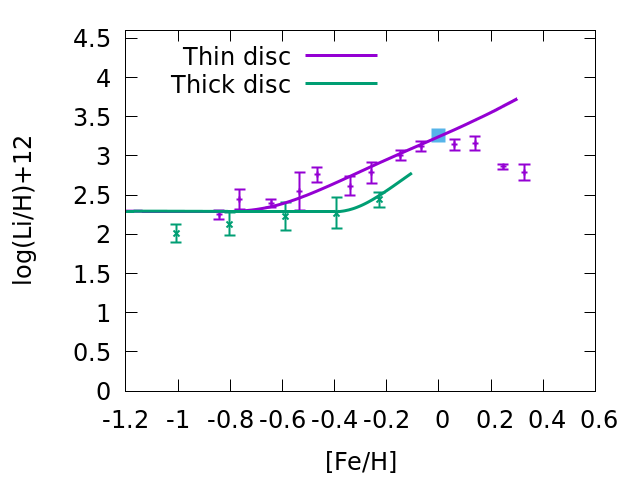}
\includegraphics[width=0.5\textwidth]{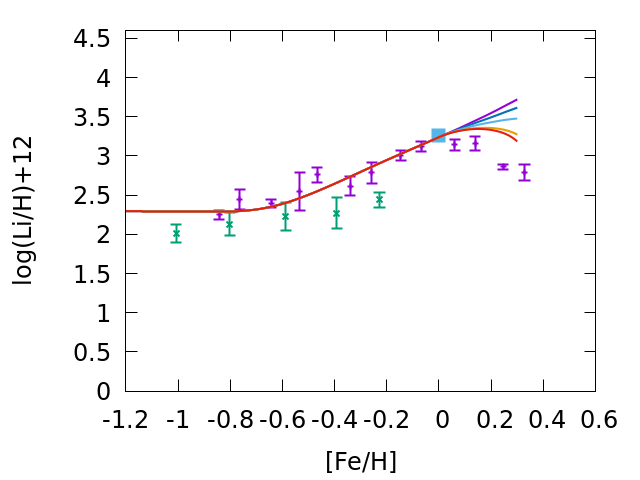}
  \caption{The relation A(Li)=log(Li/H) +12 vs.\ [Fe/H] for the solar vicinity thick and thin disk stars. {\it Left panel}: the evolution of Li abundance in the thick (green line) and thin (purple line) disk. {\it Right panel}: the evolution of Li in the thin disk under various assumptions about the fraction of nova system decreasing with metallicity. The data are from AMBRE (\citealt{Guiglion2016}). Image reproduced with permission from \cite{Grisoni2019}, copyright by the authors.}
    \label{fig:litio}       % Give a unique label
\end{figure}

From the analysis of the previous figures, we can see that the Li abundance in halo stars is roughly constant  except perhaps for very metal poor stars, while in disk stars the Li abundance is increasing. Recently, Li data for thick and thin disk stars have become available and in thin disk stars is evident a bending of the Li abundance at the highest observed metallicities (\citealt{Guiglion2016, Fu2018}), although more recent data from \cite{Randich2020} do not show such a bending of Li abundance in stars of open clusters at high metallicity (up to [Fe/H]$\sim +0.3$\,dex). Several explanations have been proposed for such a bending, if real, and they are: i) stellar migration from the inner disk (\citealt{Guiglion2019}), ii) Li yields depending on metallicity (\citealt{Prantzos2017}), iii) a decreasing fraction of novae with increasing metallicity (\citealt{Grisoni2019}). Figure~\ref{fig:litio} presents the results of \cite{Grisoni2019} for the predicted evolution of $^{7}$Li in thick and thin disk stars. In this model, the two disks evolve in parallel with different SFRs, as already mentioned in the previous paragraphs. The fraction of nova systems in the IMF, the parameter $\delta$ in Eq.~\eqref{eq:novae}, is assumed to decrease with metallicity, as suggested by observations of \cite{Gao2014, Gao2017} and \cite{Yuan2015}. In Fig.~\ref{fig:litio} we show both the results with a constant fraction of novae and with a variable one. As one can see in the left panel of this figure, the thick disk does not show any bending and the Li abundance displays only a little increase starting at around [Fe/H] $\sim -0.3$\,dex.

\subsection{The chemical evolution of D}
Deuterium is only destroyed to form $^{3}$He in stars, so it is very low or completely exhausted in highly evolved systems. \cite{RomanoTosi2003} presented the predictions of a chemical evolution model concerning D in the solar vicinity. The amount of D destroyed during Galactic evolution is called \emph{astration}. The predicted astration in the solar vicinity is quite low ($X_{D,p}/X_{D,t_{\rm Gal}} \sim 1.5$), as shown in Fig.~\ref{fig:deuterio}, where the primordial D was assumed to be $(D/H)_p=2.5 \cdot 10^{-5}$, consistent with the abundance derived by WMAP. The first detailed chemical evolution models for D demonstrating a low astration factor were presented by \cite{Steigman1992} and \cite{Dearborn1996}.
Observationally, D abundances measured in QSO absorbers at high redshift provide a good lower limit to the primordial value and they agree with the WMAP estimate. 
In Fig.~\ref{fig:deuterio}, the D measured at the age of the formation of the Solar System ($\sim$ 4.5 Gyr ago) and in the ISM at the present time are reported. Two slightly different models are shown (see \citealt{RomanoTosi2003} for details),  but the astration factor of D is always the same. Therefore, chemical evolution models suggest a small astration factor for D, at least in the solar vicinity. On the other hand, the D astration in the Galactic bulge should be higher than in the solar neighbourhood, because of the higher SFR that should have occurred in the bulge, as predicted by \cite{Matteucci1999} (see later). Measurements of D in the Galactic center region are from \cite{Polehampton2002}, who observed the giant molecular cloud Sagittarius B2 and observed an abundance D/H=$(0.2-11) \cdot 10^{-6}$, from \cite{Lubowich2000} and \cite{Jacq1999} who found D/H= $(1.7 \pm 0.3) \cdot 10^{-6}$, a factor of ten lower than in the local ISM.
Concerning the D abundance in the local ISM, it should be said that there is a spread relative to measurements in different lines of sight. The interpretation of such a spread is not clear and it can be related to the different degree of dust condensation for D in different spatial regions.

\begin{figure}[htbp]
  \centering
  \includegraphics[width=0.9\textwidth]{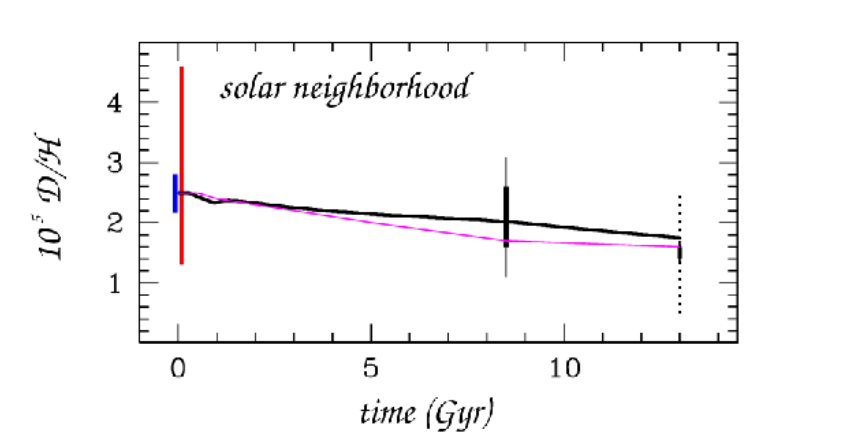}
  \caption{The temporal evolution of the abundance of D in the solar vicinity. The models are from \cite{RomanoTosi2003}. The data are from WMAP (black bar) and high redshift QSO (red bar) for the primordial D and the bars represent the uncertainty in the data. The bar at the time of the formation of the Solar System reflects the measurements of \cite{Geiss1998}, while the little bar at 13 Gyr represents ISM data from \cite{Linsky1998}. The other points at 13 Gyr reflect the measurements from Copernicus HST-GHRS, IMAPS, STIS, and FUSE (see \citealt{Vidal1998, Jenkins1999, Moos2002, Hoopes2003}),  show a large spread. Image reproduced with permission from \cite{RomanoTosi2003}, copyright by RAS.}
    \label{fig:deuterio}       % Give a unique label
\end{figure}

\subsection{The chemical evolution of $^{3}$He and $^{4}$He}

The abundance of $^{3}$He is strictly related to that of D: $^{3}$He is produced during H-burning in low and intermediate mass stars, but is also destroyed to produce heavier elements. The stellar production of $^{3}$He is still uncertain since it depends on the treatment of extra-mixing and subsequent Cool Bottom Process in low mass stars (\citet{Charbonnel2010}). From a theoretical point of view we remind a few seminal papers on the evolution of $^3$He by \citet{Galli1995}, \citet{Dearborn1996} and \citet{Lagarde2012}. From an observational point of view, it is not easy to measure the $^{3}$He abundance: \cite{Balser1997} and \cite{Balser1999} measured its abundance in Galactic planetary nebulae (PNe), and \cite{Bania2002} in Galactic HII regions. Data and model results show a light increase of the $^{3}$He abundance during Galactic lifetime.

Also $^{4}$He is produced and destroyed in stars of all masses and its abundance increases slightly in time. Starting from a primordial value of $Y_P= \sim 0.24$ , its abundance reaches $\sim$ 0.28 in the Sun and it can reach 0.35 in the Galactic bulge (\citealt{Renzini1994}), although there is no real consensus about this high value.

\begin{figure}[htbp]
  \includegraphics[width=\textwidth]{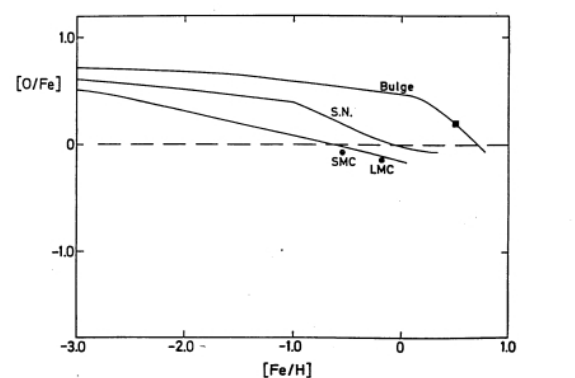}
  \caption{Predicted behaviour of the [O/Fe] ratio vs.\ [Fe/H] in three different galaxy types: the solar vicinity, the Galactic bulge and the Magellanic Clouds (LMC and SMC). The difference between the chemical models describing these galaxies is the SFR. The SFR is highest for the bulge and lowest for LMC and SMC. Reported in the figure are also some observational points relative to the bulge and SMC and LMC, available at that time. Image reproduced with permission from \cite{Matteucci1990}, copyright by AAS.}
    \label{fig:MB90}       % Give a unique label
\end{figure}

\begin{figure}[htbp]
\includegraphics[width=0.5\textwidth]{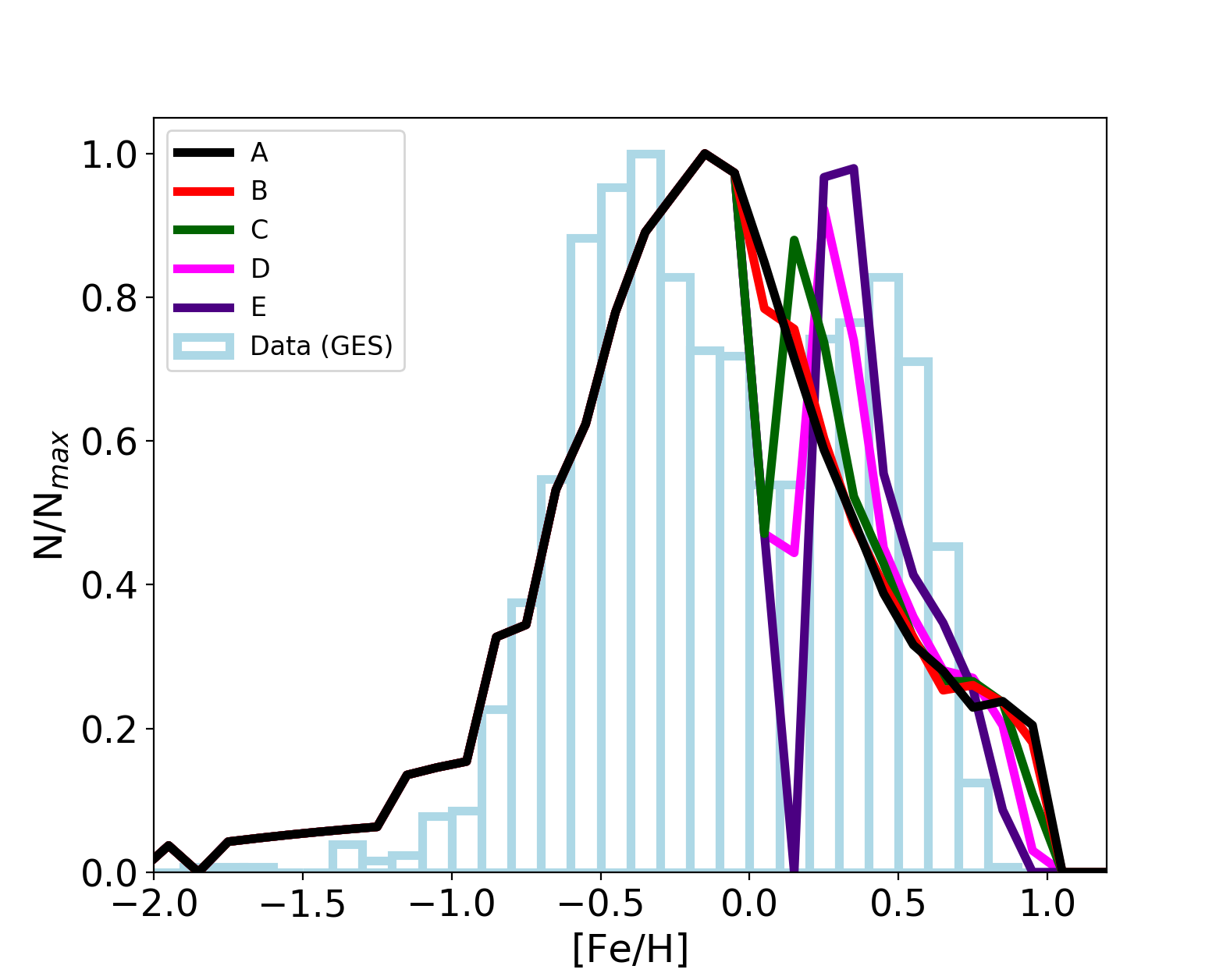}  
\includegraphics[width=0.5\textwidth]{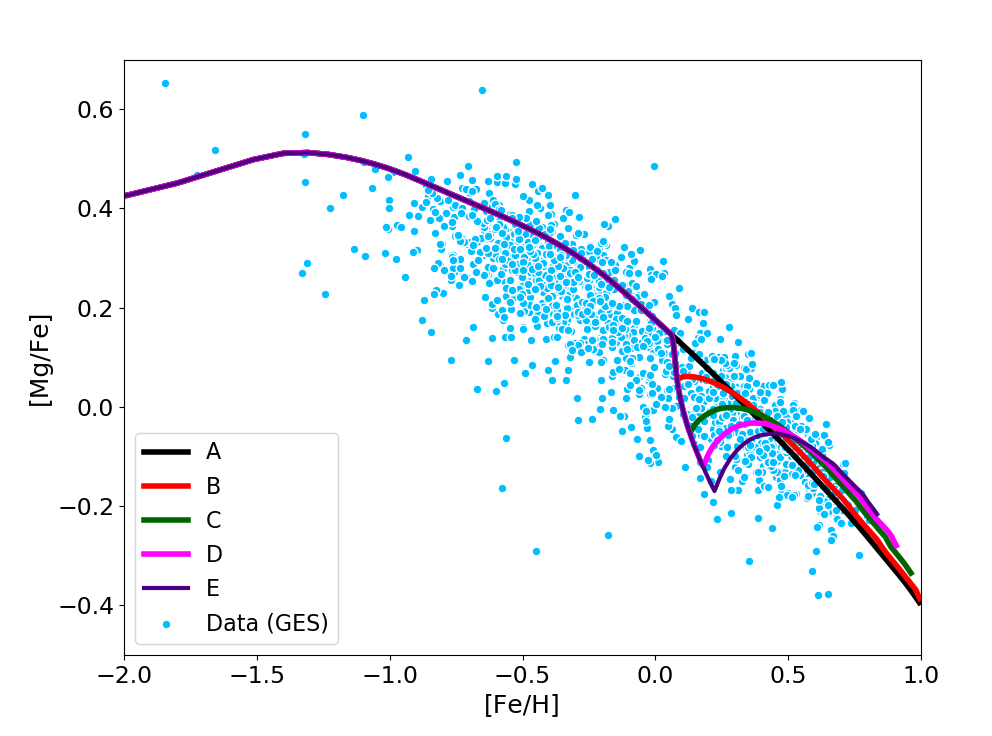}
\includegraphics[width=0.5\textwidth]{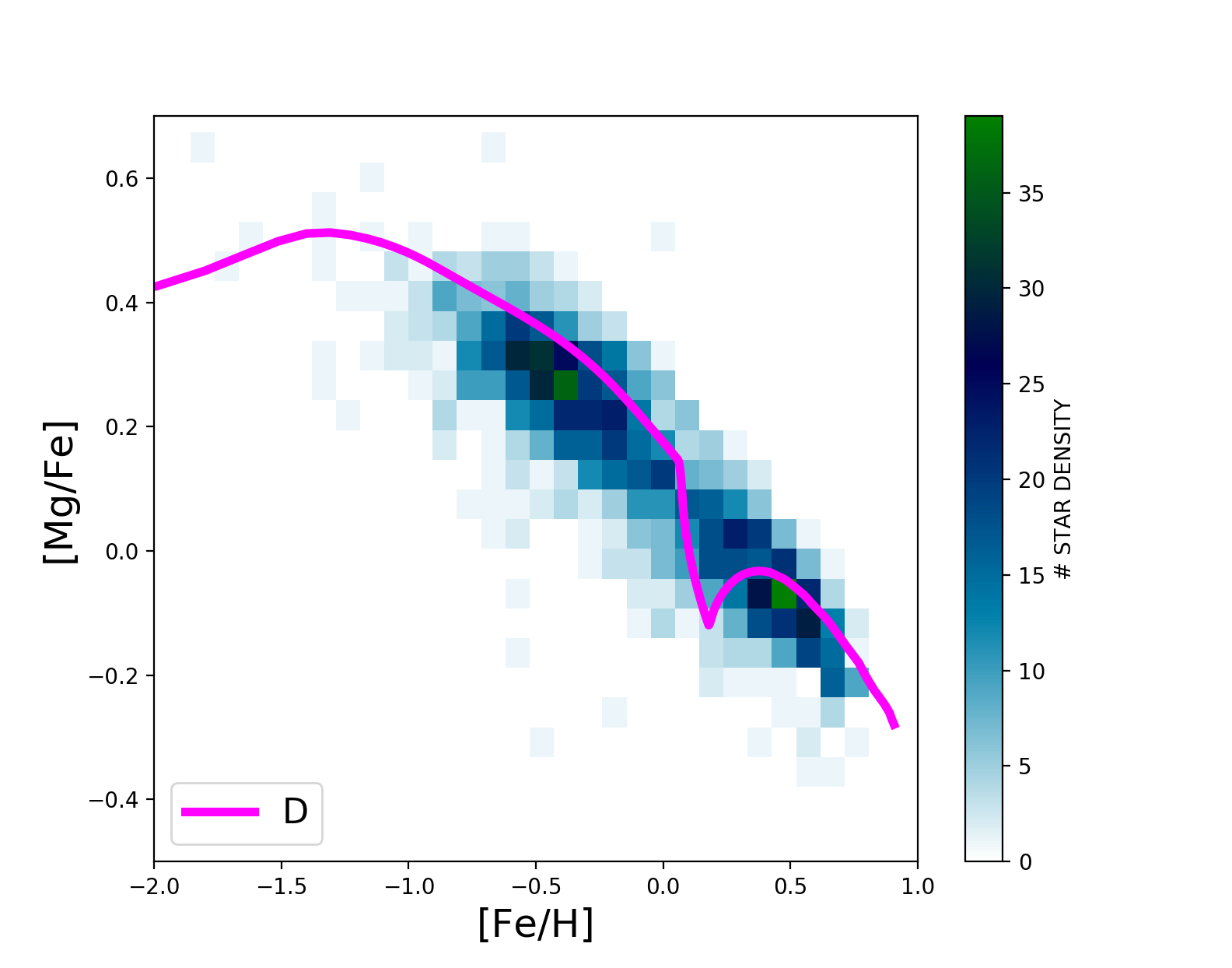}
\includegraphics[width=0.5\textwidth]{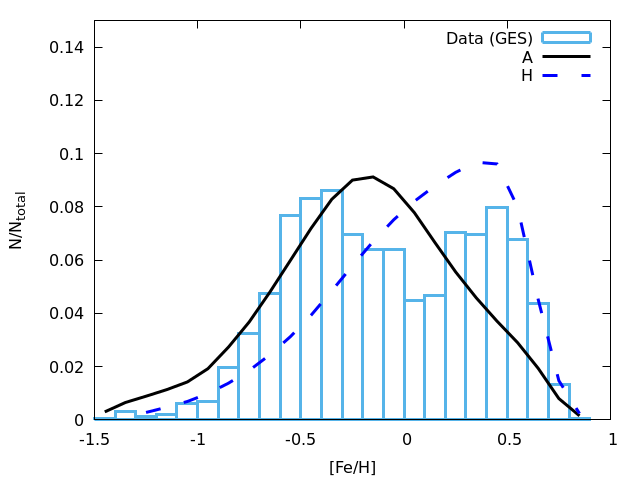}
\caption{\textit{Upper panel left}: Predicted MDF in the Galactic bulge for models with continuous star formation with stops of the duration of 50, 150, 250 and 350 Myr, compared to Gaia-ESO data. As one can see, longer is the stop in star formation and deeper is the dip between the two populations. The model which best reproduces the data is the one with a stop of 250 Myr. \textit{Upper panel right}: Predicted [Mg/Fe] vs.\ [Fe/H] in the Galactic bulge, for the same models, also compared to Gaia-ESO data.
  \textit{Lower panel left}: a density plot for the Gaia-ESO data compared to the results of the model with 250 Myr stop in star formation. \textit{Lower panel right}: predicted MDF in the case in which the second peak is due to accreted inner disk stars. The data are the same as in the upper left panel. Image reproduced with permission from \cite{Matteucci2019}, copyright by the authors.}
 \label{fig:Matt19}
\end{figure}

\section{The chemical evolution of the Galactic bulge}
\label{sec:bulge}

In this section, we will describe what we have learned up to now about the chemical evolution of the Galactic bulge and Galactic center region.
In the last years, several spectroscopic and photometric surveys have been devoted to the understanding of the formation and evolution of the bulge: in particular, Gaia-ESO (\citealt{Gilmore2012}), APOGEE (\citealt{Majewski2017}), Argos (\citealt{Freeman2013}), GIBS (\citealt{Zoccali2014}), VVVX (\citealt{Minniti2010}), (\citealt{Perryman2001}), BRAVA(\citealt{Johnson2011}) and COMBS
(\citealt{Lucey2019}). From these surveys we can obtain information on chemical abundances, kinematics and ages of bulge stars.

\cite{Wyse1992}
summarized different scenarios proposed for the bulge formation, they are: a) accretion of extant stellar systems which settled
in the center of the Galaxy by dynamical friction, with negligible star formation in situ; b) the bulge formed from gas accumulated at the center of the Galaxy, either by mergers or simply
reflecting initial conditions, evolving independently of the other Galactic components. In this case, the star formation was either rapid or slow; c) the bulge formed from gas supply from inflow of metal-rich gas from the thick disk or halo. Also in this case, the formation process could have been rapid or slow.
For a recent review on the chemodynamical history of the Galactic bulge, we address the reader to \cite{Barbuy2018}. The main conclusions of this review can be summarized as: i) the bulk of bulge stars are old and span a metallicity range $-1.5 \le {\rm [Fe/H]} \le +0.5$. The analysis of the stellar populations suggest that the overall bulge formed on a timescale $<$2 Gyr and that is barred and follows cylindrical rotation, and the more metal rich stars trace a Box/Peanut structure.

\subsection{The evolution of $\alpha$-elements in the bulge}

From a chemical evolution point of view, the best tool to derive the timescale of bulge formation is represented by the abundances and abundance ratios in bulge stars. In particular, the stellar MDF and the plots [X/Fe] vs.\ [Fe/H] can impose important constraints on the history of star formation in the bulge.
\cite{Matteucci1990} were the first trying to reproduce the MDF obtained by \cite{Rich1988}, and concluded that one should assume a fast bulge formation with intense SFR and IMF flatter than in the solar vicinity. Under these conditions, they predicted that the bulge stars should have shown a longer plateau in the [$\alpha$/Fe] vs.\ [Fe/H] plot than in the solar neighbourhood, as shown in Fig.~\ref{fig:MB90}. In other words, the knee in such a plot, due to the intervention of Type Ia SNe in producing Fe (time-delay model), should occur at a higher metallicity than for stars in the solar vicinity. The reason for that relies in the fact that a high SFR enriches the ISM very fast in Fe by means of CC-SNe, so that when the SNe Ia start occurring the Fe abundance in the ISM is already high and due only to the contribution of CC-SNe. The contrary occurs in systems with low SFR, such as irregular and dwarf spheroidal galaxies (see \citealt{Matteucci2012}).
Such a prediction for the Galactic bulge  was confirmed later by other theoretical papers (e.g., \citealt{Ballero2007, Cescutti2011, Grieco2012b}), suggesting a fast formation for the majority of bulge stars (0.3--0.5\,Gyr).
Since this is a precise prediction of the time-delay model coupled with the star formation history, if observed, it represents a confirmation of the time-delay model itself.
A different position of the knee in the [$\alpha$/Fe] ratio in the bulge stars has been observed both in Gaia-ESO (\citealt{Rojas2017}) and APOGEE (\citealt{Zasowski2019}) surveys. In both cases, the knee in the [Mg/Fe] ratio is found at $[Fe/H]_{\rm knee} =-0.37 \pm 0.09$\,dex, which is $\sim0.06$\,dex higher than the knee found in the thick disk in the solar vicinity. Although this difference is not remarkable, this is an indication that the bulge formed faster than the solar neighbourhood.
However, this knee refers to the so-called metal poor bulge, because recently two main bulge stellar populations have been found, one metal poor and one metal rich. In fact, a bimodal metallicity distribution was observed by \cite{Hill2011} and then confirmed by more recent papers (e.g., \citealt{Rojas2017, Schultheis2017, Zoccali2017}). On the other hand, \cite{Bensby2011, Bensby2013, Bensby2017} found a multi-modal distribution with the last population formed 3 Gyr ago (see also \citealt{Ness2016}). Several studies of the ages of bulge stars (\citealt{Clarkson2011, Johnson2011, Valenti2013, Renzini2018}) suggested that  the majority of them is old (age $>$ 10 Gyr). \cite{Bernard2018} inferred the history of star formation in the bulge and concluded that only 10\% of bulge stars are younger than 5\,Gyr, but this fraction raises to 20--25\% in the metal rich peak.
Chemical models should be able to reproduce the MDF as well as the [$\alpha$/Fe] vs.\ [Fe/H] plot.
In Fig.~\ref{fig:Matt19} we show the predicted and observed MDF of bulge stars from Gaia-ESO data and a good agreement is reached for a model assuming  a stop of 250\,Myr duration in the SFR. Also in this figure, the plot of [Mg/Fe] vs.\ [Fe/H] for the same model is shown: here, the discontinuity due to the stop in star formation is not so evident as in the MDF, but is probably present, as shown in the left lower panel of Fig.~\ref{fig:Matt19}, where a density plot for [Mg/Fe] is reported and a clear lower stellar density is visible in correspondance to the dip in the predicted [Mg/Fe] ratio. An alternative explanation for the two peaks in the MDF is that the more metal rich and younger population could have been accreted from the inner disk. Such a population should be younger than the one of the first peak, with lower average [$\alpha$/Fe] ratios and more metal rich. As we have already pointed out, from a kinematical point of view the metal rich population is associated with the Boxy/Peanut X-shaped bulge (\citealt{Zoccali2017}), while the more metal poor population seems to be isotropically distributed. This kinematical characteristics is consistent with either the accreted inner disk stars or with the stop in the star formation (see \citealt{Debattista2017}).
\cite{Matteucci2019} found that a good agreement with data is reached if the metal poor population formed very fast, on a timescale lower than 0.5 Gyr and with an IMF containing more massive stars (see also \citealt{Johnson2014}) than in the solar vicinity, thus confirming all the previous results. Moreover, they found that the fraction of stars younger than 5\,Gyr should be not higher than 10\%, in agreement with the observational estimate of \cite{Bernard2018}, but at variance with other suggestions (e.g., \citealt{Bensby2017, Haywood2016}).
In conclusion, chemical evolution models cannot help in understanding the dynamical evolution of the bulge but they can impose important constraints on the formation timescales of the bulge stellar populations. The suggestion of a very fast formation for the metal poor spheroidal bulge favors a formation in situ. On the other hand, the metal rich population could have been accreted from the thin or thick disk but we cannot exclude that it could have also formed ``in situ''.

\subsection{The evolution of n-capture elements in the bulge}

The n-capture elements, in particular r-and s-process elements, are characterized by a large spread in their abundances in solar vicinity stars, especially in halo stars. This spread can be explained by assuming inhomogeneous mixing, as discussed in Sect.~6.1.2. Abundances of Eu, Ba, La and Zr  have been measured also in bulge stars, as shown in Fig.~\ref{fig:Barbuy}, where we show the [X/Fe] vs.\ [Fe/H] plots for such elements. As one can see, concerning Eu, a pure r-process element, the [Eu/Fe] follows the behaviour of $\alpha$ -elements, being overabundant relative to the Sun at low metallicity and decreasing for higher metallicities, as it appears to be the average trend in solar vicinity stars. As we have already discussed, this fact suggests a fast Eu production at early times which can be obtained either if the Eu producers are MNS or CC-SNe or both. In the case of MNS, this is achieved either by assuming that all MNS have very short coalescence
times, or that the coalescence timescales are distributed in a large range of values, but the fraction of MNS is higher at early times (\citealt{Simonetti2019}).
The only difference between bulge and halo data is the smaller spread observed in the former.
For what concerns Ba, Sr, La and Zr, they are mainly produced by low mass stars  as s-process elements but also with a contribution from r-process in massive stars, although recently it has been suggested that they can originate also from s-process occurring in  fastly rotating massive stars (\citealt{Chiappini2011}). In Fig.~\ref{fig:Barbuy} we can see that Ba, La and Zr in bulge stars tend to decrease  with increasing metallicity. This is different to what happens in the solar vicinity (see Fig.~\ref{fig:Cesc08}) where [Ba/Fe] increases up to [Fe/H]$\sim -2.0$\,dex and then decreases slighlty for higher metallicity. \citet{Cescutti2018} applied the inhomogeneous model to the bulge and the results are also shown in Fig.~\ref{fig:Barbuy}. For the evolution of the abundances of Zr, La, Nd and Eu in the bulge see also \citet{Johnson2012}.

\begin{figure}[htbp]
  \includegraphics[width=\textwidth]{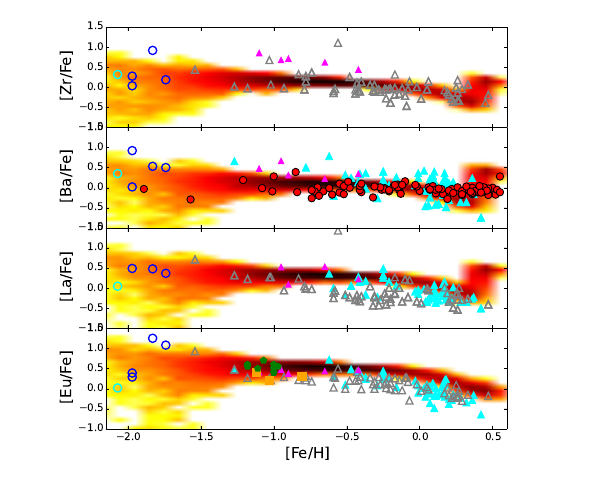}
  \caption{Observed and predicted [Zr,Ba,La,Eu/Fe] vs.\ [Fe/H] relations. The data are a compilation from \cite{Barbuy2018} and the predictions of the inhomogeneous model of \cite{Cescutti2018} are described by the orange-yellow shaded area. Image reproduced with permission from \cite{Barbuy2018}, copyright by Annual Reviews.}
    \label{fig:Barbuy}       % Give a unique label
\end{figure}

\begin{figure}[htbp]
  \includegraphics[width=\textwidth]{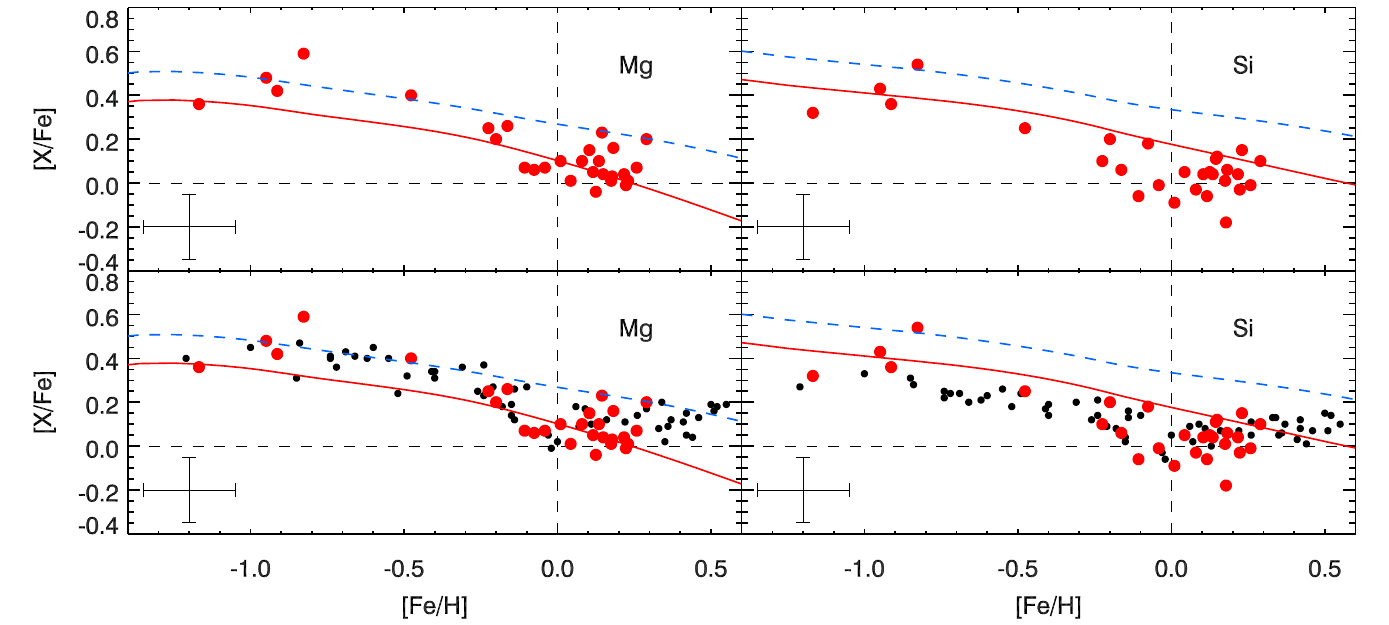}
  \caption{Predicted and observed [Mg/Fe] and [Si/Fe] vs.\  [Fe/H] for the inner 500 pc of the Galactic bulge.  The abundance ratios for 28 M giants derived by  \cite{Ryde2016} are  shown  with  red  dots.  The  predictions  are  represented  by  the  two  curves:  the  dotted curve represents a  model adopting standard yields for the Mg and Si production, whereas the continuous line is the prediction of the same model except for the yields of Mg and Si which have been decreased by a factor of 1.35 (\citealt{Grieco2015}).  In the lower panels are also reported the abundances based on micro-lensed dwarfs in the `outer' bulge by \cite{Bensby2013}, marked with black smalldots. Image reproduced with permission from \cite{Ryde2016}, copyright by AAS.}
    \label{fig:Grieco}       % Give a unique label
\end{figure}

\subsection{Chemical evolution of the Galactic centre}
In  \cite{Grieco2015} and \cite{Ryde2016}, the evolution of the central region of the Galaxy ($\sim$ 500 pc in radius) was studied by comparing chemical models to spectroscopic stellar data and estimates of the SFR.  The adopted chemical model for the Galactic centre is very similar to that of the entire bulge for what concerns the SF and infall rates. The main conclusion was that the Galactic central region should have evolved as the rest of the bulge, with intense SFR (a suggested SFR efficiency of $\sim 25 \, {\rm Gyr}^{-1}$) and therefore on a short timescale, confirming all the previous chemical works. The spectroscopic stellar data for stars in this very central region were derived from very high resolution data taken in three different fields. These studies have suggested that the stars in the central region lie in a large range of metallicities ($-1.2 < {\rm [Fe/H]} < +0.3$). They also found a lack of a [$\alpha$/Fe] gradient, similarly to what has been found in rest of the bulge. This result again points towards the conclusion that the whole bulge formed very quickly (see also \citealt{Rich2012}). Figure~\ref{fig:Grieco} shows the models for the Galactic centre region compared to the data.

\section{The chemical evolution of the Galactic halo}
\label{sec:halo}

The Galactic halo is formed by field stars and GCs and it contains only a small fraction of the total stellar mass of the Milky Way.
\cite{Hartwick1976} derived the MDF of 60 Galactic GCs, and showed that is quite different from the metallicity distribution of the G-dwarfs in the solar neighbourhood. He compared the data with the predictions of the Simple Model (see Sect.~\ref{sec:anal}) and concluded that this model predicts too few very metal poor stars. He solved this problem in two ways: i) by lowering the yield per stellar generation by a factor of 13, or ii) by gas loss, with the gas going later to form the disk. However, the first assumption resulted in a model with a very large halo mass/disk mass ratio, thus leaving only the second assumption as feasible. A recent estimate of the halo stellar mass is from \cite{Mackereth2020} and is $M_{\rm *halo}= 1.3 ^{+0.3}_{-0.2} \cdot 10^9\,M_{\odot}$.

\cite{Prantzos2003} also attempted to model the halo chemical evolution and concluded that an outflow model can well reproduce the peak and the shape of the halo MDF, but it fails in describing the lowest metallicity region, especially if I.R.A.\ is relaxed. He suggested, as a solution, an early gas infall phase for the halo. Such a conclusion was shared lately by \cite{Brusadin2013} , who suggested a halo model including infall and outflow at the same time. All of these models assumed that all halo stars formed ``in situ''. In Fig.~\ref{fig:halo} we report the results of \cite{Brusadin2013} in the framework of the two-infall model, where the halo forms by infall but without outflow and in the case where both infall and outflow are present.

\begin{figure}[htbp]
  \includegraphics[width=\textwidth]{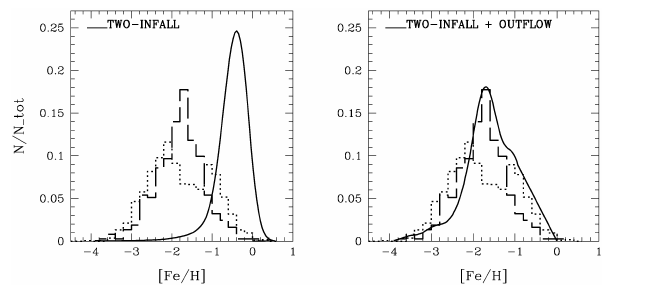}
  \caption{Stellar halo metallicity distribution function in the framework of the two-infall model (\citealt{Chiappini1997}), where no outflow is considered during halo formation (left panel, solid curve) and in the two-infall plus outflow framework (right panel, solid curve). The data are represented by the histograms, where the dashed one represents data from \cite{Ryan1991} and the dotted one the data from \cite{Schoerck2009}. The theoretical predictions have been smoothed by a Gaussian function with a variance equal to the data error of 0.2\,dex. Image reproduced with permission from \cite{Brusadin2013}, copyright by ESO. }
    \label{fig:halo}       % Give a unique label
\end{figure}

A recurrent question about the Galactic stellar halo, and not yet completely understood, is: \emph{Has the Galactic halo formed in situ or some or all stars have been accreted from dwarf satellite galaxies?}. May be the answer is that part of the halo stars formed in situ and part have been accreted (\citealt{Carollo2007, Carollo2012}). In fact, it is likely that in the past there have been many tidal interactions between the Milky Way and its satellites and a large fraction of these satellites are seen today as stellar streams or tidal debris (e.g., \citealt{Helmi1999, Simion2019}). The best diagnostic for understanding the origin of the stars in the halo is to look at their chemical abundance ratios and compare them with those observed in dwarf galaxies, which are expected to show particular abundance patterns. For example, dwarf spheroidal (dSphs) and ultra faint dwarf galaxies (UfDs) around the Milky Way show [$\alpha$/Fe] ratios at low metallicity lower than in the Galactic halo stars (see \citealt{Matteucci2012}).

\begin{figure}[htbp]
  \includegraphics[width=\textwidth]{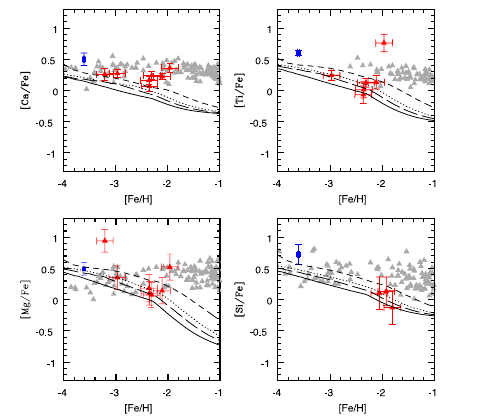}
  \caption{Comparison between [$\alpha$/Fe] ratios in the stars of Bo\"otes I (red and blue points) and halo stars (grey triangles). Overimposed are the results of models for Bo\"otes I from \cite{Vincenzo2014}. The blue square is from \cite{Norris2010} and the red triangles from \cite{Gilmore2013}. The data for halo stars are from \cite{Gratton2003, Reddy2003},\cite{Reddy2006} and
    \cite{Cayrel2004}. Image reproduced with permission from \cite{Vincenzo2014}, copyright by the authors.}
    \label{fig:ufd}       % Give a unique label
\end{figure}

It is worth noting that Nissen \& Schuster (2010) showed that the metal rich tail of the Galactic halo is made of two distinct stellar populations. They suggested that the population with a lower [$\alpha$/Fe] should have been accreted from dwarf galaxies. In Fig.~\ref{fig:ufd} we show a comparison between [$\alpha$/Fe] ratios in UfDs satellites of the Milky Way and Galactic halo stars, together with models for UfDs. It is easy to recognize that the [$\alpha$/Fe] ratios overlap at very low metallicity but many halo stars have ratios larger than those in UfD stars.
\cite{Spitoni2016}, compared the abundance patterns ([$\alpha$/Fe], [Ba/Fe] vs.\ [Fe/H]) of dSphs with those of Galactic halo stars and suggested that the majority of halo stars should have formed in situ.

Recently, it has been revealed a metal-rich component in the inner Galaxy halo, showing a peculiar elongated shape (\citealt{Belokurov2018}) like a ``sausage''. This object represents the record of a head-on major collision that the Galaxy should have experienced more than 10\,Gyr ago with a dwarf galaxy rather massive. The progenitor, now disrupted,  of this ``sausage'' is called Gaia-Enceladus or Gaia Sausage.
A sample of stars in the ``sausage'' were selected by \cite{Helmi2018}, and show a [$\alpha$/Fe] vs.\ [Fe/H] pattern similar to those of dSphs, namely they show lower [$\alpha$/Fe] ratios than metal poor halo stars of the same metallicity.
\cite{Vincenzo2019} modeled the chemical evolution of Gaia-Enceladus, assuming that this object evolved as a dwarf spheroidal galaxy with lower SFR than in the Galaxy. By fitting the [$\alpha$/Fe] vs.\ [Fe/H] relation, as well as the MDF of Gaia Sausage, they predicted a median age for its stars of $12.33^{+0.92}_{-1.36}$ Gyr and a total stellar mass at the time of merging ($\sim$ 10 Gyr ago) of $M_{\rm *Sausage}= 5 \cdot 10^{9} \,M_{\odot}$. They also suggested that the merging event might have contributed to inhibit the gas accretion onto the Galaxy, thus producing a gap in the SFR, in agreement with predictions from chemical evolution models (e.g., \citealt{Chiappini1997, Spitoni2019}). In the context of the two-infall model, it is likely that Gaia Sausage was cannibalised by the Galaxy at the end of the first infall episode.

\cite{Mackereth2020} studied the Galactic halo in the space of [Fe/H], [Mg/Fe] and $e$ (orbital eccentricity) by means of red giant counts from APOGEE DR14, and concluded that the majority of the total stellar halo mass was accreted,  and that $\sim$ 30--50\% of this accreted mass belonged to Gaia Enceladus, whose mass was then estimated to be $M_{\rm *Sausage}= 3 \pm1 \cdot 10^{8} \,M_{\odot}$, lower than previous estimates.
On the other hand, \cite{Iorio2020} presented results about a large sample of RRLyrae detected by Gaia. Their chemo-kinematics analysis suggested  that the inner halo (inside 10 Kpc) likely contains RRLyrae formed in situ, while other RRLyrae out to 30 kpc are consistent with disk kinematics and are young and metal rich. In our opinion, a good criterion to establish if the halo stars were accreted is represented by low [$\alpha$/Fe] ratios at low [Fe/H] (see Fig.~\ref{fig:ufd}), and even better by [Ba/Fe] at low [Fe/H], as suggested by \cite{Spitoni2016}.

Finally, concerning the possibility of inhomogeneous chemical evolution of the Galactic halo, we address the reader at Fig.~\ref{fig:Cesc15}, where we already discussed the spread observed in the abundances of neutron -process elements and how it can be reproduced by assuming inhomogeneous chemical evolution.

\section{Chemo-dynamical models in cosmological context}
\label{sec:cosmo}

Up to now we have mostly discussed how to model the chemical evolution of the Milky Way by means of the so-called ``analytical chemical models'', although only a small fraction of them is really analytical, being the largest fraction constituted by numerical models. These models, that we will call \emph{pure chemical models}, both analytical and numerical, although possessing a nice predictive power, do not assume any cosmological paradigm for galaxy formation. The need for a cosmological framework for galaxy formation has led to the adoption of more complex chemo-dynamical models for the  formation and evolution of galaxies.
As we will see, in the majority of cases the conclusions of the latter models confirm what found by pure chemical models but adding information on the stellar and gas kinematics.
The cosmological galaxy formation models belong to two categories: i) semi-analytical models, and ii) numerical simulations. The cosmological framework is always the $\Lambda$CDM one (a cold dark matter Universe with a cosmological constant), which assumes a hierarchical formation of galaxies, a scenario where the largest galaxies form by accretion of smaller ones. 

The basic ingredients of these models are:
\begin{itemize}
\item the cosmological model;
\item the dark matter haloes: in particular it should be given the abundance of haloes of different mass, the formation history of each halo (the merger tree) and the internal structure of each halo, in terms of radial density and angular momentum.
  \end{itemize}
Then, other more complicated ingredients which are related to the physics of the gas, and they are far more uncertain than the treatment of gravitational instability, should be added. These processes are: star formation, gas cooling, stellar feedback, chemical evolution, gas and stellar dynamics, galaxy mergers.
These models followed the successful study of the evolution of dark matter in the $\Lambda$CDM scenario, with the aim of understanding also the evolution of the baryonic component, which presents a higher complexity due to the many poorly understood physical processes involved. Among the first semi-analytical approaches to galaxy formation and evolution are those of \cite{White1991, Kauffmann1993} and \cite{Cole1994}; they combine the growth of dark matter haloes with simple parametrizations of the physics of baryons. The other common approach, which is complementary to the semi-analytical one, is to adopt hydrodynamical simulations, and models of this type have been constructed either for isolated or populations of galaxies (e.g., \citealt{Katz1992, Mihos1994, Navarro1994, Steinmetz1994, Cen1999, Springel2003, Kawata2003a, Kawata2003b, Nakasato2003}).  
It should be said that initially, given the complexity of these models, the chemical evolution has often been treated in a simplistic way, by adopting for example I.R.A., that allows us to follow only the chemical enrichment from CC-SNe, and/or adopting the closed-box model assumption. \cite{Thomas1999} and \cite{ThomasKauf1999} were among the first who included enrichment by Type Ia SNe in semi-analytical models but in the framework of the closed-box model (no infall or outflow). In the following years, detailed chemical evolution was included in hydrodynamical simulations (e.g., \citealt{Kobayashi2004, Kobayashi2007, Tornatore2007, Jimenez2015}) and in semi-analytical models studying the properties of elliptical galaxies (e.g., \citealt{Nagashima2005, Arrigoni2010, Calura2011}). In most of these papers, stellar lifetimes and detailed chemical enrichment from Type Ia SNe were considered, following the basic approach of \cite{Greggio1983, MatteucciGreggio1986} and \cite{Chiappini1997}.

\subsection{The Milky Way in a cosmological context}
Here we are interested in modelling the Milky Way galaxy and therefore we will present only models whose results can be compared to our Galaxy. In the last few years, several papers dealing with chemical evolution of a Milky Way like galaxy in a cosmological context appeared in the literature (\citealt{Colavitti2008, Kobayashi2011, Brook2012, Few2014, Loebman2016, Grand2018, Mackereth2018, Haynes2019, Vincenzo2020, Clarke2019, Buck2020, Vincenzo2020}), but we will discuss only some of the results obtained by means of the above cosmological simulations.\par
\cite{Colavitti2008} aimed at finding a cosmologically motivated gas infall law for the formation of the Milky Way, since in pure chemical models such a law is assumed a priori as an exponential or gaussian function. They assumed that baryons assemble like dark matter and selected an infall law compatible with the formation of a disk galaxy like the Milky Way. The selection occurred by means of cosmological N-body simulations adopting the public code GADGET2 (\citealt{Springel2005}). Not without surprise, they concluded that the best infall law is very similar to that of the two-infall model (see Eq.~\ref{eq:twoinfall}) of \cite{Chiappini1997}. In Fig.~\ref{fig:Cola} we show their infall laws at different Galactocentric distances: as one can see, the law presents two peaks occurring at $\sim$ 1 and $\sim$ 4 Gyr and this clearly resembles to the two major infall episodes predicted by the two-infall model. Clearly, the abundance patterns for the solar vicinity, as predicted by this cosmological infall law, are very similar to the results of the two-infall model, including the gap in star formation between the two peaks. It should be noted that in this formulation no timescale for gas accretion are given a priori, as it is instead done in pure chemical models.

\begin{figure}[htbp]
  \includegraphics[width=\textwidth]{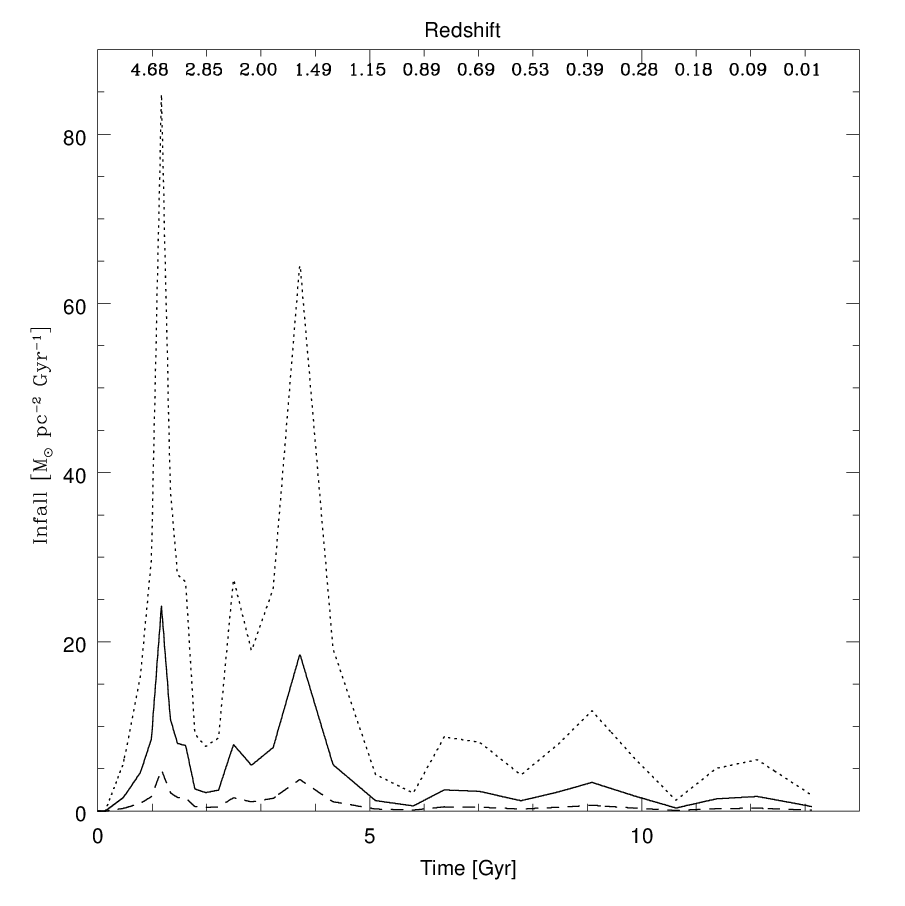}
  \caption{A cosmologically motivated gas infall law for the formation of the Milky Way. The continuous line refers to a Galactocentric distance  of 8 kpc (solar neighbourhood), the dotted line to 4 kpc and the dashed line to 14 kpc. The redshift is also indicated in the figure. Image reproduced with permission from \cite{Colavitti2008}, copyright by ESO.}
\label{fig:Cola}
\end{figure}

\cite{Kobayashi2011} performed detailed chemo-dynamical simulations of a Milky-Way like galaxy, and starting from cold dark matter conditions they included supernova (CC-SNe and Type Ia SNe) feedback as well as chemical enrichment, and  followed the evolution of the gas abundances of several elements from O to Zn.
In this kind of model, the star formation history goes like this: the CDM initial fluctuations grow into nodes and filaments and small collapsed haloes of gas and dark matter are created. Inside the haloes the gas cools radiatively and star formation takes place starting from redshift $z\sim 15$. According to the hierarchical formation of dark haloes, subgalaxies merge to form larger galaxies and the merging induces star formation. In this picture, the Galactic bulge forms during the initial starburst which is triggered by the assembly of gas-rich sub-galaxies with masses in the range $\sim (5-10) \cdot 10^{9} \,M_{\odot}$ at redshift $z\ge 3$. Due to its angular momentum, the gas then accretes onto the plane to form a disk which grows inside-out, as suggested by the chemical evolution models of the previous paragraphs (e.g., \citealt{Larson1976, Matteucci1989, Chiappini2001}). In this disk, the star formation has a longer timescale than the bulge, thanks to the self-regulation of supernova feedback.  Many satellites are successively accreted and disrupted, but there is no major merger event after $z\sim 2$, so the disk structure is retained. The assumed SN Ia progenitor model is the single degenerate one, as presented in \cite{Kobayashi2009}.  
They compared the chemical and kinematical results for bulge, disk and halo stars with observations: the [X/Fe] vs.\ [Fe/H] plots were predicted for several chemical species in the solar neighbourhood and in the bulge, as well as the MDFs for the thin disk at the solar ring, the thick disk and the halo. In Fig.~\ref{fig:Koba11}, we report the predicted [X/Fe] vs.\ [Fe/H] relations for the solar neighbourhood at $z=0$. This plot shows the effect of the time-delay model discussed before, since the SN progenitors, stellar lifetimes and stellar yields are the same as in the pure chemical evolution models. Therefore, the physical interpretation of the behaviours of the [X/Fe] vs.\ [Fe/H] diagrams is always the same, and is due to the different contributions to chemical enrichment by SNe  with different stellar lifetimes. What is different from pure chemical evolution models, is that the evolution of the ISM is not homogeneous and some spread in the abundance ratios is naturally predicted, as evident in Fig.~\ref{fig:Koba11}.

\begin{figure}[htbp]
  \includegraphics[width=\textwidth]{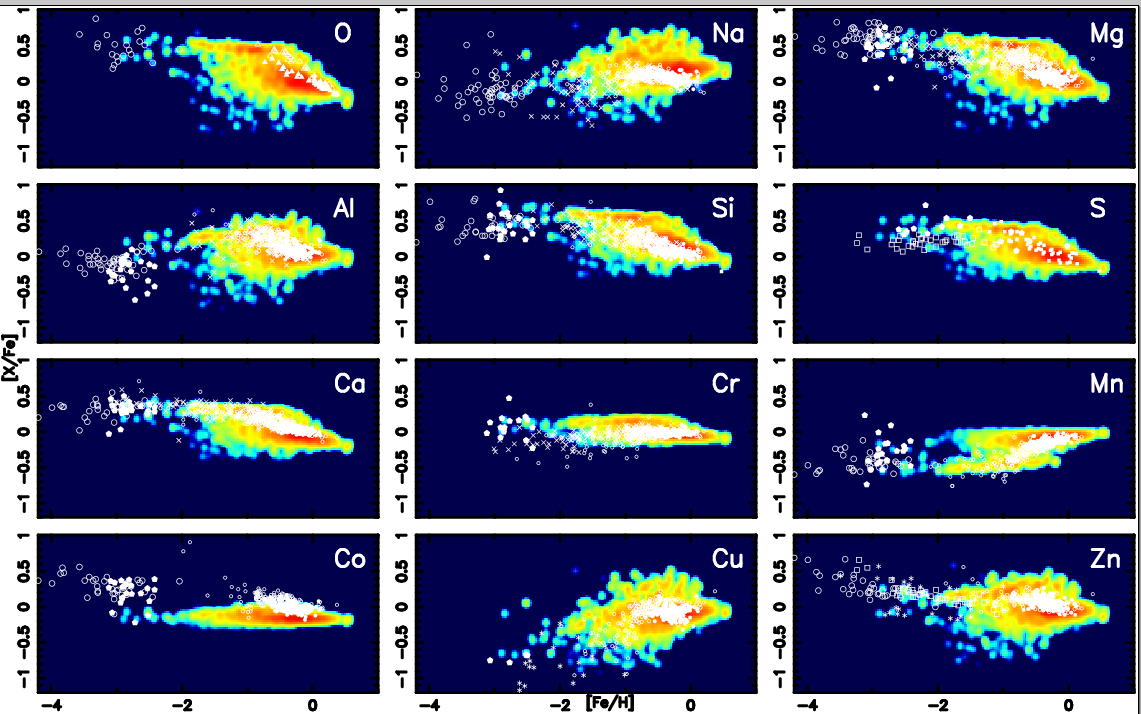}
  \caption{Predicted and observed [X/Fe] vs.\ [Fe/H] relations in the solar neighbourhood. The [Fe/H] range includes halo, thick and thin disk stars. The data are indicated with white dots, large open circles, filled pentagons, crosses, small filled and small open circles, filled triangles, open triangles, filled squares, filled pentagons, open squares and asterisks. The contours show the predicted frequency distribution of stars in the simulated Milky Way-like galaxy, and red represents the highest frequency. It is worth noting that here a bimodality is predicted for all the studied elements. See data references in \cite{Kobayashi2011}. Image reproduced with permission from \cite{Kobayashi2011}, copyright by AAS.}
  \label{fig:Koba11}
\end{figure}

%Brook et al. (2012) also performed  a  cosmological  hydrodynamical  simulation
%aimed at forming a disk galaxy with sub-component which could be assigned to a thin stellar disk, thick disk and low mass stellar halo. They found  thin- and
%thick-disk populations distinguished by their ages, kinematics and metallicities: i) thin disk stars are young ($<$ 6.6 Gyr), possess low velocity dispersion, high [Fe/H] and low [O/Fe] ratios, ii) the thick disk stars are old (6.6$<$age$<$9.8 Gyr), possess higher velocity dispersion and have relatively low [Fe/H] and high [O/Fe] ratios. iii) The halo component includes less than 4\% of stars in the solar ring, has low metallicity, a velocity ellipsoid and is formed primarily ``in situ'' during an early merger epoch. They concluded that gas rich mergers during this epoch played a major role in fuelling the formation of the thick disk stars. Their finding confirms that cold gas accretion should be the major source of the baryons lying in a disk, in agreement with suggestions of simpler chemical evolutionary models.

\begin{figure}[htbp]
  \centering
  \includegraphics[width=0.9\textwidth]{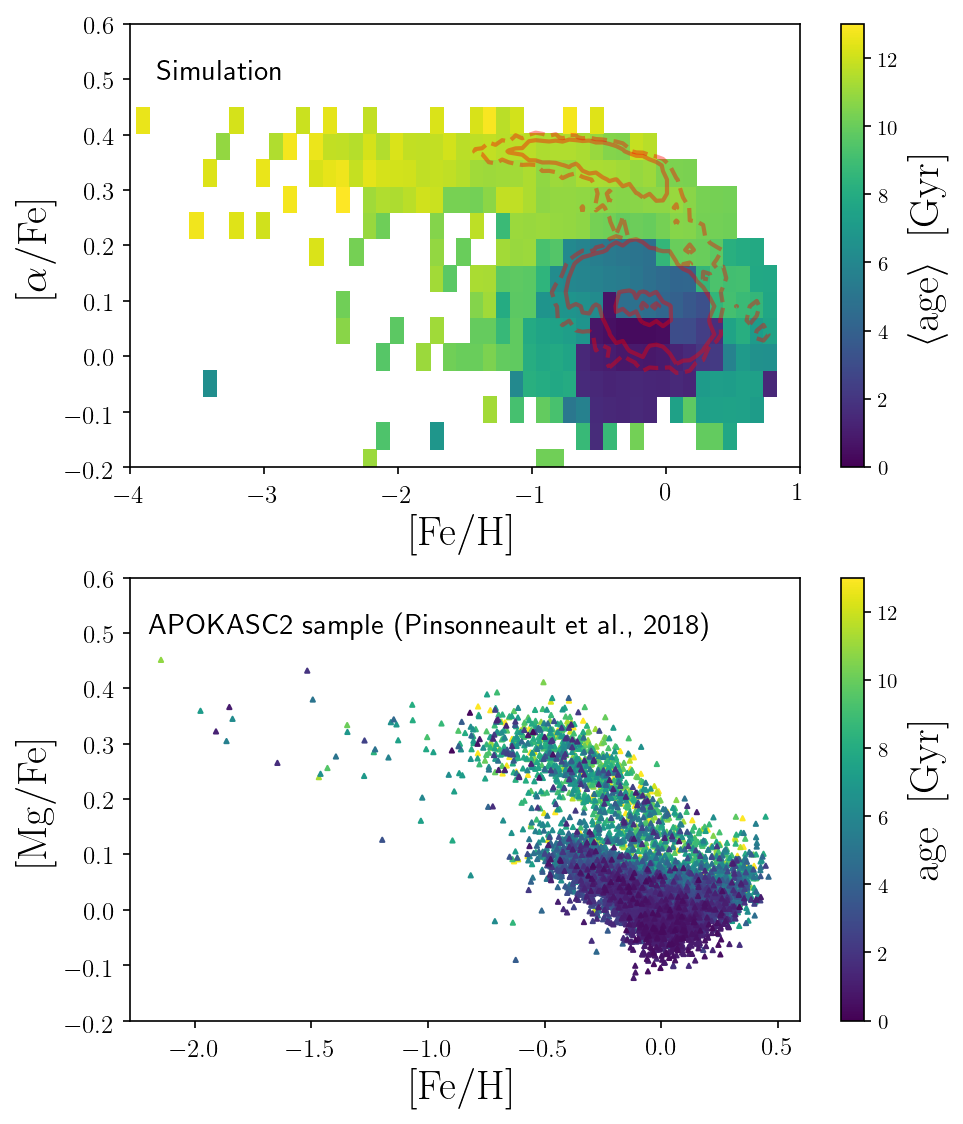}
  \caption{{\it Top panel}: predicted [$\alpha$/Fe] vs.\ [Fe/H] together with the age distribution in the simulated galaxy at the present time.
    {\it Bottom panel}: The observed age-distribution of the stars from the second APOKASC catalog (\citealt{Pinsonneault2018}) in the [$\alpha$/Fe] vs.\ [Fe/H] diagram. Image reproduced with permission from \cite{Vincenzo2020}, copyright by the authors.}
    \label{fig:Vinkoba}       % Give a unique label
\end{figure}

Several cosmological simulations of the Milky Way in the last few years, were aimed at explaining the apparent bimodal distribution of the [$\alpha$/Fe] ratios in disk stars, as already discussed previously (see Fig.~\ref{fig:Spito19}). \cite{Mackereth2018}, by means of EAGLE simulation, analysed this distribution for 133 Milky Way-like galaxies. They found that the bimodality occurring in the Milky Way is rare, since it appears only in $\sim 5\%$ of their simulated galaxies. They suggested this as the consequence of an early gas accretion episode occurring at an atypically-rapid growth. Their interpretation of the bimodal sequence is related to two different infall episodes, similar to the two-infall model of \cite{Chiappini1997} and \cite{Spitoni2019}. Clearly, the first infall episode has to be faster to ensure rapid star formation avoiding substantial pollution from Type Ia SNe, and thus creating stars with high [$\alpha$/Fe] ratios.  However,  it is worth noting that the Type Ia SN rate adopted in \cite{Mackereth2018} is quite different from those described in this review, because the majority of SNe Ia occurr almost istantaneously, thus loosing the effect of the time-delay model. Other differences are present between the two approaches (cosmological and pure chemical evolution), for example the cosmological model predicts a continuous increase of metallicity over time rather than tending to an equilibrium value (see also \citealt{Grand2018}). However, a common interpretation of the [$\alpha$/Fe] bimodality among pure chemical models and cosmological ones is related to the time-delay model for chemical enrichment, as well as the two distinct infall episodes and a hiatus in the star formation between them. We remind here that \cite{Grisoni2017} suggested also another interpretation of the bimodality, as due to two parallel and distinct episodes of gas infall, not separated by a stop in the star formation. Another common feature to all models of the Milky Way, both cosmological and purely chemical, is the inside-out formation of the thin disk. \cite{Vincenzo2020} also reproduced the bimodality in the [$\alpha$/Fe] ratios, as shown by the APOGEE-DR16 data, as well as the stellar age distribution from APOKASC-2. They also confirmed the inside-out formation of the thin disk and studied stellar migration, as already discussed (see Fig.~\ref{fig:migrVinc}). Their chemo-dynamical simulations is based on Gadget-2 (\citealt{Springel2005}) but include many relevant processes, such as radiative cooling depending on metallicity and [$\alpha$/Fe] ratios, star formation, thermal feedback from stellar winds and SNe and detailed nucleosynthesis (elements up to Zn) from Type Ia SNe and CC-SNe. \cite{Haynes2019} treated separately the evolution of elements derived from neutron-capture elements. In Fig.~\ref{fig:Vinkoba}, we show some results from \cite{Vincenzo2020}, where the bimodality in [$\alpha$/Fe] ratios is reproduced together with the stellar ages; the bimodality in their model is attributed to both infall and ouflow events that they find during the evolution of the Milky Way. Finally, as already mentioned, other cosmological simulations (e.g., \citealt{Buck2020, Sharma2020}), as well as chemical models (\citealt{Anders2017}) have attributed the bimodality to stellar migration.

In conclusion, models of the Milky Way in a cosmological context have in general confirmed the suggestions of pure chemical models and in particular the timescales for the formation of the various Galactic components, the time-delay model, the interpretation of [$\alpha$/Fe] bimodality in disk stars and the inside-out formation of the thin disk.

\section{Discussion and conclusions}
\label{sec:concl}

In this review we have described in detail the chemical modeling of the Milky Way, divided in its main components: stellar halo, thick and thin disks and bulge.  Particular attention has been devoted to the evolution of single chemical elements from H to heavy elements, and on how to impose constraints on the formation and evolution of our Galaxy by comparison theory-observations.
The detailed chemical evolution of the Milky Way has been studied in the past years mainly by means of pure chemical models, either analytical or numerical. Moreover, in recent times several attempts to model the Milky Way in a cosmological context have appeared, by means of either semi-analytical models or hydrodynamical numerical simulations. We have shown the predictive power of all of these models
as well as indicated the many uncertainties still present in modeling the Milky Way.
In the following, we will summarize the most important still open questions and how the results achieved by models have contributed to a better understanding of the formation and evolution of the Milky Way.

The most important questions are:

\begin{itemize}
  
\item {\it Did the halo stars form in situ or were they  accreted?}

  This is not yet clear, although it seems that at least a fraction of halo
  stars have been accreted, since they show lower [$\alpha$/Fe] ratios. In fact,
  the best tool
  to ascertain this point is represented by the abundance ratios versus
  metallicity relations. A recent
  important discovery has been the realization that 10 Gyr ago a massive dwarf
  galaxy, called Gaia-Enceladus or Gaia-Sausage, has fallen into the potential well of our
  Galaxy. The stars of this object might represent a large fraction of the
  stars accreted by the Galactic halo.

\item {\it How should we explain the large spread observed in some abundance
  ratios in halo stars?}

  The abundances and abundance ratios of neutron-process elements show a
  particularly large spread observed in halo stars relative to other elements,
  such as $\alpha$- and Fe-peak elements. The first explanation for
  the spread is to assume inhomogeneous evolution of the halo, but then the
  spread should be visible in all the elements. A tentative explanation was
  given by \cite{Cescutti2008}, who suggested that the different extent of
  the spread in the plot abundance ratio vs.\ metallicity is due to the
  different nucleosynthesis and stellar progenitors of different elements,
  coupled with inhomogeneous mixing.

\item {\it How did the two disks form? Thick disk formed fast whereas the thin
  formed slowly? Did the thin disk form inside-out?}

  The two Galactic disks could have formed in a sequential way but
  with a halt in the star formation between the two, and by means of different
  gas infall events (e.g., \citealt{Chiappini1997}). Alternatively, the two disks could have formed in
  parallel,  out of two independent infall events but occurring at the same
  time and at different rates (e.g., \citealt{Grisoni2017}). In both
  scenarios, the rate of chemical evolution must have been different in the two
  disks, with the thick disk evolving faster than the thin disk.
  Most of chemical models as well as chemo-dynamical cosmological simulations agree that
  the thin disk should have formed on a longer
  time scale than the thick disk and the halo. Late time major or minor mergers for the
  formation of the thin disk seem to be excluded since they would have
  cancelled the abundance gradients. On the other hand, abundance gradients
  are favored by an inside-out formation of the thin disk, although other
  processes such as inward gas flows and decreasing efficiency of star
  formation with the Galactocentric distance should also be present
  (e.g., \citealt{Palla2020, Spitoni2011}).

\item {\it What is the cause of bimodality in the [$\alpha$/Fe] ratios in the
  thick and thin disks, if real?}

  The cause of bimodality, if confirmed to be real, is certainly a consequence of the mechanisms and timescales for the formation of the two Galactic disks. The bimodality shows that thick disk stars have larger [$\alpha$/Fe] ratios than the thin disk stars and part of them lie in the same [Fe/H] range of thin disk stars, so that these abundance ratios appear in two sequences. The sequences look parallel in Gaia-ESO and AMBRE data, whereas in APOGEE data the low [$\alpha$/Fe] sequence appears rather as a plateau. This bimodal effect is indeed interesting, and it seems to be a common characteristic up to large Galactocentric distances (\citealt{Queiroz2020}). In any case, many have been the explanations suggested for the bimodality. Several authors suggested that a framework like that of the two-infall model can explain the bimodality, since there is a stop in the star formation between the formation of the two disks with consequent dilution and decrease of the Fe abundance. This effect had been found first by \cite{Chiappini1997}, who showed that a gap in star formation of less than 1 Gyr was expected because of the second infall coupled with a threshold gas density for the
 star formation. In order to explain APOGEE data, \cite{Spitoni2019} proposed a longer gap of $\sim$ 4.3 Gyr, while \cite{Grisoni2017} proposed a parallel disk formation to explain the data of AMBRE survey. Other authors (e.g., \citealt{Anders2017, Kubryk2015, Buck2020, Sharma2020}) have suggested that the bimodality is due to stellar migration. Also a late infall event occurring in the thin disk has been suggested (e.g., \citealt{Calura2009}),
  as well as the possibility that Gaia Enceladus can have influenced the evolution of the thick disk (\citealt{Grand2020}).

\item {\it How important is stellar migration?}

  Stellar migration seems indeed to exist and most of the studies suggest that it should occur mainly from the inner to the outer Galactic thin disk regions. However, there is not a general agreement on how really important is stellar migration.  It has been invoked to solve several problems including the observed spread in the abundance patterns observed in the solar neighbourhood, the [$\alpha$/Fe] bimodality as well as the existence of the thick disk itself. On the other hand, models without stellar migration can still reproduce the majority of the observed features in the solar vicinity, except for the presence of stars with super solar metallicity, for which a 10--20\% of migrated stars could be enough (see \citealt{Spitoni2015}).
  Anyway, the exact  amount of solar vicinity stars which have migrated from other regions is still difficult to establish.

\item {\it How did the bulge form? How many different stellar populations are in the bulge?}

Most of the chemical studies, including chemo-dynamical cosmological simulations, relative to the Galactic bulge have suggested that
 it formed quickly, as a consequence of a strong burst of star formation lasting less than 1 Gyr. With high star formation efficiency, short infall timescale and an IMF with more massive stars than in the solar neighbourhood, it is possible to reproduce the MDF and the [X/Fe] vs.\ [Fe/H] relations for a large part of bulge stars, as first shown by \cite{Matteucci1990}. However, there is a fraction of bulge stars which are more metal rich and associated to the Boxy/Peanut X-shaped bulge (\citealt{Zoccali2017}), and might have been accreted from the inner disk. The true bulge stars seem to be old and  the fraction of stars younger than 5 Gyr to be no more than 10\%  (e.g., \citealt{Bernard2018}, although other studies have suggested a larger fraction of young stars, such as \citealt{Bensby2017}).

\item {\it How did abundance gradients along the thin disk formed? Which is the role of radial gas flows?}

  Abundance gradients are present along the thin disk: they have been derived from young stars, PNe and HII regions. The abundance gradients generally indicate the present time abundances along the disk, except perhaps for data of PNe which can refer also to older objects.  Chemical evolution models predict abundance gradients if there is a gradient in the SFR along the disk. This can be obtained in several ways: i) by assuming an inside-out formation of the disk by means of gas infall, with the infall timescale increasing with Galactocentric distance (e.g., \citealt{Matteucci1989, Chiappini2001}, ii) with an efficiency of star formation decreasing with Galactocentric distance (e.g., \citealt{Prantzos2000, Colavitti2008}), iii) by assuming a gas threshold for star formation (\citealt{Chiappini2001}), iv) by varying the IMF with Galactocentric distance, although this variation should imply a smaller number of massive stars at larger Galactocentric distances, at variance with the \cite{Jeans1902} criterion for star formation; v) by assuming radial gas flows (e.g., \citealt{Lacey1985, Portinari2000, Schonrich2009, Spitoni2011}). All of these processes can be at work at the same time but while we could avoid ii), iii) and iv), we cannot exclude i) and v).
  In fact, radial gas flows and inside-out formation seem to be unavoidable physical processes: in particular, the inward radial gas flows have a strong effect on the formation of abundance gradients and they are the natural consequence of gas infall.  The inside-out process derives from a faster accretion in the inner denser disk regions relative to the less dense outer ones (see \citealt{Larson1976}).
In addition, most of the chemo-dynamical cosmological simulation found that the disk of a Milky Way-like spiral forms inside-out.

  Another still open question is whether the abundance gradients have steepened or flattened in time, depending mostly on the assumptions on the SFR. For example, by assuming a constant star formation efficiency leads to a steepening of the gradients in time while a variable efficiency as a function of Galactic radius induces a flattening of gradients with time. This problem will be solved only when we will have more data on gradients shown by old stars.

\item {\it Which stars are the main contributors of r-process elements? Merging
  neutron stars or supernovae or both?}

  The r-process elements, such as Eu, are mainly formed in massive stars either by means of explosive nucleosynthesis during the CC-SN events and/or by merging neutron stars. The second channel seems to be favored, as shown by the heavy elements which arose from the merging neutron star event associated to the detection of GW170817 (\citealt{Abbott2017}). Chemical evolution studies have explored the possibility that Eu can be produced only by merging events of compact objects (neutron stars and black holes) as well as by CC-SNe or both. In fact, CC-SNe alone seem to be not able to produce the right quantity of Eu to reproduce its solar abundance. The plot of [Eu/Fe] vs.\ [Fe/H] clearly shows that Eu behaves like an $\alpha$-element, namely it shows a plateau of [Eu/Fe] at low metallicity followed by a decrease of this ratio at higher metallicities. This behaviour is interpreted by means of the time-delay model, in which elements formed on short timescales (by massive stars), such as $\alpha$-elements, at low metallicity show an overabundance relative to Fe, which is mainly formed and ejected by Type Ia SNe on longer timescales.
  Therefore, from a chemical point of view, the mergers of compact objects can be the main source of Eu, if the merging timescale is very short. \cite{Matteucci2014} suggested that all systems should merge in a time of 1 Myr, otherwise one should assume a distribution of merging times together with  CC-SNe also contributing to Eu. This conclusion was shared by papers in the following years (e.g., \citealt{Cescutti2015, Cote2018, Wehmeyer2019, Kobayashi2020}), unless one assumes that the fraction of merging systems should have been higher at early times (e.g., \citealt{Simonetti2019, Beniamini2019}).

\item {\it Why the origin of some elements is not yet understood? }
  
When we compare the predicted [X/Fe] vs.\ [Fe/H] relations with observations in the solar vicinity and bulge, we find that for some elements there is no agreement and these elements mainly belong to the Fe-peak group. Some of them do not behave like Fe, as one would expect on the basis of the time-delay model, and this fact suggests that we have not yet understood the nucleosynthesis of such elements, which are $^{51}$V, $^{52}$Cr, $^{55}$Mn, $^{59}$Co and $^{64}$Zn. Other elements, which do not belong to the Fe-peak element group, and whose nucleosynthesis is not yet well understood are $^{41}$K and $^{48}$Ti. Generally, in chemical evolution models, the yields of the above mentioned species need to be changed (increased or decreased) relative to standard nucleosynthesis calculations, in order to reproduce their abundance patterns (e.g., \citealt{Francois2004, Matteucci2020}). Future improvements in stellar  and nucleosynthesis models are expected to shed light on the origin of these elements.
  \end{itemize}

\begin{acknowledgements}
I would like to thank several of my past and present collaborators, whose work represents an important part of this review. They are: F. Calura, G. Cescutti, C. Chiappini, V. Grisoni, M. Molero, M. Palla, S. Recchi, D. Romano, P. Simonetti, E. Spitoni and F. Vincenzo. I am very grateful to the referees, Monica Tosi and Chiaki Kobayashi, for carefully reading the paper and giving me many useful suggestions.

\end{acknowledgements}

% BibTeX users please use one of
\bibliographystyle{spbasic-FS-etal}      % basic style, author-year citations
\bibliography{bibliography}   % name your BibTeX data base

\end{document}